\documentclass[fleqn,usenatbib]{mnras}

\usepackage{newtxtext,newtxmath}

\usepackage[T1]{fontenc}
\usepackage{caption}
\usepackage{subcaption}
\usepackage{enumitem}

\definecolor{amaranth}{rgb}{0.9, 0.17, 0.31}

\usepackage{xcolor}

\setlist[itemize]{align=parleft,left=0pt}
\DeclareRobustCommand{\VAN}[3]{#2}
\let\VANthebibliography\thebibliography
\def\thebibliography{\DeclareRobustCommand{\VAN}[3]{##3}\VANthebibliography}

\usepackage{graphicx}	
\usepackage{amsmath}	
\usepackage{caption}

\usepackage{orcidlink}
\usepackage{comment}
\usepackage{threeparttable}
\usepackage{lineno}
\newcommand{\angstrom}{\text{\normalfont\AA}}

\title[Benchmark Brown Dwarfs as Chemical Laboratories]{Benchmark Brown Dwarfs as Chemical Laboratories: Linking System Bulk Properties to Atmospheric Retrievals}
\author[V. Kecskem\'ethy et al.]{Vikt\'oria Kecskem\'ethy\orcidlink{0000-0002-9511-0901},$^{1}$\thanks{E-mail: v.kecskemethy@herts.ac.uk}
Ben Burningham,$^{1}$ Fei Wang,$^{1}$ Austin J. Rothermich,$^{2,3,4,5}$\newauthor
Jacqueline K. Faherty,$^{2,3}$ Genaro Su\'arez\orcidlink{0000-0002-2011-4924},$^{2}$ Caprice L. Phillips\orcidlink{0000-0001-5610-5328}\thanks{NASA Sagan Fellow},$^{6}$,
Melanie Rowland\orcidlink{0000-0003-4225-6314},$^{7}$\newauthor Channon Visscher\orcidlink{0000-0001-6627-6067},$^{8,9}$
Johanna M. Vos\orcidlink{0000-0003-0489-1528},$^{10,2}$ and Daniella C. Bardalez Gagliuffi\orcidlink{0000-0001-8170-7072}$^{11}$
\\ \\
$^{1}$Centre for Astrophysics Research, Department of Physics, Astronomy and Mathematics,
University of Hertfordshire, Hatfield, UK\\
$^{2}$Department of Astrophysics, American Museum of Natural History, Central Park West at 79th Street, New York, NY 10034, USA\\
$^{3}$Department of Physics, The Graduate Center City University of New York, New York, 10016, NY, USA\\
$^{4}$Department of Physics and Astronomy, Hunter College, City University of New York, 695 Park Avenue, New York, NY, 10065, USA\\
$^{5}$Backyard Worlds: Planet 9\\
$^{6}$Department of Astronomy \& Astrophysics, University of California, Santa Cruz, CA 95064, USA\\
$^{7}$University of Texas at Austin, Department of Astronomy, 2515 Speedway C1400, Austin, TX 78712, USA\\
$^{8}$Chemistry \& Planetary Sciences, Dordt University, Sioux Center, IA\\
$^{9}$Center for Extrasolar Planetary Systems, Space Science Institute, Boulder, CO\\
$^{10}$School of Physics, Trinity College Dublin, The University of Dublin, Dublin, Ireland\\
$^{11}$Department of Physics \& Astronomy, Amherst College, 25 East Drive, Amherst, MA 01003, USA
}

\date{Accepted  2026 August 25. Received 2026 August 25; in original form 2026 March 20}

\pubyear{\the\year{}}

\begin{document}
\label{firstpage}
\pagerange{\pageref{firstpage}--\pageref{lastpage}}
\maketitle


\begin{abstract}
We present the first atmospheric retrieval analysis of two compositional benchmark mid-L dwarfs -- SDSSJ141659.78+500626.4 and GJ 499 C -- using the \textit{Brewster} retrieval framework, where the wide benchmark nature of these systems provides independent constraints on age and bulk composition from their stellar primaries. These targets were observed with JWST using NIRSpec Prism and MIRI LRS, providing low-resolution ($R$ $\sim$ 100) spectra between 0.6–14.0 $\mathrm{\mu}$m with high signal-to-noise ratios (SNR $\sim$100 – 600). For SDSSJ141659.78+500626.4, the retrieved cloud combination of a high-altitude enstatite slab and low-altitude iron deck clouds matches phase-equilibrium predictions based on the primary star's Mg/Si ratio. We retrieve a super-solar C/O = 0.71$^{+0.01}_{-0.01}$ and slightly metal-rich [M/H] = 0.22$^{+0.03}_{-0.03}$. This C/O ratio can only be reconciled with the value inferred for the primary if additional oxygen sequestration beyond the retrieved cloud mass is present, or if there are uncertainties in the adopted opacities or other model deficiencies. The inferred [C/H] and [O/H] are 0.31$\pm$0.03 and 0.19$\pm$0.03, respectively, which are consistent within the relatively large uncertainties of the host star abundances. For GJ 499 C, the retrieved silicon-monoxide and forsterite slab clouds are difficult to explain with simple phase-equilibrium assumptions, yielding Mg/Si $\sim$ 1.9. We estimate C/O = 0.69$^{+0.02}_{-0.02}$ and [M/H] = 0.13$^{+0.04}_{-0.06}$. For both objects, the inferred radii of 0.85$^{+0.01}_{-0.01}$ $R_{\mathrm{Jup}}$ and 1.00$^{+0.02}_{-0.02}$ $R_{\mathrm{Jup}}$ and masses of 73.7$^{+4.9}_{-9.2}$ $M_{\mathrm{Jup}}$ and 68.0$^{+8.5}_{-12.7}$ $M_{\mathrm{Jup}}$ are consistent with evolutionary models and system ages, highlighting the plausibility of our results.

\end{abstract}

\begin{keywords}
stars: brown dwarfs 
\end{keywords}



\section{Introduction}

Brown dwarfs (hereafter BDs) are substellar objects with masses below the hydrogen burning limit  (13 M$_{\mathrm{J}}$ $\lesssim$  $M$ $\lesssim$ 80 M$_{\mathrm{J}}$) \citep{RevModPhys.65.301, refId0, refId1}, filling the gap between giant planets and low-mass stars. In terms of spectral type, BDs can  be classified as late-M, L, T, or Y dwarfs (e.g.,
\citeauthor{2005ARA&A..43..195K} \citeyear{2005ARA&A..43..195K}, \citeauthor{2011ApJ...743...50C} \citeyear{2011ApJ...743...50C}), based on spectral features produced by chemical species in their atmospheres, which are primarily governed by temperature and age. 
L dwarfs may exhibit similar properties (e.g. bolometric luminosity, radius, temperature, and infrared colors) to those of directly imaged exoplanets \citep{2013MmSAI..84..955F}. Hence, by improving our understanding of these dwarfs, we can also gain valuable insights into the atmospheric processes of directly imaged exoplanets, enhancing our abilities to characterise their properties and evolution. L dwarfs are expected to be cloudy as their thermal profiles favor the formation of atmospheric condensates, primarily consisting of silicates (e.g. enstatite - MgSiO$_{3}$, forsterite - Mg$_{2}$SiO$_{4}$, quartz - SiO$_{2}$), silicon monoxide (SiO), aluminum oxides (e.g. corundum Al$_{2}$O$_{3}$), and iron (Fe) \citep{2001ApJ...556..357A}. These clouds -- depending on their particle size and location -- can considerably shape the observed infrared spectra (e.g., \citeauthor{2014A&ARv..22...80H} \citeyear{2014A&ARv..22...80H}, \citeauthor{2015ARA&A..53..279M} \citeyear{2015ARA&A..53..279M}, \citeauthor{2025AJ....170...43T} \citeyear{2025AJ....170...43T}). The most direct evidence of these clouds is a mid-infrared silicate absorption feature at $8-10\,\mu$m, which is on average the strongest among L4$-$L6 dwarfs \citep{2022MNRAS.513.5701S}. Silicate clouds sink into deeper atmospheric layers as temperatures decrease, causing their absorption to weaken and disappear beyond L8, indicating cloud sedimentation \citep{2004AJ....127.3553K}.

The \textit{Sinking Silicates} \citep{2023jwst.prop.3670B} James Webb Space Telescope (JWST) Cycle 2 (ID. \#3670, PI: Burningham) program targeted 12 benchmark BDs across the L-T sequence. 
Benchmark systems are composed of BDs that orbit a well-characterised star, where the properties of the primary star provide constraints that are otherwise difficult to determine. Different benchmark systems provide different types of constraints. For example, some benchmark BDs have their ages constrained through membership in clusters or coeval binary systems, while others offer constraints on fundamental properties such as mass or metallicity. The targets of the \textit{Sinking Silicates} program are compositional benchmarks: wide  binary systems with projected separations of between 254 AU and 38344 AU in which the primary star provides external constraints on age and bulk chemical composition, enabling robust tests of atmospheric retrieval frameworks. Due to the large orbital distance and high mass of the companion, they very likely formed coevally via gravitational collapse (e.g., \citeauthor{2006MNRAS.368.1281P} \citeyear{2006MNRAS.368.1281P}). Such benchmark BDs have been detected and analysed many times previously (e.g., \citeauthor{2006MNRAS.368.1281P} \citeyear{2006MNRAS.368.1281P};
\citeauthor{2008ApJ...689..436L} \citeyear{2008ApJ...689..436L};
\citeauthor{2009MNRAS.395.1237B} \citeyear{2009MNRAS.395.1237B}, \citeyear{2013MNRAS.433..457B};
\citeauthor{2011MNRAS.414..575M} \citeyear{2011MNRAS.414..575M};
\citeauthor{2012MNRAS.422.1922P}, \citeyear{2012MNRAS.422.1922P};
\citeauthor{2014ApJ...781...29C} \citeyear{2014ApJ...781...29C},
\citeyear{2016ApJ...831..136C}, \citeyear{2018ApJ...853..192C};
\citeauthor{2020A&A...635A.203R} \citeyear{2020A&A...635A.203R}; \citeauthor{2022AJ....163..189W} \citeyear{2022AJ....163..189W}; \citeauthor{2025A&A...701A..78C} \citeyear{2025A&A...701A..78C}). The benchmark objects are most commonly identified among widely separated companions to Sun-like stars due to the observational advantage of being individually resolved \citep{2010AJ....139..176F,2014ApJ...792..119D,2024AJ....167..253R}. The targets of the \textit{Sinking Silicates} program are in wide binary systems with FGK-type primary stars, some of which have well-characterised chemical composition \citep{2026arXiv260709851P} and are expected to share similar age, metallicity [M/H], and composition with their primaries. Thus, they allow us to critically test our ability to use retrievals to infer bulk properties shared with the primary and to test predictions for cloud species with the benefit of compositional constraints. The theoretical work by \citet{2024ApJ...963...67C} showed that the magnesium-to-silicon (Mg/Si) ratio can be used to predict the dominant silicate cloud composition under phase equilibrium assumptions in benchmark BDs with F, G, and K-type primary stars, as follows:
\begin{itemize}
    \item if $\mathrm{Mg} / \mathrm{Si} \lesssim 0.4$, then SiO$_2$ (quartz) clouds are likely to dominate.
    \item if $\mathrm{Mg} / \mathrm{Si} \gtrsim 0.4$ and $\mathrm{Mg} / \mathrm{Si} \lesssim 0.9$, then MgSiO$_3$ (enstatite) and SiO$_2$ clouds are likely to dominate.
    \item if $\mathrm{Mg} / \mathrm{Si} \gtrsim 0.9$ and $\mathrm{Mg} / \mathrm{Si} \lesssim 1.4$, then MgSiO$_3$ and Mg$_2$SiO$_4$ (forsterite) clouds are likely to dominate.
    \item if $\mathrm{Mg} / \mathrm{Si} \gtrsim 1.4$, then Mg$_2$SiO$_4$ clouds are likely to dominate.
\end{itemize}

Atmospheric retrievals of exoplanets and BDs involves determining the atmospheric characteristics based on their observed spectrum (see detailed review by \citeauthor{2018haex.bookE.104M} \citeyear{2018haex.bookE.104M}). In this paper, we present our retrieval analyses of two \textit{Sinking Silicates} targets, namely SDSSJ141659.78+500626.4 (an L4.5 dwarf; hereafter SDSS1416) and GJ 499 C (an L5 dwarf). These JWST observations cover the 0.6--14.0 $\mu$m wavelength range with high signal-to-noise (SNR $\sim$ 100--600) ratio. The two targets for this paper were chosen as they showed the strongest silicate feature at $\sim$ 9 $\mu$m among the 12 observed BDs. 
We use \textit{Brewster}, an atmospheric retrieval framework, which has been optimised and utilised for cloudy and cloud-free L and T dwarfs \citep{2017MNRAS.470.1177B, 2020ApJ...905...46G,2021ApJ...923...19G, 2021MNRAS.506.1944B, 2022ApJ...938...56G, 2022ApJ...940..164C, 2023ApJ...944..138V, 2023MNRAS.521.5761G, 2024ApJ...972..172P, 2024Natur.628..511F, 2025Natur.645...62F}. In Section \ref{sec:targets}, we describe our targets, while in Section \ref{sec:obs} we present our spectroscopic data. Section \ref{sec:brewster} explains the \textit{Brewster} retrieval framework and in Section \ref{sec:results} our retrieval results are shown which we discuss in Sec. \ref{sec:disc}. Finally, we conclude our findings in Section \ref{sec:conc}.

\section{Targets}

\label{sec:targets_sec}

In Table \ref{tab:targets}, we summarise the relevant known information about our target systems.
\label{sec:targets}

\begin{table}
\begin{threeparttable}
\caption{  Summary of the target BDs and their primary stars.}
\renewcommand{\arraystretch}{1.3}
\label{tab:targets}
\begin{tabular}{l|ll}
\hline 
\textbf{Target Name}  & \textbf{SDSS1416} & \textbf{GJ 499 C} \\ \hline
Primary Name          & HD 125141 & GJ 499 AB\\ 
Target Spectral Type  & L4.5$_a$& L5$_g$ \\
Primary Spectral Type & G5$_h$ & A: K5$_h$, B: M4$_h$ \\
$d_{\mathrm{primary}}$ (pc)    & 47.11$\pm$0.06$_c$ & 18.80 $\pm$ 0.61$_c$\\
$d_{\mathrm{companion}}$ (pc)    & 45.15$\pm$1.57$_d$ & 19.91$\pm$0.29$_{i}$\\
Proj. Separation  (AU)     & 26951$_c$& 9708$_c$\\
Age (Gyr)                  & 4.48$^{+2.80}_{-2.71}$$_e$ & --\\
Primary Mg/Si         & 0.95$\pm$0.07$_f$ & --\\
Primary C/O         & 0.56$\pm$0.16$_f$ & --\\
Primary [M/H]         & -0.084$\pm$0.10$_f$ & --\\
Primary [Fe/H]        & -0.08$\pm$0.04$_f$ & -0.394$\pm$0.036$_j$\\
Primary [C/H]         & 0.00$\pm$0.18$_f$ & --\\
Primary [O/H]        & 0.03$\pm$0.23$_f$ & --\\
\hline
\textbf{Results from this work}                    &  & \\
\hline
Target Mg/Si         & $\sim$ 1.0 & $\sim$ 1.9 \\
Target C/O         & 0.71$\pm$0.01 & 0.69$\pm$0.02 \\
Target [M/H]         & 0.22$\pm$0.01 & 0.13$^{+0.04}_{-0.06}$\\
Target [Fe/H]        &-1.80$^{+1.47}_{-1.07}$  & -1.05$^{+0.90}_{-0.68}$\\
Target [C/H]         & 0.31$\pm$0.03 & 0.22$^{+0.05}_{-0.06}$\\
Target [O/H]        & 0.19$\pm$0.03 & 0.12$^{+0.04}_{-0.05}$\\
Target $T_{\mathrm{eff}}$ (K) & 1643.3$^{+5.6}_{-6.4}$ & 1611$^{+13.8}_{-13.2}$ \\
Target $\log L/L_{\odot}$   &  -4.30$^{+0.03}_{-0.03}$   &  -4.19$^{+0.02}_{-0.02}$\\
Target $R$ (R$_{\mathrm{Jup}}$)   & 0.85$\pm$0.01   & 1.00$\pm$0.02 \\
Target $M$ (M$_{\mathrm{Jup}}$)   & 73.7$^{+4.9}_{-9.2}$   & 68$^{+8.5}_{-12.7}$\\
Target age (Gyr)  &  4.3 -- 12.9 & 0.5 -- 0.9\\
\hline 
\end{tabular}

\begin{tablenotes}
 
\item References: 
$a$: \citet{2006AJ....131.2722C},
$b$: \citet{1993yCat.3135....0C},
$c$: \citet{2021AA...649A...9G},
$d$: \citet{2012ApJ...752...56F},
$e$: \citet{2011A&A...530A.138C},
$f$: \citet{2026arXiv260709851P},
$g$: \citet{2013MNRAS.431.2745G},
$h$: \citet{2004AJ....128..463R},
$i$: \citet{2020yCat.1350....0G},
$j$: \citet{2026yCat.5162....0L}
\end{tablenotes}

\end{threeparttable}

\end{table}

SDSS1416, discovered by \citep{2006AJ....131.2722C}, is an L4.5 companion to the G5 primary HD 125141 \citep{1993yCat.3135....0C}, located at a distance of 47.11 $\pm$ 0.06 pc with a projected separation of 26951 AU \citep{2021AA...649A...9G}. The binary nature of the system was discovered by \citet{2019A&A...623A..72K} using proper motion anomaly. The system has an estimated age of 4.48$^{+2.80}_{-2.71}$ Gyr that was obtained via Bayesian isochrone fitting \citep{2011A&A...530A.138C}, and the primary star has near-solar metallicity, with [Fe/H] = -0.08 $\pm$ 0.04 and [M/H] = -0.084 $\pm$ 0.10. The primary also has measured elemental abundance ratios of Mg/Si = 0.95 $\pm$ 0.07 and C/O = 0.56 $\pm$ 0.16 \citep{2026arXiv260709851P}. 

\citet{2026arXiv260709851P} use high-resolution~(R = 130,000) optical data from the Potsdam Echelle Polarimetric and Spectroscopic Instrument (PEPSI; \cite{Strassmeier2015}, and the Brussels Automatic Code for Characterizing High accUracy Spectra (BACCHUS, \cite{Masseron2013} to  derive spectroscopic parameters and elemental abundance for HD 125141. In this work, we adopt the values from \citet{2026arXiv260709851P}, as it provides fuller wavelength coverage (4800 $-$ 5441 $\angstrom$ \& 7419 $-$ 9140 $\angstrom$) and a higher resolution dataset to access key lines used in this analysis to determine the C/O and Mg/Si ratios. The fundamental parameters for HD 125141 from \citet{2026arXiv260709851P} produce a $T_\mathrm{eff}$ = 5749 $\pm$ 78 K and $\log g$ = 4.49 $\pm$ 0.57 dex.
\par
Previous stellar characterisation of HD 125141 focused on fundamental  spectroscopic parameters from the Geneva-Copenhagen Survey~\citep{2011A&A...530A.138C}. \cite{2011A&A...530A.138C} found a $T_\mathrm{eff}$ = 5930 $\pm$ 89 K, $\log g$ = 4.35 and [Fe/H] = $-0.02$. The Genera-Copenhagen survey utilised the  CORAVEL spectrometer, which had a $R = 20,000$ and 375 $-$ 640 nm wavelength coverage. HD 125141 was also observed in \textit{Gaia} DR3, with $T_\mathrm{eff}$ $\sim$ 5759 K, $\log g$ = 4.24, and [M/H] = $-0.20$.

GJ 499 C is an L5 companion \citep{2013MNRAS.431.2745G} to the GJ 499 AB binary system, whose components have spectral types of K5 and M4, respectively \citep{2004AJ....128..463R}. \citet{2013MNRAS.431.2745G} identified GJ 499 ABC by cross-matching L dwarf candidates from 2MASS and WISE with known nearby Hipparcos and Gliese main-sequence stars. The system is at 18.80 $\pm$ 0.61 pc with a projected separation of 9708 AU. The age of GJ 499 AB is unknown, however, its far-infrared spectrum at 70 $\mu$m shows emission excess \citep{2006ApJ...644L.125S}, indicating a debris disk and possibly a young age. Furthermore, the [Fe/H] of the K primary is known to be -0.394$\pm$0.036 \citep{2026yCat.5162....0L}.

Both BD targets are wide-separation companions to Sun-like primaries in terms of spectral type. Given their spectral types (L4.5 -- L5), both BDs are expected to exhibit a prominent silicate feature at $\sim$ 9 -- 10 $\mu$m wavelength region \citep{2022MNRAS.513.5701S}.

Based on the preliminary results of \citet{2026arXiv260709851P}, the primary of SDSS1416 has a Mg/Si of 0.95$\pm$0.07, therefore, the most expected cloud in this target's atmosphere is enstatite \citep{2024ApJ...963...67C}.

\section{The Spectra}
\label{sec:obs}
\subsection{Data reduction}
We analyse the low-resolution ($R \sim 100$) JWST spectra of the targets between 1.0--12.0 $\mu$m. The calculation of the resolving power is described in Sec. \ref{sec:R}. The original observations span a broader range, obtained with the Near Infrared Spectrograph (NIRSpec, \citeauthor{2022A&A...661A..80J} \citeyear{2022A&A...661A..80J}), covering 0.55--5.37 $\mu$m, and the Mid-Infrared Instrument (MIRI, \citeauthor{2015PASP..127..665R} \citeyear{2015PASP..127..665R}), covering 3.46--14.01 $\mu$m. The raw, unprocessed data was downloaded from the Mikulski Archive for Space Telescopes (MAST) and processed using version 1.18.1 of the official JWST reduction pipeline \citep{bushouse_2022_7229890}. We ran the pipeline with default parameters and the Calibration Reference Data System (CRDS) context file jwst\_1364.pmap, correcting for cosmic rays, dark currents, count rate non-linearity, bad pixel flagging, and other detector level corrections with the ``calwebb\_detector1'' module. The ``calwebb\_spec2'' module is then used to produce calibrated individual exposures using the corrected data. The exposures are then combined, and the final spectrum is extracted. We note that for the MIRI LRS data scattered light artifects is a known possible effect, which is not automatically removed by the pipeline. However, this issue is present in bright objects, and our BD targets are isolated and light scattering was not visible in their images.

\begin{figure*}
     \centering
     \begin{subfigure}[b]{0.49\textwidth}
         \centering
         \includegraphics[width=\textwidth]{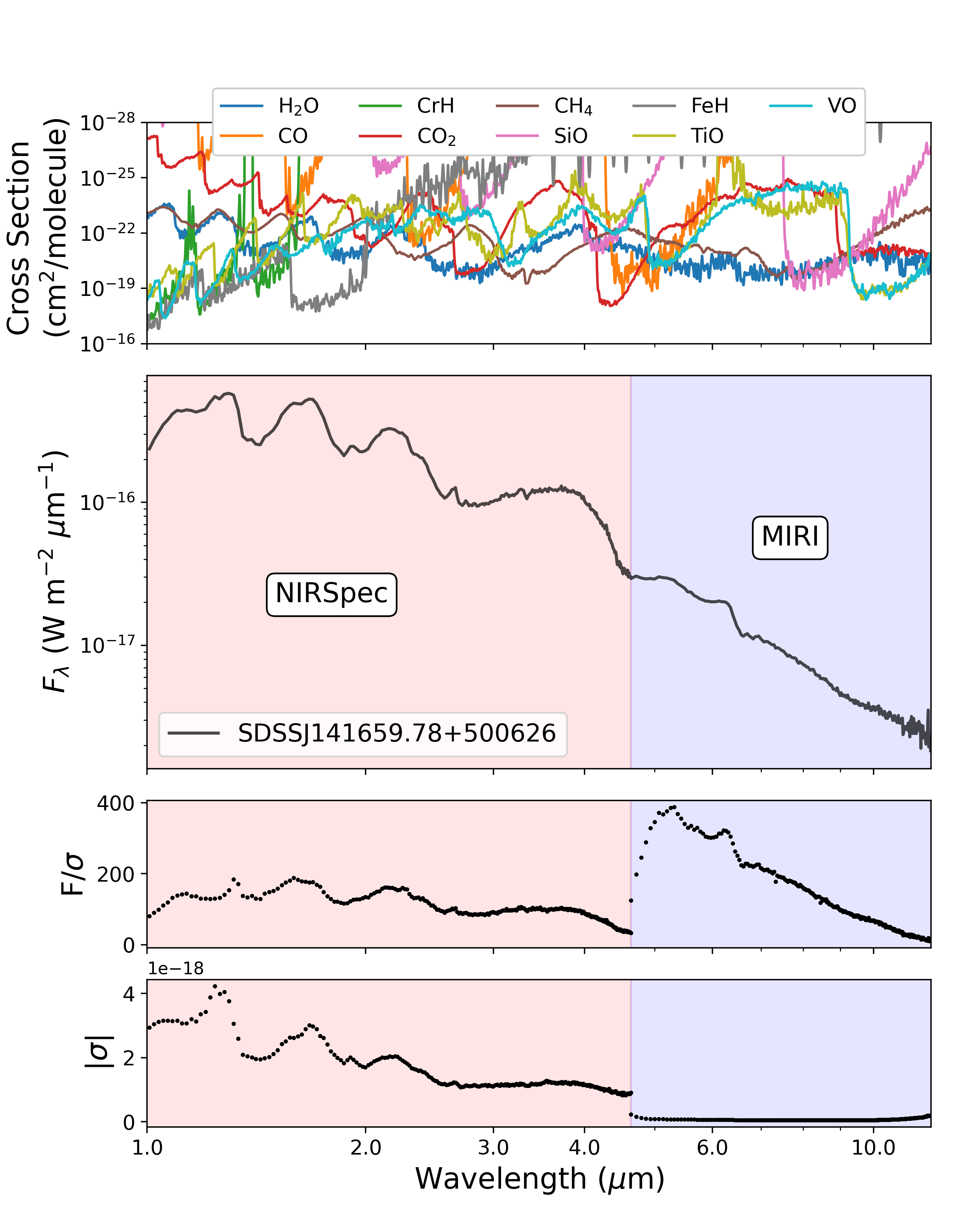}

     \end{subfigure}
     \hfill
     \begin{subfigure}[b]{0.49\textwidth}
         \centering
         \includegraphics[width=\textwidth]{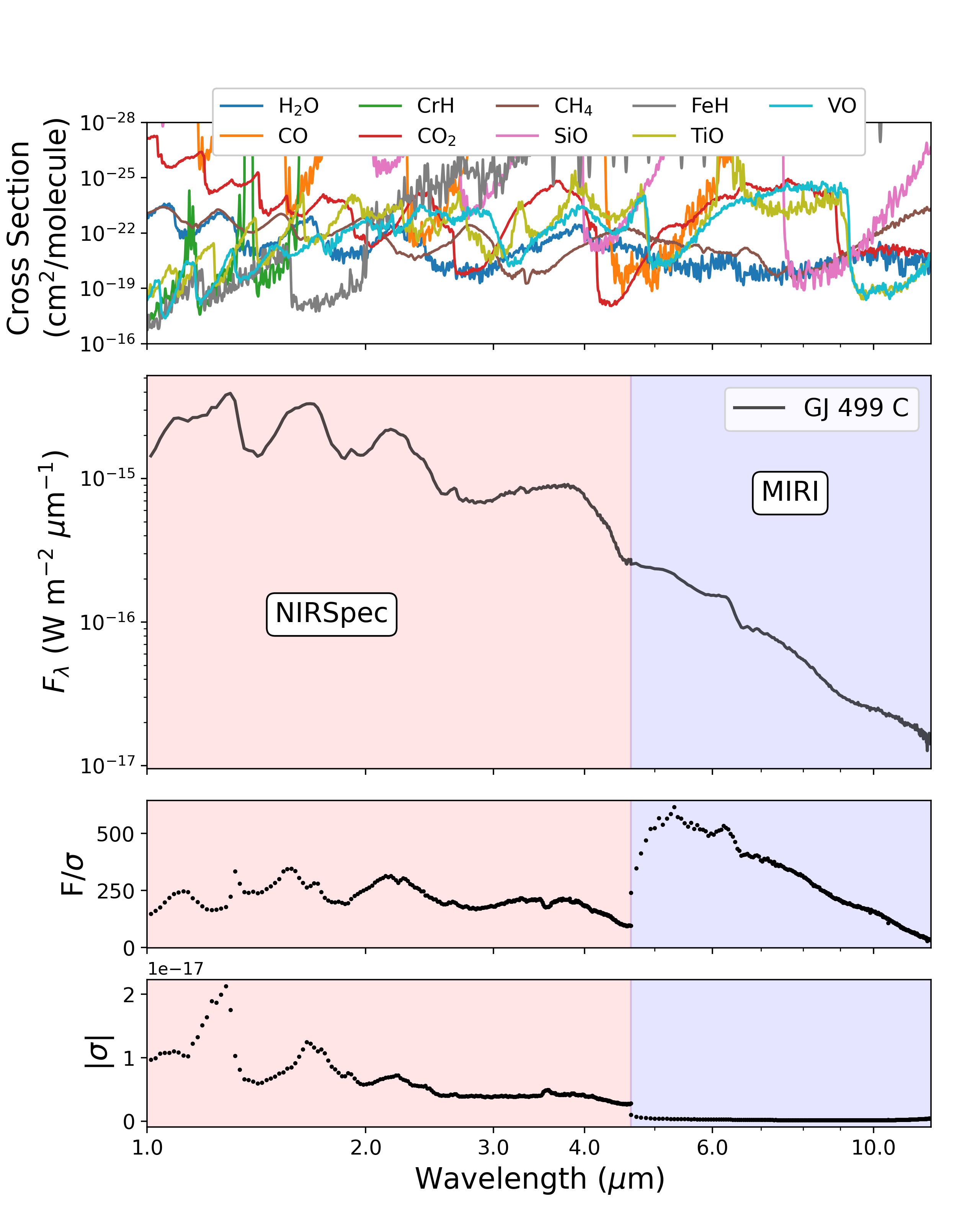}
     \end{subfigure}
     \caption{  The JWST spectra in units of $F_{\lambda}$ of SDSS1416 (left side) and GJ 499 C (right side) between 1.0 and 12.0 $\mu$m obtained with the NIRSpec and MIRI instruments. The top panel shows the cross section of the most prominent species. These are shown at a temperature of 1600 K  and a pressure of 1 bar, as this is the expected temperature of an L dwarf at this pressure level. Note, that the cross sections are convolved with a Gaussian for clarity. In the two lower panels, the SNR and the absolute errors are plotted, respectively. The red shaded regions indicate the wavelength range in which the NIRSPec data is used, whereas the blue shaded area represents the MIRI observation. The figures share the x-axis.}
     \label{fig:spectra}
\end{figure*}

We restrict our analysis to the 1.0–12 $\mu$m region, where the SNR is highest. In the case of SDSS1416, the SNR is between around 100 $-$ 400, whereas for GJ 499 C it is roughly between 200 and 600.
We merge the observational data of the two JWST instruments and keep the one with less noise in the region where they overlapped. The final spectra that are used in this study can be seen in Figure \ref{fig:spectra}.

In general, JWST provided data with extremely small observational errors, especially for MIRI (see blue shaded region in the lower panels in Figure \ref{fig:spectra}. The signal-to-noise ratio of SDSS1416 and GJ 499 C reaches $\sim$ 400 and $\sim$ 600 at around 5 $\mu$m, which is shown in the middle panel. The mean SNR for SDSS1416 is $\sim$ 110, while for GJ 499 C is $\sim$ 210.

\subsection{Data analysis}
We estimate the luminosity by integrating the observed spectrum over the available wavelength range and scaling by $4\pi D^{2}$. We infer a luminosity of $\log L/L_{\odot}$ $\sim$ -4.30$\pm$0.03 for SDSS1416 and $\log L/L_{\odot}$ $\sim$  -4.19$\pm$0.02  for GJ 499 C.

\begin{figure*}
     \centering
     \begin{subfigure}[b]{0.48\textwidth}
         \centering
         \includegraphics[width=\textwidth]{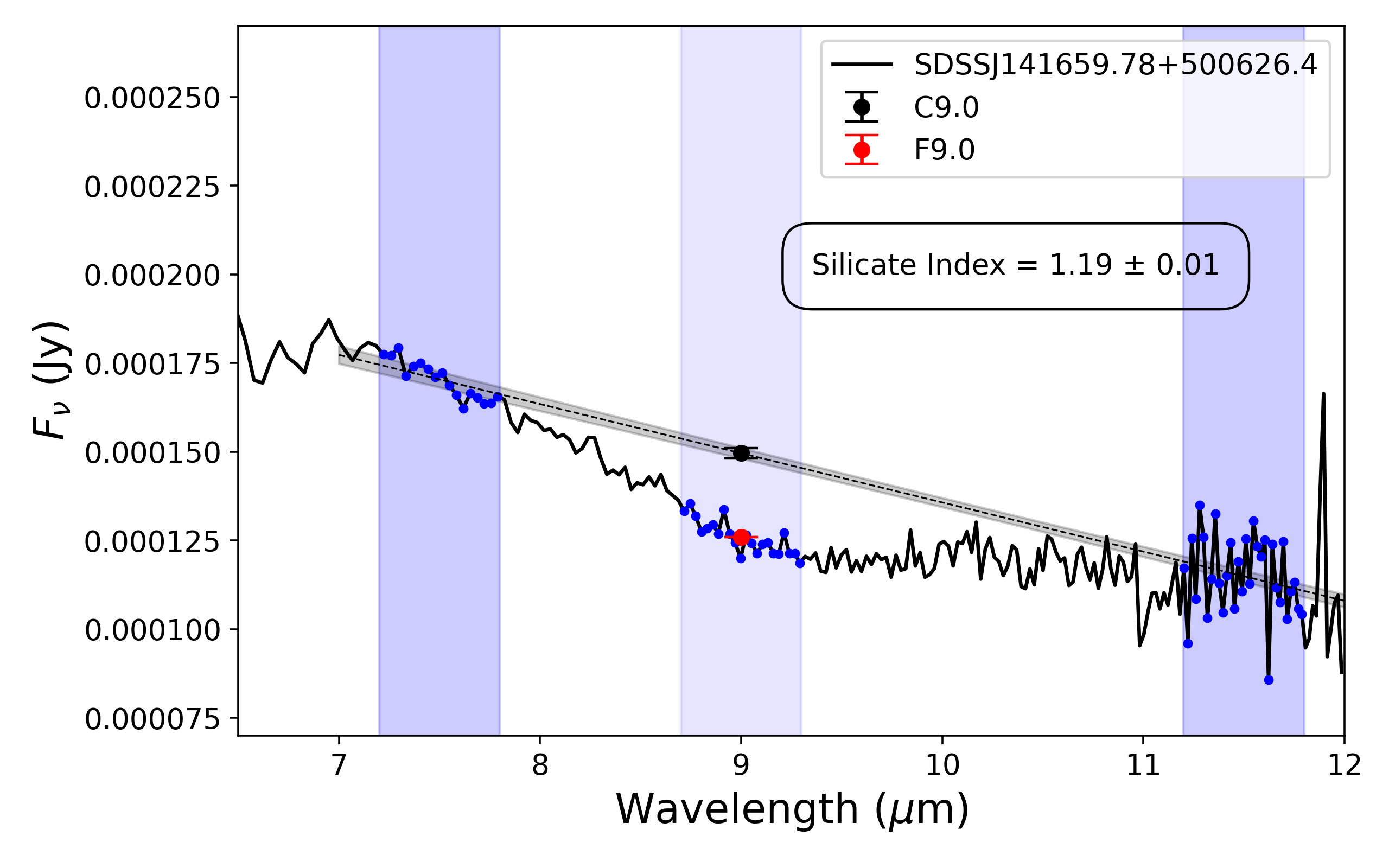}

     \end{subfigure}
     \hfill
     \begin{subfigure}[b]{0.48\textwidth}
         \centering
         \includegraphics[width=\textwidth]{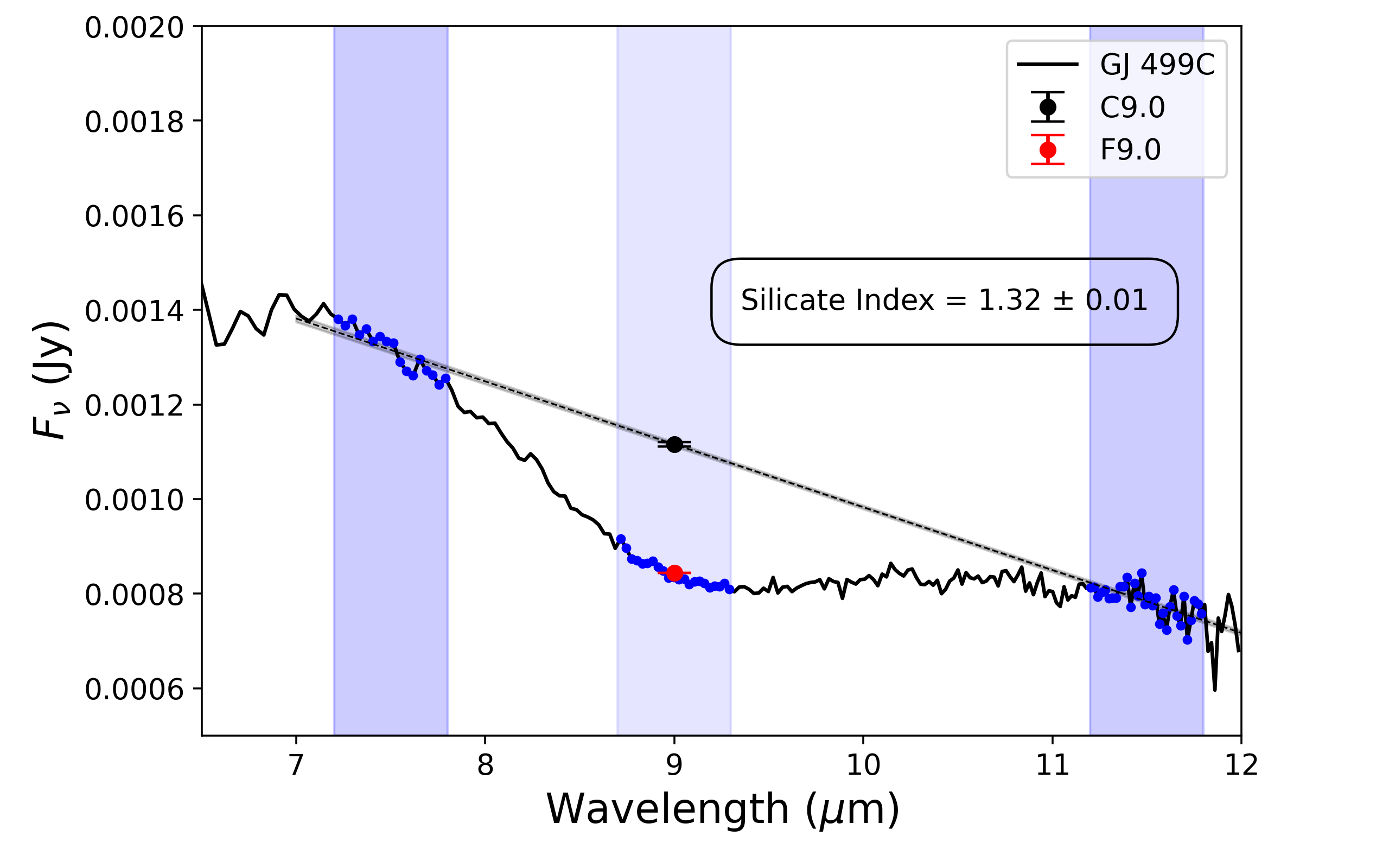}
     \end{subfigure}
     \caption{  The silicate index of SDSS1416 (left side) and GJ 499 C (right side) following \citet{2022MNRAS.513.5701S}. The darker shaded area indicates the windows around 7.5 and 11.5 $\mu$m that are used to obtain the best linear fit (black dashed line with uncertainties shown in gray) and the interpolated flux at 9.0 $\mu$m (black dot; C9.0). The lighter shaded area corresponds to the window that is used to calculate the average flux at 9.0 $\mu$m (red dot; F9.0). Note that the fluxes are in $F_{\nu}$.}
     \label{fig:silicate_index}
\end{figure*}

The mid-infrared spectra of both objects exhibit silicate absorption, indicating the presence of silicate clouds in their atmospheres (e.g. see Fig. 9. of \citeauthor{2006ApJ...648..614C} \citeyear{2006ApJ...648..614C}). We quantified the depth of these features by determining the silicate index of our targets using the method of \citet{2022MNRAS.513.5701S}. In short, we calculate the interpolated continuum flux at 9.0 $\mu$m (C9.0), which is obtained by line fitting between the fluxes at wavelength of 7.5 and 11.5 $\mu$m in a 0.6 $\mu$m window. Then, we calculate the average flux at 9.0 $\mu$m (F9.0) using a 0.6 $\mu$m window. Finally, we divide C9.0 with F9.0 to obtain the silicate index. These are shown in Fig. \ref{fig:silicate_index}. These targets provided the largest silicate indices in our sample. SDSS1416 has a silicate index of 1.19$\pm$0.01, whereas GJ 499 C has a silicate index of 1.32$\pm$0.01. Therefore, it is expected that their photospheres have silicate clouds. These results fall within the expected range of a mid-L dwarf as found by \citet{2022MNRAS.513.5701S} (see their Fig. 6 bottom panel).

\begin{figure*}
     \centering
     \begin{subfigure}[b]{0.48\textwidth}
         \centering
         \includegraphics[width=\textwidth]{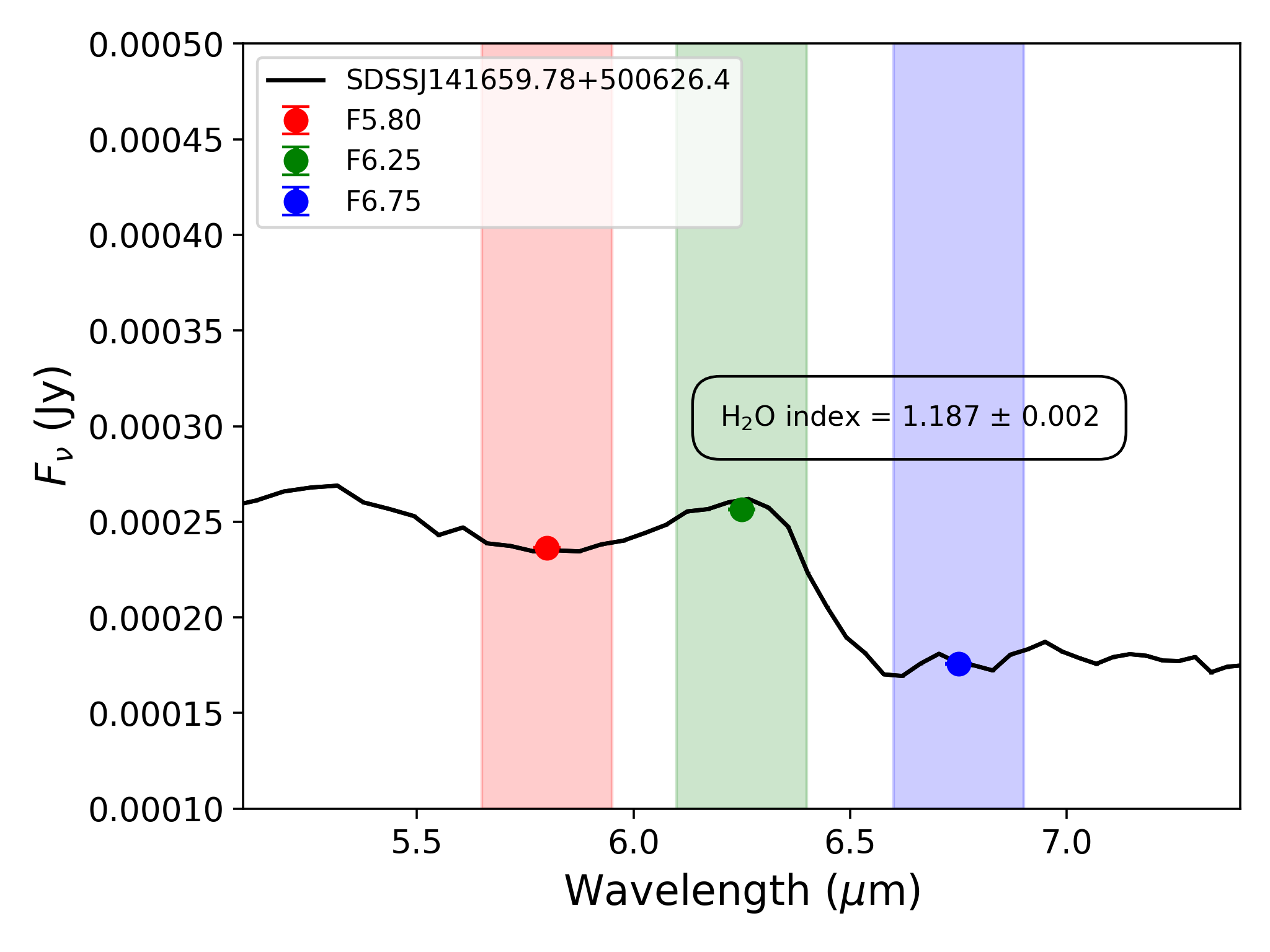}

     \end{subfigure}
     \hfill
     \begin{subfigure}[b]{0.48\textwidth}
         \centering
         \includegraphics[width=\textwidth]{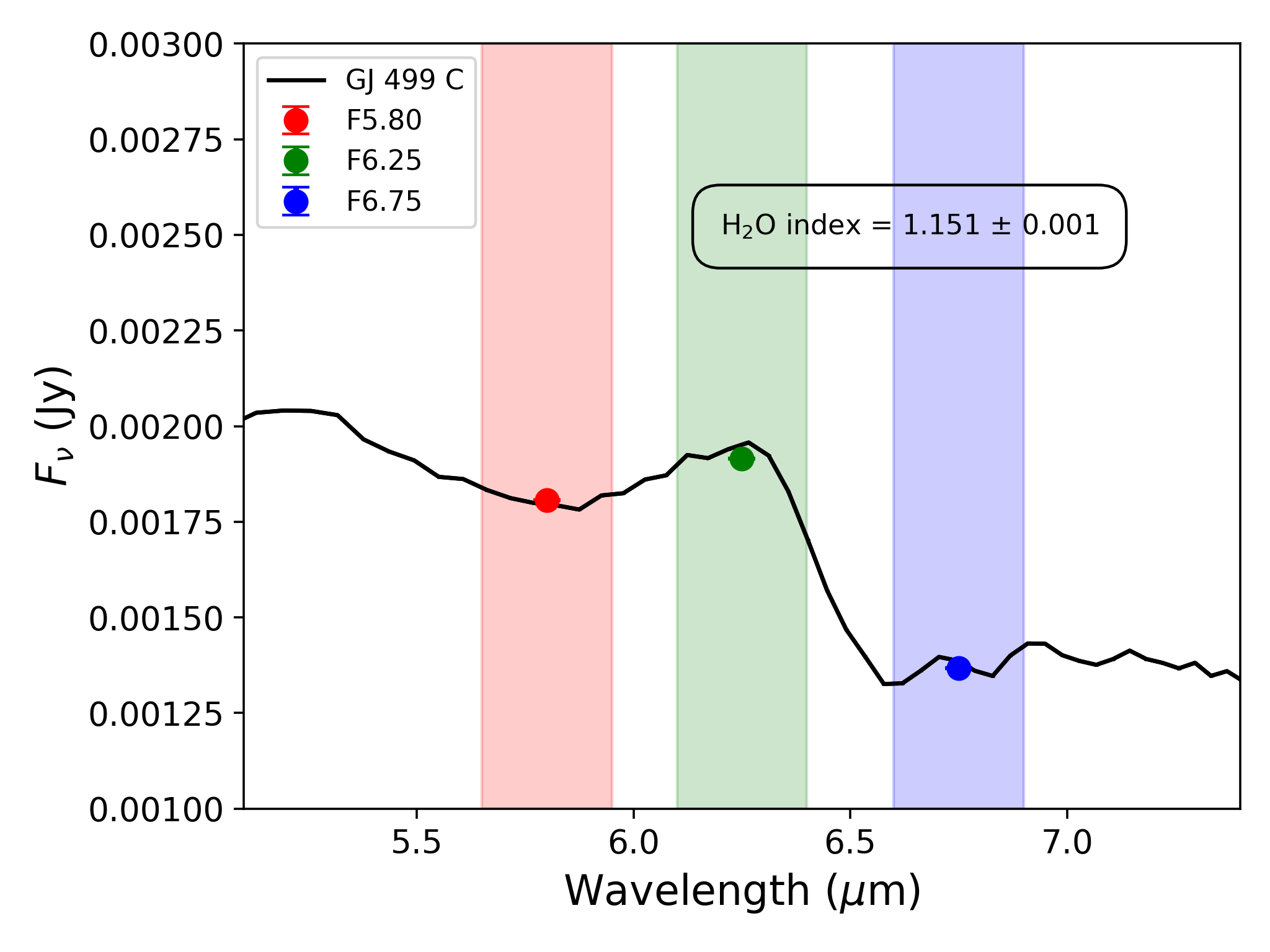}
     \end{subfigure}
     \caption{  The water index of SDSS1416 (left side) and GJ 499 C (right side) as measured in \citet{2006ApJ...648..614C}. The red, green, and blue dots indicate the mean flux at 5.8, 6.25, and 5.75 $\mu$m, respecitvely. The shaded areas show the 0.3 $\mu$m windows used to calculate the mean flux. Note that the fluxes are in $F_{\nu}$.}
     \label{fig:water_index}
\end{figure*}

We also calculate the water index that was defined by \citet{2006ApJ...648..614C} as

\begin{equation}
\mathrm{H}_2 \mathrm{O} \text { Index }=\frac{F_{6.25}}{0.562 F_{5.80}+0.474 F_{6.75}},
\end{equation}

where the subscript indicates the mean flux of the specific wavelength around a 0.3 $\mu$m window. The water index for SDSS1416 is 1.187$\pm$0.002, while GJ 499 C has a water index of  1.151$\pm$0.001. These are illustrated on Fig. \ref{fig:water_index}.  The calculated water indices are consistent with the findings of \citet{2022MNRAS.513.5701S} (see their Fig. 6 top panel).

\section{The retrieval framework}
\label{sec:brewster}
For this study, we utilise the second version of the retrieval framework model, \textit{Brewster\footnote{\href{https://github.com/fwang23nh/brewster_v2}{https://github.com/fwang23nh/brewster\_v2}}}. The main difference between versions 1 and 2 is the underlying structure of the codes: version 2 uses Python dictionaries and classes, which makes further development easier. Additionally, while in version 1 the gases were assumed to be uniform-with-altitude, version 2 allows for vertically varying gas abundances, as discussed in Sec. \ref{sec:gasop}.

\textit{Brewster} \citep{2017MNRAS.470.1177B, 2021MNRAS.506.1944B} includes a forward model that generates a synthetic spectrum based on the atmospheric parameters. Then it utilises a Bayesian sampler to explore the parameter space and estimate the posterior distribution of these parameters using the observed data. In the following sections, we describe the forward model and sampler in more detail.

\subsection{Forward model}

The forward model calculates the 1D radiative transfer via the two stream approximation \citep{1989JGR....9416287T} with 64 uniformly distributed layers in log-pressure, from log$P$ = -4 bar to log$P$ = 2.3 bar. 64 layers are used because experiments with higher layer numbers did not improve the model fitting, however, it did increase the run time significantly.  The radiative transfer model also includes scattering \citep{1989Icar...80...23M}. The output of the forward model corresponds to the flux at the top of the atmosphere, which we scale with $R^{2}/D^{2}$, where $R$ and $D$ are the radius and the distance of the target, respectively. From the retrieved scaling parameter, we can calculate the radius, given the known distance. Once obtaining the radius, the mass can be easily estimated using the retrieved log $g$.

\subsubsection{Thermal structure}
\label{sec:tp}



For our retrievals we use a ``hybrid" thermal profile, originally developed by \citet{2020A&A...640A.131M}. It assumes that the atmosphere is divided into three different regions, i.e. the high altitude, the middle altitudes (photosphere), and the low altitudes (troposphere). The higher altitudes are between P = 10$^{-4}$ bar and $\tau$ = 0.1 and split into four different regions, where the three upper layers' temperature (T$_{1}$, T$_{2}$, T$_{3}$) are free parameters. For some of our runs, we are not allowing inversion, hence T$_1$ < T$_2$ < T$_3$, as originally implemented by \citet{2020A&A...640A.131M}. However, recent findings showed inversion in BD atmospheres \citep{2024Natur.628..511F, 2025A&A...702A...1N}, therefore, we also perform retrieval runs in which inversion is allowed.  The fourth layer's temperatures, which is at $\tau$ = 0.1 and also the start of the photosphere, is calculated using the following Eddington approximation:

\begin{equation}
     T(\tau)^{4} = \frac{3}{4}T_{0}^{4}\left ( \frac{2}{3}+\tau \right ),
\end{equation}
where 
\begin{equation}
    \tau = \delta P^{\alpha}.
\end{equation}
T$_{0}$, $\alpha$, and $\delta$ are free parameters. The transition between the photosphere and troposphere is at the radiative-convective boundary. This boundary is determined by comparing the atmospheric temperature gradient ($\frac{\partial lnT}{\partial lnP} $) from the Eddington approximation with the dry adiabatic temperature gradient  that assumes H$_{2}$ and He dominated atmosphere \citep{2024ApJ...972..172P}.

\subsubsection{Gas opacity}
\label{sec:gasop}
The following absorbing gases are used for this study: H$_{2}$O, CO, CO$_{2}$, CH$_{4}$, TiO, VO, CrH, FeH, K, Na and SiO. Their opacity data are taken from \citet{2008ApJS..174..504F, 2014ApJS..214...25F} as described in \citet{2017MNRAS.470.1177B} (see Table \ref{tab:opacities} for references). The layer optical depths for these gases are computed using opacities sampled at a resolving power ($R$) of 10 000. Using $R$ = 30 000 provided nearly identical results, while increasing the runtime approximately three-fold. The line opacities are tabulated over a grid spanning 75-4000 K in temperature and in 0.5 dex pressure intervals, and are linearly interpolated onto our working temperature-pressure grid.

The gas fraction is assumed to be either vertically constant or non-uniform. In the non-uniform case, we have two extra parameters: the pressure at which rainout occurs ($P_{\mathrm{ref}}$) and its gradient ($\alpha_{g}$). 
FeH and CrH are chosen to be non-uniform as they are expected to rainout in L dwarfs and impact the spectral features (e.g., \citeauthor{2023ApJ...947....6R} \citeyear{2023ApJ...947....6R}). Additionally, we tested models with uniform FeH and CrH, however, these produced worse fits.

We use the median of the retrieved gas fractions (in the case of FeH and CrH the non-depleted gas fractions) to calculate the carbon-to-oxygen ratio, C/O, and  metallicity, [M/H].

The C/O ratio is calculated by summing the carbon bearing species and dividing by the sum of the oxygen bearing species:

\begin{equation}
    \frac{C}{O} = \frac{X_{CO} + X_{CO_2} + X_{CH_4}} {X_{H_2O} + X_{CO} + 2 \times X_{CO_2} + X_{SiO} + X_{TiO} + X_{VO}}
\end{equation}
where X$_{i}$ is volume mixing ratio of species $i$. We obtain [M/H] using the following equations:

\begin{equation}
    f_{\mathrm{H_2}} = 0.84(1 - f_{\mathrm{gas}})  
\end{equation}

\begin{equation}
    N_{\mathrm{H}} = 2f_{\mathrm{H_2}}N_{\mathrm{tot}}
\end{equation}
Here, $f_{\mathrm{H_2}}$ is the gas fraction of H$_2$, and $f_{\mathrm{tot}}$ is the gas fraction of all other gases. $N_{\mathrm{H}}$ and $N_{\mathrm{tot}}$ are the number of atoms of H and the total number of gas molecules, respectively. Then, we calculate $N_{\text {element }}$,  defined as the total number of atoms of each element of interest:

\begin{equation}
N_{\text {element }}=\sum_{\text {molecules }} n_{\text {atom }} f_{\text {molecule }} N_{\text {tot }},
\end{equation}

where $n_{\text {atom }}$ indicates how many atoms of that element are present in one molecule and $N_{\mathrm{tot}}$ represents the total number of gas molecules contributing to the composition. Finally, we calculate the metallicity following 

\begin{equation}
\label{eq:mh}
N_{\mathrm{M}}=\sum_{\text {element }} \frac{N_{\text {element }}}{N_{\mathrm{H}}}
\end{equation}

 and
 
\begin{equation}
[\mathrm{M} / \mathrm{H}]=\log \frac{N_{\mathrm{M}}}{N_{\odot}}.
\end{equation}

$N_{\odot}$ is calculated the same way as Eq. \ref{eq:mh}, but with the solar abundances from \citet{2007SSRv..130..105G}.

\begin{table*}
\caption{  Opacity sources of the species used from the compendium of \citet{2008ApJS..174..504F, 2014ApJS..214...25F}.} 
\begin{tabular}{ |l|l| }\hline
 Species & References   \\\hline
 CH$_4$ & \citet{2013JMoSp.291...69Y}; \citet{2014MNRAS.440.1649Y}; \citet{1994Icar..111..174K};\\ 
 CO & HITRAN 2010 \citep{2010JQSRT.111.2139R}; \citet{1976JMoSp..61..272T} \\
 CO$_2$ & \citet{2014JQSRT.147..134H, 2013JQSRT.130..134H}\\
 CrH & \citet{2002ApJ...577..986B}\\
 FeH &  \citet{2003ApJ...594..651D}; \citet{2010AJ....140..919H}  \\
 H$_2$O & \citet{2018MNRAS.480.2597P}
 \\
 K & \citet{2007AA...465.1085A,2007EPJD...44..507A} \\ 
 Na & \citet{2007AA...465.1085A,2007EPJD...44..507A} \\ 
 SiO & EXOPLINES \citep{2021ApJS..254...34G}; ExoMol \citep{2013MNRAS.434.1469B}\\
 TiO & \citet{1998FaDi..109..321S}; \citet{2000ApJ...540.1005A} \\
 VO &  \citet{1998AA...330.1109A}\\
 \hline
\end{tabular}
\label{tab:opacities}
\end{table*}

Continuum opacities for H$_2$-H$_2$ and H$_2$-He collisionally induced absorption (CIA) are also included using the cross sections from \citet{2012JQSRT.113.1276R} and \citet{2012ApJ...750...74S}, as well as Rayleigh scattering due to H$_{2}$ and He, and continuum opacities due to bound-free and free-free absorption by H$^{-}$ \citep{1988A&A...193..189J, 1987JPhB...20..801B} and free-free absorption by H$^{-}_{2}$ \citep{1980JPhB...13.1859B}.

\subsubsection{Cloud models}
\label{sec:cloudspecies}
In \textit{Brewster}, clouds can be defined as either ``slab" or ``deck" clouds. The slab cloud has a well-defined bottom and total optical depth, while the deck cloud quickly becomes optically thick at a certain pressure and has an undefined total optical depth below that level, which means we cannot see the bottom of it. The parameters of the different types of clouds are listed in \autoref{tab:cloud_params}. In our retrievals, we test Fe deck clouds with either one or two of the following slab clouds: SiO, SiO$_2$, MgSiO$_3$, and Mg$_2$SiO$_4$. The clouds are modelled assuming Mie scattering with wavelength-dependent optical properties for different condensate species as described in \citet{2021MNRAS.506.1944B}.


\begin{table*}
\caption{  Summary of the parameters of the slab and deck cloud in which  $\Phi = \frac{P_{\mathrm{top}}\left(10^{\Delta \log P} - 1\right)}{10^{\Delta \log P}}
$.}
\centering
\footnotesize
\begin{tabular}{lll}
\hline
\textbf{Cloud Type} & \textbf{Parameter}                  & \textbf{Description}                                    \\ \hline
\textbf{Slab Cloud} & Cloud base pressure ($\log$ $P$\(_\text{base}\)) & Log-pressure where the cloud starts (\(\tau_\text{cloud} = 0\)). \\
                    & Thickness (\(\Delta \log P_s\))        & Extent of the cloud in log-pressure.                   \\
                    & Total optical depth (\(\tau_\text{cloud}\)) & Total optical depth at the cloud base at 1 $\mu$m.                \\ \hline
\textbf{Deck Cloud} & Cloud top pressure ($\log$ $P$\(_\text{top}\)) & Log-pressure where \(\tau_\text{cloud} = 1\) (at 1 $\mu$m).   \\
                    & Decay height (\(\Delta \log P_d\))     & Pressure range in log-scale over which $\tau$ decreases as  $\frac{d\tau}{dP} \propto \exp\left(\frac{P - P_{\mathrm{deck}}}{\Phi}\right)$,\\ 
                    & &with the effective decay limited by the atmospheric scale height  . \\ \hline
\end{tabular}

\label{tab:cloud_params}
\end{table*}

\subsubsection{Spectral Convolution with Wavelength-Dependent Resolving Power}
\label{sec:R}

In order to match the observational data from JWST, we convolve the model spectra with a wavelength-dependent Gaussian kernel, where the kernel width is determined by the instrument's non-uniform resolving power ($R$). This ensures that the model spectra are degraded to the same resolving power as the observations before performing retrieval analysis. 

We assume that the NIRSpec instrument is slit limited, while MIRI LRS is source limited \citep{2023ApJ...951L..48B}. In case of a slit limited instrument, $R$ can be calculated via:

\begin{equation}
    R = \frac{\lambda}{2\delta\lambda},
\end{equation}
where $\delta\lambda$ is the pixel width.

\begin{figure}
     \centering
     \begin{subfigure}[b]{0.4\textwidth}
         \centering
         \includegraphics[width=\textwidth]{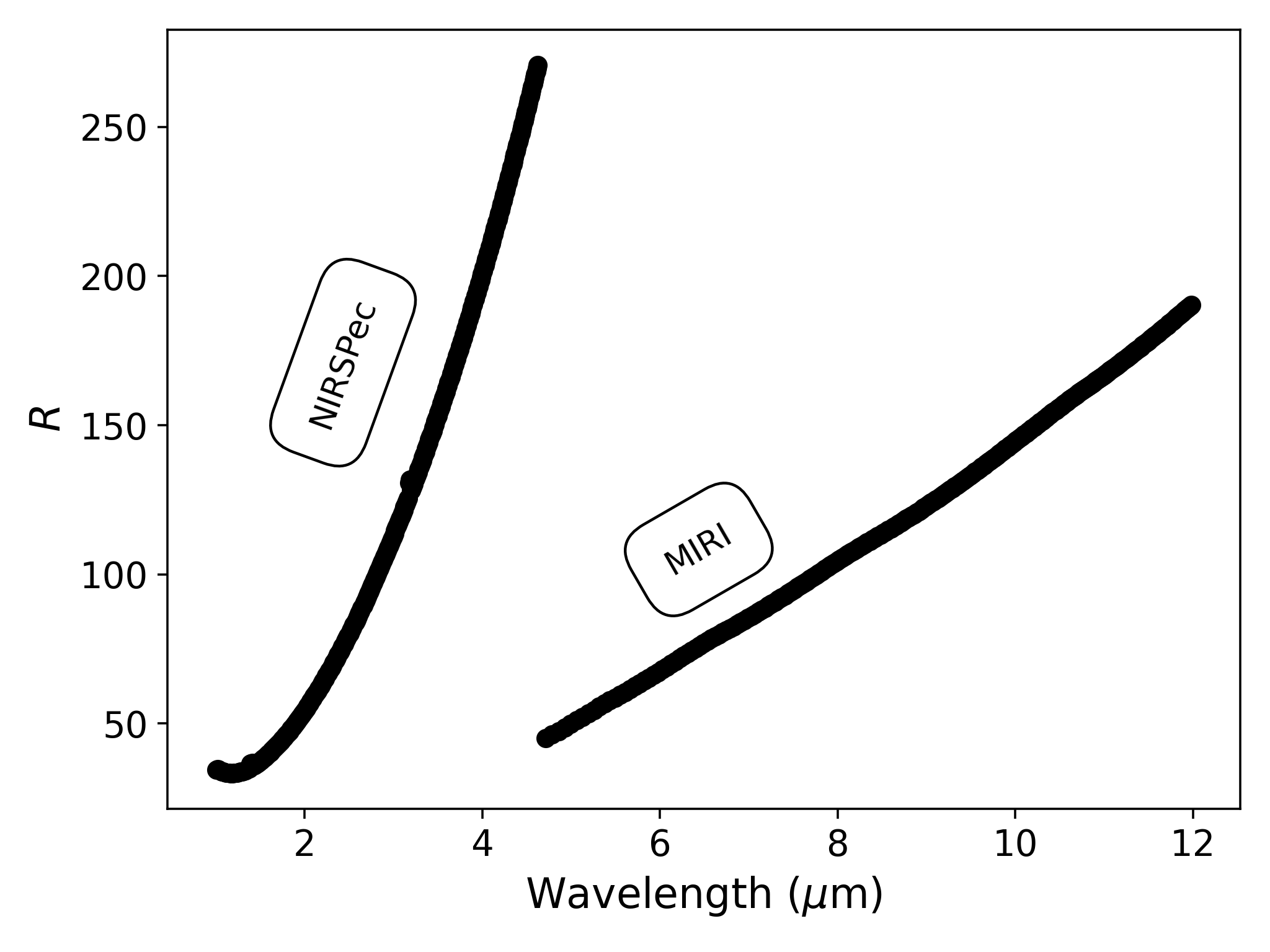}

     \end{subfigure}
     \caption{  The resolving power as a function of wavelength for both NIRSpec and MIRI instruments.}
     \label{fig:r_vs_wl}
\end{figure}

In the source limited case, $R$ is given by:

\begin{equation}
    R=\frac{\lambda}{\delta \lambda\left[\theta^{\prime \prime}(\lambda) / \theta_{\mathrm{pix}}^{\prime \prime}\right]},
\end{equation}

where $\theta^{\prime \prime}(\lambda)$ is the primary's FWHM in arcseconds and can be expressed as $(648000 / \pi) 1.028 \lambda / D$. $D$ is the 6.5m primary mirror of JWST and $\theta_{\mathrm{pix}}=0^{\prime \prime}. 11$ is the angular size of a pixel. For further details, we refer the reader to \citet{2023ApJ...951L..48B}. Firstly, we calculate the wavelength-dependent $R$ of SDSS1416, we then interpolate $R$ to the wavelength of GJ 499 C. The wavelength-dependent $R$ can be seen in Fig. \ref{fig:r_vs_wl}, spanning $\sim$30 -- $\sim$280.

\subsection{Sampler and Model Selection}
\label{sec:priors}

To constrain the model parameters, the \textsl{emcee} Python package  \citep{2013PASP..125..306F} is utilised, which is a widely used Markov Chain Monte Carlo (MCMC) ensemble sampler. MCMC samples the model parameters based on user-defined priors and gradually builds up the posterior probability distribution for those parameters. For each model run, typically we use 50,000 iterations with 10,000 burn-in and 16 walkers for each parameter. Then, we reinitialise the parameters based on the median retrieved values for another 30,000 - 50,000 iterations for a better convergence. The convergence is validated via visual inspection. The \textsl{emcee} sampler also gives us the maximum likelihood, which can be used to obtain the Bayesian Information Criterion
(BIC) \citep{https://doi.org/10.1111/j.1467-9574.2012.00530.x} that also takes into account how many parameters the model has via: 

\begin{equation}
    \mathrm{BIC} = -2 \ln \mathcal{L}_{\max,wl[::2]}+k \ln (N/2),
\end{equation}
where $\mathcal{L}_{\max,wl[::2]}$ is the maximum likelihood using every second wavelength points due to the pixel resolution element of $\sim$ 2, $k$ is the number of model parameters, and $N$ is the number of wavelength elements. We only take into account every second wavelength point due to the resolution element \citep{2023ApJ...951L..48B} to ensure that the BIC is computed assuming the correct number of independent flux measurements, avoiding correlations between pixels within the same resolving element  (see Sec. \ref{sec:R}). In general, a model with more parameters will be less favored compared to one with fewer parameters, unless it provides a significantly better fit. The model corresponding to the lowest BIC value is selected as the best fit. By comparing the BIC values of the different models ($\Delta$BIC), we can also find the winning model (see significant thresholds in \autoref{tab:bic_strength}). 

The tolerance parameter in the \textit{emcee} sampler mitigates overly optimistic uncertainty estimates in parameter inference by incorporating an additional error term  in the likelihood function. This term inflates the effective uncertainties during the fit.
Since the data come from two instruments with different SNR ranges, we assign separate tolerance parameters to each. The priors of the model parameters adopted from \citet{2017MNRAS.470.1177B}, are shown in Table \ref{tab:priors}. These priors are chosen to be broad while excluding physically unfeasible or numerically unstable regions of parameter space. The lower limit of the gas mixing ratio is $\log f_{\mathrm{gas}}$ as the chosen species are not expected to produce detectable spectral features in the data below this abundance. The prior on $\log g$ is restricted by the derived mass, ensuring that the retrieved surface gravity and radius (through the $R^{2}/D^{2}$ scaling parameter) fall within the expected mass range of a substellar object. We adopt an upper mass prior of $80,M_{\mathrm{Jup}}$, slightly above the expected hydrogen mass-burning limit of approximately $75$--$79,M_{\mathrm{Jup}}$ \citep{refId0, refId1}. This higher limit allows the retrieval to include objects straddling the stellar--substellar boundary, as the luminosities of the companions are consistent with both high-mass BDs and very-low-mass stars. The pressure priors span the full atmospheric pressure range of the model grid ($-4.0 \le \log P$ (bar) $\le 2.3$), allowing cloud opacity and the reference pressure of the rainout to be placed throughout the observable atmosphere. The priors on the temperatures are intentionally broad, with an upper limit of 6000.0 K, to avoid biasing the retrieved thermal structure, while remaining within physically and numerically stable limits. However, as mentioned in Sec. \ref{sec:tp}, inversion is allowed in some of our runs. Furthermore, the priors on the clouds' optical depth allows both optically thin and thick clouds, whereas the vertical extent allow both compact and extended cloud solutions. The Hansen particle size priors cover a wide range of condensate grain sizes, with a log-uniform prior on effective radius (-3.0 $< \log a(\mu m) <$ 3.0) to avoid favouring any particular scale, and a uniform prior on the distribution spread (0.0 $< b < $1.0).

\begin{table}
\centering
\caption{  Interpretation of \(\Delta\)BIC values.}
\begin{tabular}{ll}
\hline
\textbf{\(\Delta\)BIC} & \textbf{Strength of evidence}                 \\ \hline
0 to 2                & Not significant    \\
2 to 6                & Moderate evidence                          \\
6 to 10               & Strong evidence                          \\
$>$ 10                & Very strong evidence                      \\ \hline
\end{tabular}

\label{tab:bic_strength}
\end{table}

\begin{table*}
\caption{  Parameters and priors of the retrieval models.}
\begin{tabular}{|l|l|}
\hline \hline \textbf{Parameter} & \textbf{Prior} \\
\hline \hline Gas mixing ratio & uniform, $\log f_{\text {gas }} \geq-12.0, \sum_{\text {gas }} f_{\text {gas }} \leq 1.0$ \\
\hline Reference pressure of rainout ($P_r$) & uniform,  -4.0  $\leq$ $\log P$  $\leq$  2.3\\
\hline Non-uniform gradient ($\alpha_g$)  & uniform, 0.0  $<$ $\alpha_g$   $\leq$  1.0\\
\hline Thermal profile temperatures & uniform, $1.0 \mathrm{~K}<T_i<6000.0 \mathrm{~K}$ \\
\hline Thermal profile other parameters ($\alpha$, $\delta$) & uniform, $1.0<\alpha<2.0$ and $-20.0<\mathrm{ln}(\delta)<0.0 $\\
\hline Gravity ( $\log g$ ) & uniform, restricted by $1 M_{\text {Jup }} \leq g R^2 / G \leq 80 M_{\text {Jup }}$ \\
\hline Slab cloud base pressure ($P$\(_\text{base}\)) & uniform, -4.0 $\leq$ log($P$\(_\text{base}\)) $\leq$ 2.3\\
\hline Slab cloud thickness (\(\Delta \log P_s\))& uniform,  log($P$\(_\text{base}\)) $\leq$ log($P$\(_\text{base}\) + $\Delta$ $P$) $\leq$ 2.3\\
\hline Slab cloud total optical depth (\(\tau_\text{cloud}\)) & uniform, 0.0 $\leq$ \(\tau_\text{cloud}\) $\leq$ 100.0 \\
\hline Deck cloud top pressure ($P$\(_\text{top}\))  & uniform,   -4.0 $\leq$ log($P$\(_\text{top}\)) $\leq$ 2.3 \\
\hline Deck cloud decay height (\(\Delta \log P_d\))   & uniform, 0.0 $\leq$ \(\Delta \log P_d\) $\leq$ 0.44  \\
\hline Hansen distribution effective radius ($a$)   & log-uniform, -3.0 <$\log a (\mu m)$ < 3.0 \\
\hline Hansen distribution spread ($b$)  & uniform, 0.0 < $b$ < 1.0 \\
\hline Wavelength shift ($\Delta\lambda$)  & uniform, -0.01 < $\Delta\lambda$ < 0.01 \\
\hline Tolerance parameter & uniform, $\log \left(0.01 \times \min \left(\sigma_i^2\right)\right) \leq b \leq \log \left(100 \times \max \left(\sigma_i^2\right)\right)$ \\
\hline
\end{tabular}
\label{tab:priors}
\end{table*}

\section{Results}
\label{sec:results}

We test different cloud models as mentioned in Sec. \ref{sec:cloudspecies} for each target with and without thermal inversion. The top ten best-fitting models for SDSS1416 and GJ 499 C are listed in Table \ref{tab:summary}. For both targets, the $\Delta$BIC value between the first and second best model are much larger than 10, indicating very strong evidence favoring the models with the lowest BIC values. In addition, we also calculate the reduced $\chi^{2}$ of the top three models using the spectra observed by only either the NIRSpec or MIRI instruments (Table \ref{tab:summary_topthree}). The reduced $\chi^{2}$ values are consistent with the $\Delta$ BIC values in the case of SDSS1416. The winning model of GJ 499 C also provides the best fit in the MIRI region with a reduced $\chi^{2}$ of 26.50. However, the reduced $\chi^{2}$  in the NIRSpec region is the best in the case of a model that includes an enstatite slab and iron deck cloud with no thermal inversion, which is the third best model based on $\Delta$BIC values. Its NIRSpec region has a reduced $\chi^{2}$ of 20.23, however its MIRI region performed much worse with a reduced $\chi^{2}$ of 112.66. This indicates that performing retrievals over a narrower wavelength region might provide a better fit within that region, however, that specific model might fail to reproduce the full spectrum.

In general, SDSS1416 preferred models without thermal inversion. Enstatite (MgSiO$_3$) slab cloud with an iron (Fe) deck cloud provided the best overall fit based on the BIC value, which is expected due to the known Mg/Si ratio of the primary (see Sec \ref{sec:targets_sec}). As for GJ 499 C, the winning model consists of silicon monoxide (SiO) and forsterite (Mg$_{2}$SiO$_{4}$) slab clouds and an Fe deck cloud. Furthermore, it favours the presence of a thermal inversion in the upper atmosphere starting at $\log P \sim -2.2$ bar. We conclude that the inversion is not an artifact of the tested thermal model as the model without inversion but with the same cloud combination provides a worse fit with a BIC value of 110.54.  We further discuss the thermal inversion of GJ 499 C in Sec. \ref{sec:inv}.

\begin{table*}
\caption{  Summary of the ten best-fitting retrieval models. 
The columns list: target name; cloud configuration; whether a thermal inversion was allowed (see Sec. \ref{sec:tp} for details); the label used for the figures; the $\Delta$BIC relative to the best model (where 0 denotes the preferred model); the reduced $\chi^{2}$ of the silicate feature between 7.5 -- 11.5 $\mu$m; and the reduced $\chi^{2}$ of the NIRSpec and MIRI region, respectively.}
\begin{tabular}{lllllll}
\hline
Target & &  &  &  &\\
\hline
SDSS1416 & Clouds & Inversion & Figure Reference Name & $\Delta$BIC & Silicate $\chi^{2}$& \\
\hline
 & Enstatite + Iron & No &$\mathrm{MgSiO}_{3} + \mathrm{Fe}\ (\mathrm{noInv})$ & 0 & 8.97 \\
 & Silicon Monoxide + Iron & No & $\mathrm{SiO} + \mathrm{Fe}\ (\mathrm{noInv})$ & 43.06 & 9.54 \\
 & Silicon Monoxide + Iron & Yes & $\mathrm{SiO} +\mathrm{Fe}$ & 45.76 & 11.37 \\
 & Forsterite + Silicon Monoxide + Iron & No & $\mathrm{SiO} + \mathrm{Mg}_{2}\mathrm{SiO}_{4} + \mathrm{Fe}\ (\mathrm{noInv})$ & 56.83 & 7.19  \\
 & Forsterite + Iron & No & $\mathrm{Mg}_{2}\mathrm{SiO}_{4} +\mathrm{Fe}\ (\mathrm{noInv})$ & 70.27 & 19.84 \\
 & Enstatite + Quartz + Iron & No & $\mathrm{MgSiO}_{3}+\mathrm{SiO}_{2} +\mathrm{Fe}\ (\mathrm{noInv})$ & 75.48 & 9.89 \\
 & Enstatite + Forsterite + Iron & No & $\mathrm{MgSiO}_{3}+\mathrm{Mg}_{2}\mathrm{SiO}_{4} + \mathrm{Fe}\ (\mathrm{noInv})$ & 78.10 & 13.67 \\

 & Enstatite + Forsterite + Iron & Yes & $\mathrm{MgSiO}_{3}+\mathrm{Mg}_{2}\mathrm{SiO}_{4} + \mathrm{Fe}$& 80.46& 9.03  \\

 & Forsterite + Iron & Yes & $\mathrm{Mg}_{2}\mathrm{SiO}_{4} + \mathrm{Fe}$ & 96.30 & 14.90\\
 & Enstatite + Quartz + Iron & Yes & $\mathrm{MgSiO}_{3}+\mathrm{SiO}_{2} + \mathrm{Fe}$ & 125.74 & 8.49\\ \hline
GJ 499 C     & Clouds & Inversion & Figure Reference Name & $\Delta$BIC & Silicate $\chi^{2}$ & \\ \hline 
     & Forsterite + Silicon Monoxide + Iron & Yes & $\mathrm{SiO} + \mathrm{Mg}_{2}\mathrm{SiO}_{4} + \mathrm{Fe}$ & 0 & 40.40 \\
    & Silicon Monoxide + Iron & Yes & $\mathrm{SiO} + \mathrm{Fe}$ & 68.12& 57.42 \\
    & Enstatite + Iron & No & $\mathrm{MgSiO}_{3} + \mathrm{Fe}\ (\mathrm{noInv})$ & 94.02& 33.56  \\
    & Silicon Monoxide + Iron & No & $\mathrm{SiO} + \mathrm{Fe}\ (\mathrm{noInv})$ & 109.56 & 63.24 \\
    & Forsterite + Silicon Monoxide + Iron & No & $\mathrm{SiO} + \mathrm{Mg}_{2}\mathrm{SiO}_{4} + \mathrm{Fe}\ (\mathrm{noInv})$ & 110.54 & 59.72  \\
    & Quartz + Iron & No & $\mathrm{SiO}_{2} + \mathrm{Fe}\ (\mathrm{noInv})$ & 160.32 & 56.06 \\
     & Forsterite + Quartz  + Iron & No & $\mathrm{SiO}_{2} + \mathrm{Mg}_{2}\mathrm{SiO}_{4} + \mathrm{Fe}\ (\mathrm{noInv})$ & 183.89 & 71.47 \\
    & Enstatite + Iron & Yes & $\mathrm{MgSiO}_{3} + \mathrm{Fe}$ & 194.13 & 54.81\\
    & Enstatite + Silicon Monoxide + Iron & No & $\mathrm{SiO} + \mathrm{MgSiO}_{3} + \mathrm{Fe}\ (\mathrm{noInv})$ & 204.01 & 39.27 \\
    & Enstatite + Silicon Monoxide + Iron & Yes & $\mathrm{SiO} + \mathrm{MgSiO}_{3} + \mathrm{Fe}$ & 273.96 & 198.02 \\ \hline
\end{tabular}
\label{tab:summary}
\end{table*}

\begin{table*}
\centering
\caption{Comparison of the three best-fitting  retrieval models for each target. Columns are: the cloud species included in each model, whether a thermal inversion is allowed, and the reduced $\chi^{2}$ values calculated separately for the NIRSpec and MIRI spectra.}
\label{tab:summary_topthree}
\begin{tabular}{llccc}
\hline
Target & Clouds & Inversion & NIRSpec $\chi^{2}$ & MIRI $\chi^{2}$ \\
\hline
SDSS1416 
& Enstatite + Iron 
& No 
& 6.01 
& 7.83 \\

& Silicon Monoxide + Iron 
& No 
& 6.99 
& 8.93 \\

& Silicon Monoxide + Iron 
& Yes 
& 7.16 
& 8.63 \\
\hline
GJ 499 C 
& Forsterite + Silicon Monoxide + Iron 
& Yes 
& 30.17 
& 26.50 \\

& Silicon Monoxide + Iron 
& Yes 
& 30.05 
& 30.88 \\

& Enstatite + Iron 
& No 
& 20.23 
& 112.66 \\
\hline
\end{tabular}
\end{table*}

\subsection{Retrieved Spectra}
\begin{figure*}
     \centering
     \begin{subfigure}[b]{0.48\textwidth}
         \centering
         \includegraphics[width=\textwidth]{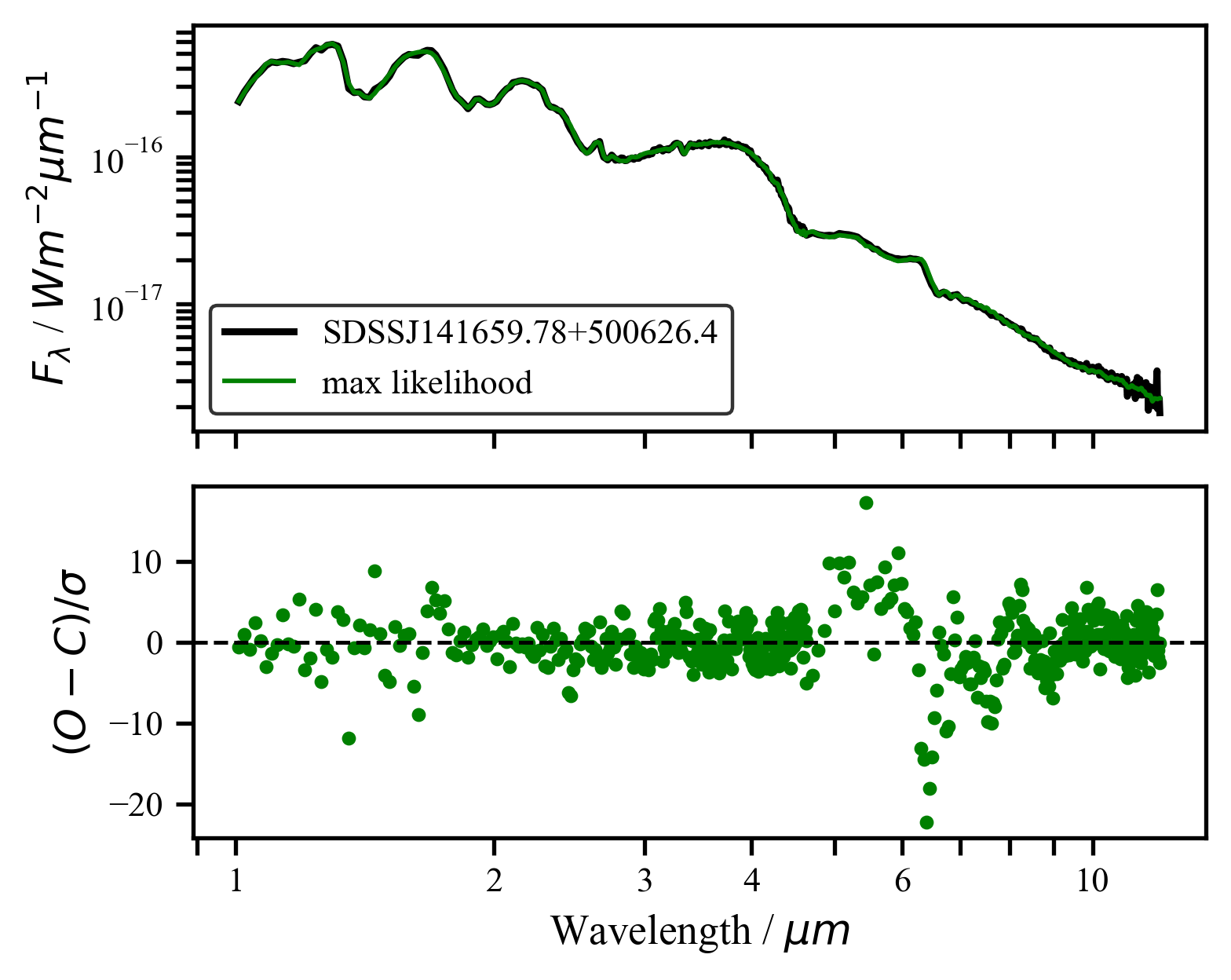}

     \end{subfigure}
     \hfill
     \begin{subfigure}[b]{0.48\textwidth}
         \centering
         \includegraphics[width=\textwidth]{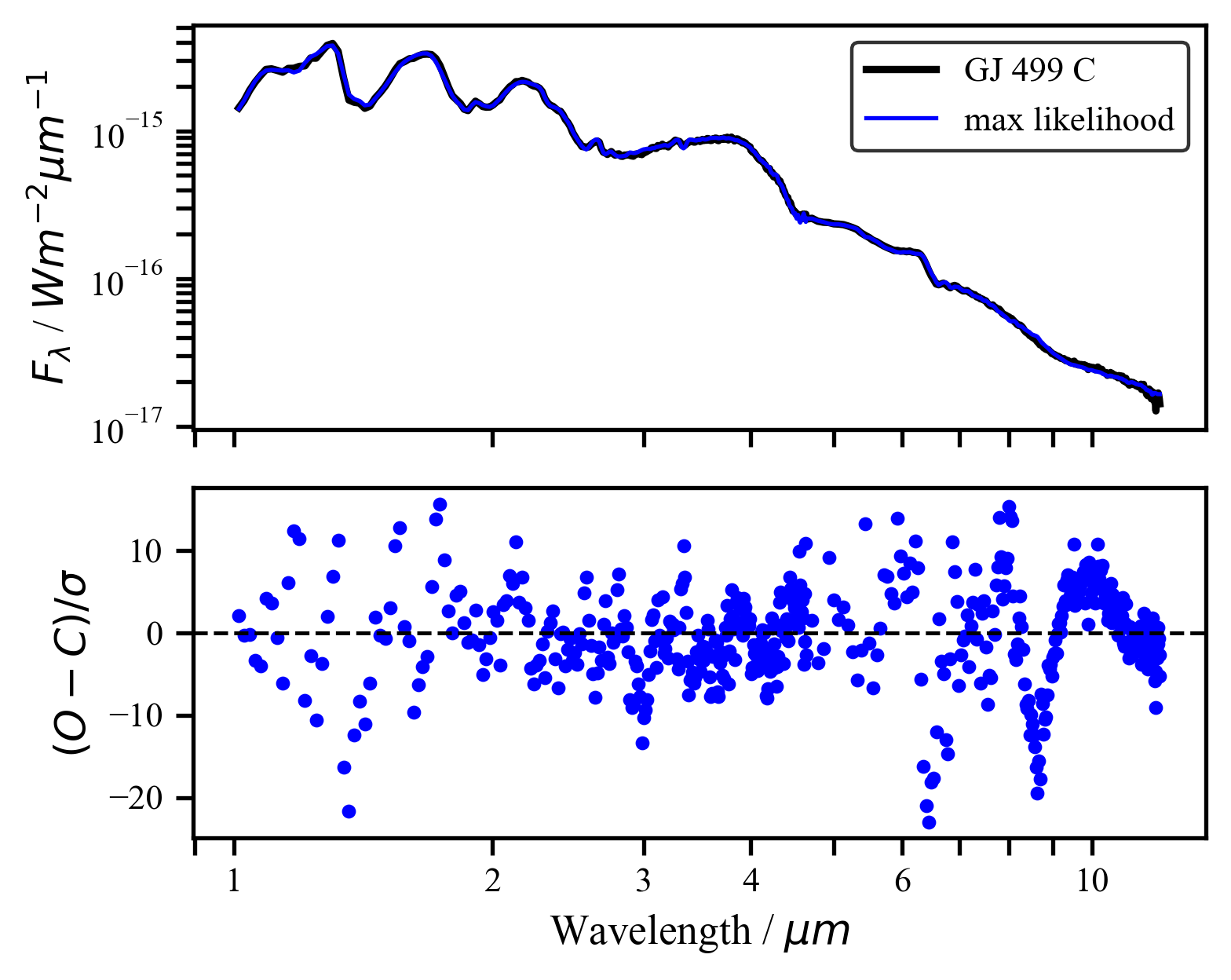}
     \end{subfigure}
     \caption{  The maximum likelihood spectra on the top figure for SDSS1416 (left) and GJ 499 C (right) and the residuals normalised by the observational error on the lower figure.}
     \label{fig:retrieved_spectra}
\end{figure*}

The maximum likelihood spectrum with residuals of our top-ranked model based on the BIC values are shown in Fig. \ref{fig:retrieved_spectra}.

The maximum likelihood spectrum is defined as the model spectrum that maximises the likelihood across all iterations. In general, the model spectrum fits the whole 1 $-$ 12 $\mu$m region of SDSS1416 well, however, it struggles to provide a perfect fit in the J and H band. This might be due to incorrect way of modelling the non-uniform FeH abundance, which we discuss later in Section \ref{sec:disc}. 

The normalised residuals peak between $\sim$5.5$-$6.5 $\mu$m for both targets. Within this wavelength region, the slope is caused by H$_{2}$O (see top panel of Fig. \ref{fig:spectra}).  While the models reproduce the shape of this feature, they struggle to fit it due to a wavelength shift. Furthermore, the residuals are normalised by the observational errors and within this wavelength range, the SNR is exceptionally high, which can be seen in Fig. \ref{fig:spectra}.

In the case of GJ 499 C, the residuals in the H and J band are more prominent. Additionally, the winning model for GJ 499 C struggles to find the shape of the silicate feature, especially at $\sim$ 8.8 $\mu$m.

\subsection{Retrieved Thermal Profiles \& Cloud Properties}
\begin{figure*}
     \centering
     \begin{subfigure}[b]{0.48\textwidth}
         \centering
         \includegraphics[width=\textwidth]{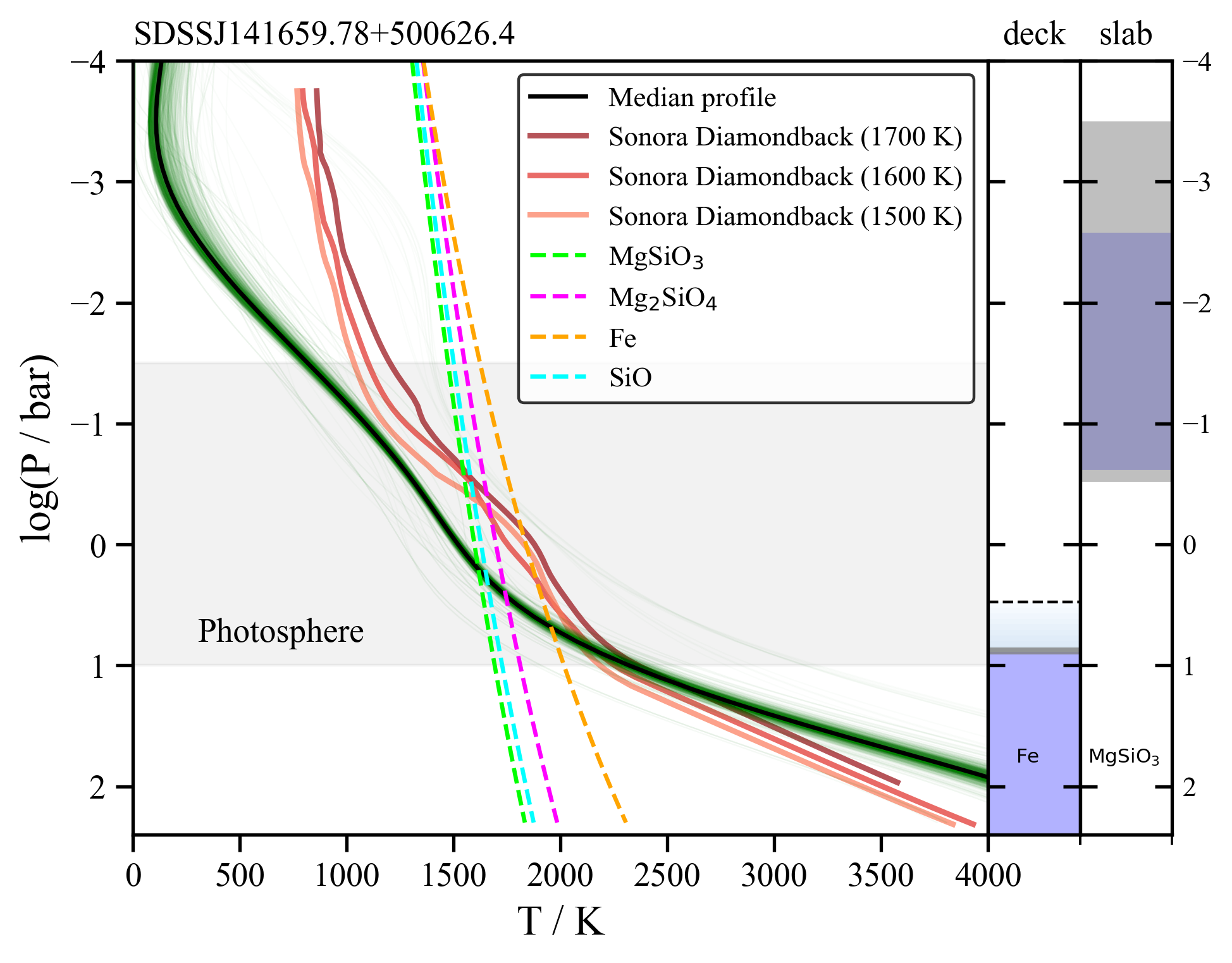}

     \end{subfigure}
     \hfill
     \begin{subfigure}[b]{0.48\textwidth}
         \centering
         \includegraphics[width=\textwidth]{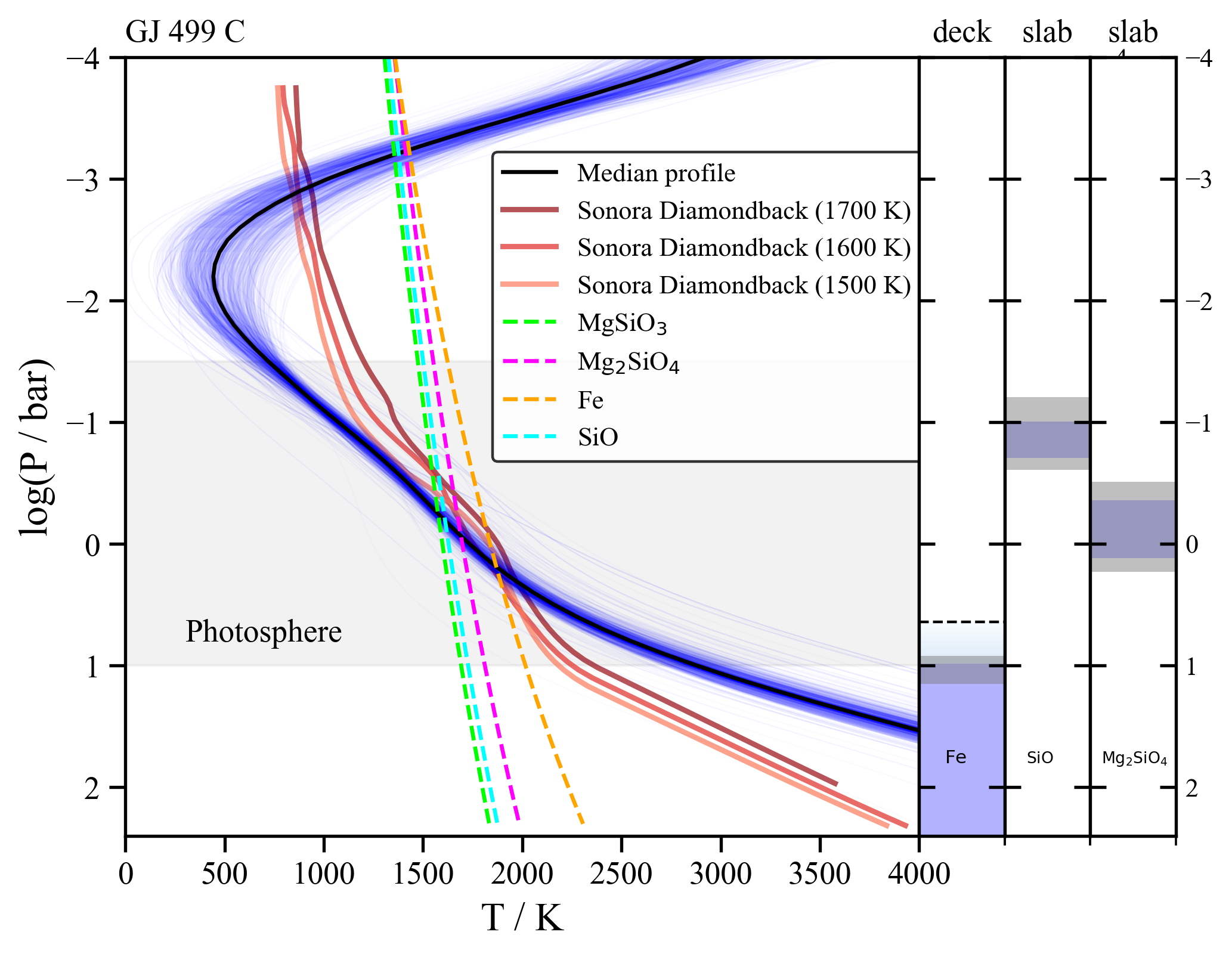}
     \end{subfigure}
     \caption{  The retrieved median temperature–pressure profile is shown in black, with 1000 randomly drawn samples from the MCMC posterior overplotted in green for SDSS1416 (left) and in blue for GJ 499 C (right). Condensation curves are also included in dashed lines as labeled, together with self-consistent Sonora Diamondback cloudy atmosphere models \citep{2024ApJ...975...59M} at effective temperatures of 1500, 1600, and 1700 K, assuming solar metallicity and C/O ratio, log $g$ $\sim$ 5.5, and a sedimentation efficiency of $f$$_{\rm sed}$ = 2. The shaded gray area indicate the approximate location of the photosphere. The right hand side of the figures show the vertical location of the clouds and the 1$\sigma$ error in purple and gray, respectively. The graduated blue shading on the top of the deck cloud is the decay height.}
     \label{fig:retrieved_TP}
\end{figure*}

We show the retrieved thermal profile with cloud locations in Fig. \ref{fig:retrieved_TP}. 

For comparison, we overplot several self-consistent cloudy Sonora Diamondback models \citep{2024ApJ...975...59M} with log $g$ $\sim$ 5.5 and sedimentation efficiency f$_{sed}$=2, spanning effective temperatures of 1500 and 1700 K. This surface gravity value is chosen as our top four best models of each target overall predicted a value closer to $\log g$ $\sim$ 5.5 than 5.0, which we later discuss in Sec. \ref{sec:age}. Our top model predicts a $\log g$ of  5.40$^{+0.03}_{-0.06}$ and 5.23$^{+0.05}_{-0.09}$ for SDSS1416 and GJ 499 C, respectively. Additionally, the chosen sedimentation efficiency and effective temperatures are expected values for mid L dwarfs \citep{2021MNRAS.506.1944B}. These models predict higher temperatures everywhere within the atmosphere of SDSS1416, except below the location of the expected photosphere where they underestimate the temperature. Interestingly, the cloudless Sonora Bobcat models so far showed lower temperature in the photosphere of L dwarfs than the retrieved thermal profile \citep[e.g. ][]{2017MNRAS.470.1177B,2020ApJ...905...46G}. This suggests that the inclusion and treatment of clouds has a considerable impact on the resulting thermal profiles, with self-consistent cloudy models predicting warmer atmospheres than both cloudless models and our retrieved profile in the photospheric region. In addition, \citet{2024ApJ...972..172P} also found that the Sonora Diamondback models poorly match the retrieved thermal profile. Fig. \ref{fig:retrieved_TP} also shows the condensation curves $-$ that are obtained using phase equilibrium calculations and thermochemical model
grids $-$ of some of the cloud species \citep{2010ApJ...716.1060V,2011ApJ...738...72V,2023ApJ...956L..32G}. The left side of these curves (i.e. lower temperatures) indicate the regions where clouds might form, but it is not necessarily guaranteed.

\subsubsection{Cloud Properties of SDSS1416}

\begin{figure}
     \centering
     \begin{subfigure}[b]{\columnwidth}
         \centering
         \includegraphics[width=\textwidth]{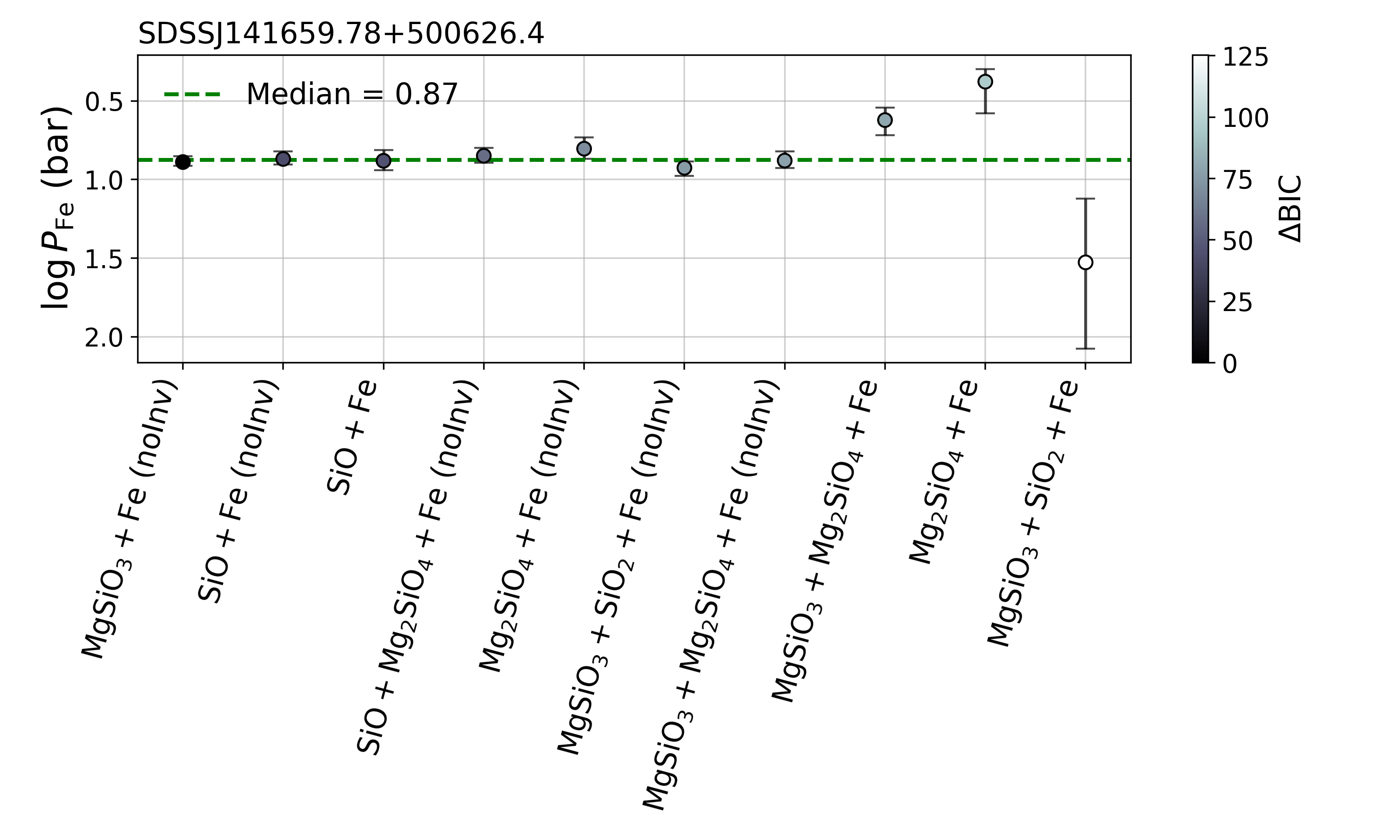}

     \end{subfigure}
     
    \begin{subfigure}[b]{\columnwidth}
         \centering
         \includegraphics[width=\textwidth]{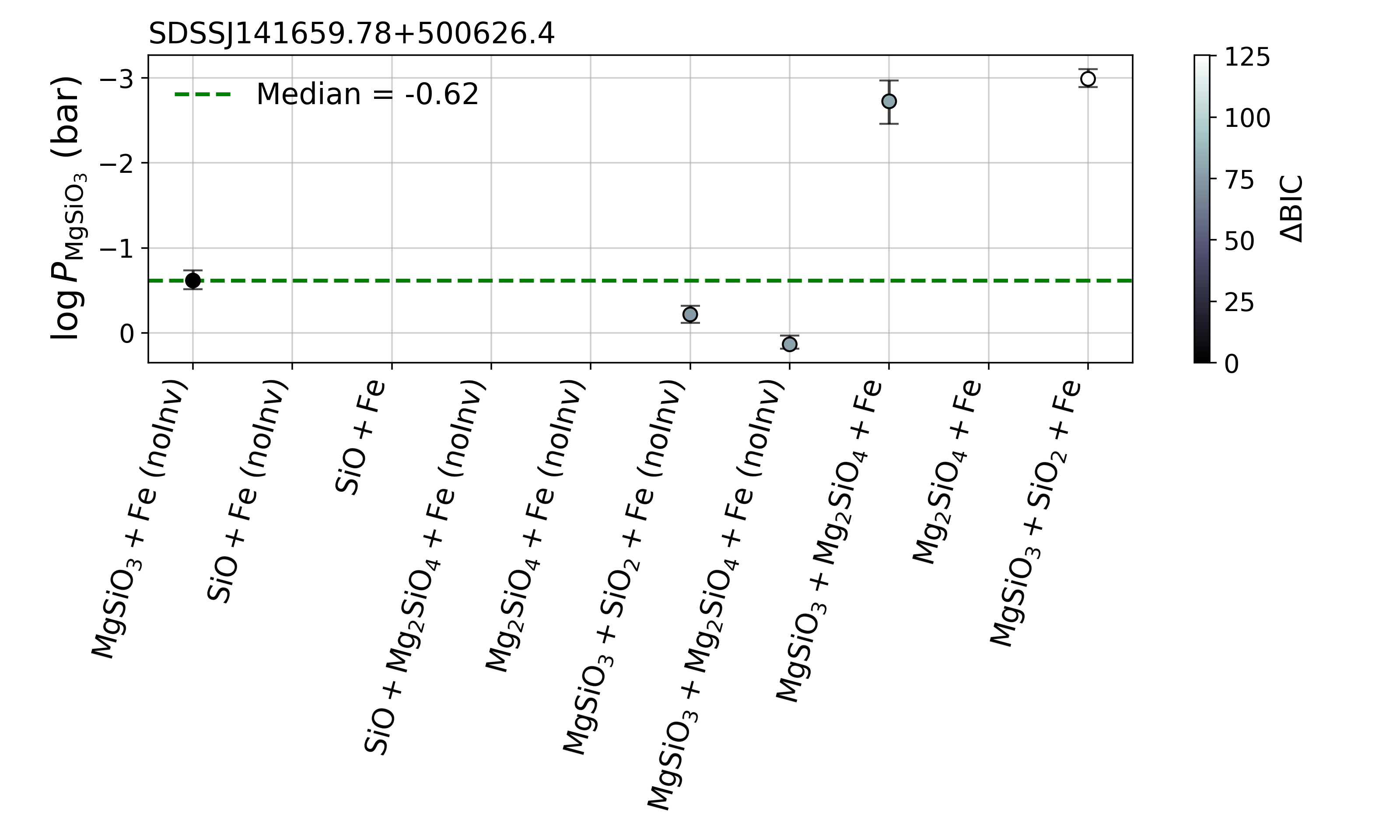}

     \end{subfigure}
     \caption{  The retrieved top pressure of the iron deck (top figure) and cloud base of enstatite slab (bottom figure) clouds of our top ten best-fitting models for SDSS1416. Model results are colour coded based on their $\Delta$BIC values and are labeled on the x axis as mentioned in Table \ref{tab:summary}.}
     \label{fig:SDSS_logp}
\end{figure}

In the case of SDSS1416, the median pressure of the enstatite slab base (log$P_{\mathrm{MgSiO_{3}}}$$\sim$-0.61 bar) and the top of the iron deck where it becomes optically thick (log$P_{\mathrm{Fe}}$$\sim$0.89 bar) can be found left to their respective condensation curves, as expected (Fig. \ref{fig:retrieved_TP}). Fig. \ref{fig:SDSS_logp} shows the retrieved cloud base pressure levels for the top 10 best models based on their BIC values. The top panel clearly indicates that the best-fitting models agree on the retrieved values of log $P_{\mathrm{Fe}}$, which are well constrained. On the other hand, the spread of values for log $P_{\mathrm{MgSiO_{3}}}$ are more significant, which is shown in the lower panel.

\begin{figure}
     \centering
     \begin{subfigure}[b]{\columnwidth}
         \centering
         \includegraphics[width=\textwidth]{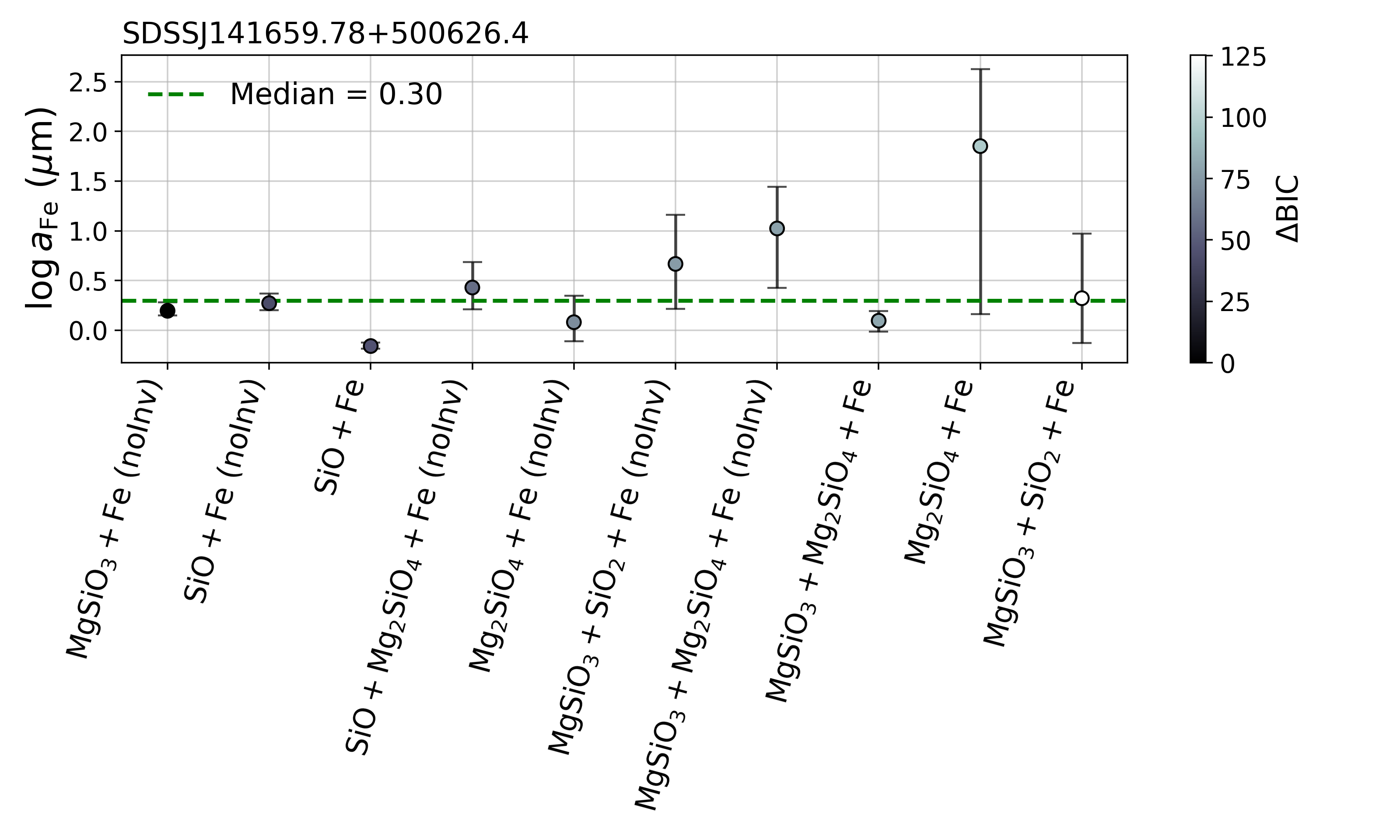}

     \end{subfigure}
     
    \begin{subfigure}[b]{\columnwidth}
         \centering
         \includegraphics[width=\textwidth]{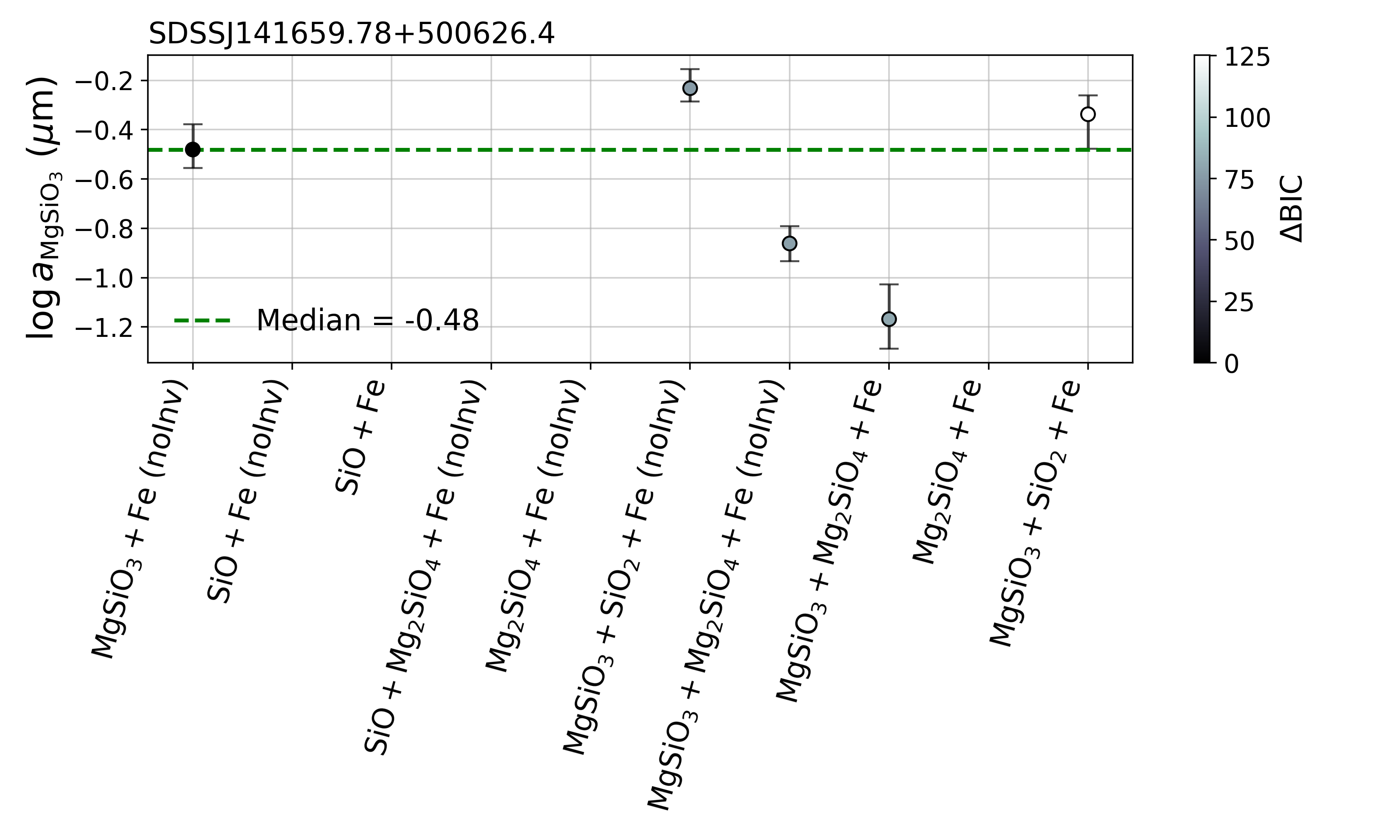}

     \end{subfigure}
     \caption{  The retrieved iron deck (top figure) and enstatite slab (bottom figure) Hansen $a$ parameter of our top ten best-fitting models for SDSS1416. Model results are colour coded based on their $\Delta$BIC values and are labeled on the x axis as mentioned in Table \ref{tab:summary}.}
     \label{fig:SDSS_hansan_a}
\end{figure}

The retrieved Hansen $a$ particle effective radii for the enstatite cloud is log$a_{\mathrm{MgSiO_{3}}}$=-0.48$^{+0.10}_{-0.08}$ $\mu$m and for the iron cloud is log$a_{\mathrm{Fe}}$=0.20$^{+0.08}_{-0.05}$ $\mu$m. We plot the retrieved iron deck Hansen $a$ parameter for our top ten best-fitting models in Fig. \ref{fig:SDSS_hansan_a} top panel. While most of the models predicted larger particle radii than the winning model with a mean of log$a_{\mathrm{Fe}}$$\sim$0.48 $\mu$m, their uncertainties are much larger as well. However, the best two models share similar retrieved Hansen $a$ radii with extremely small uncertainties.  The corresponding figure for the enstatite Hansen $a$ radii can be found in the lower panel of Fig. \ref{fig:SDSS_hansan_a}, showing the five models that included this cloud species. Each model has significantly different result, indicating a strong model dependence. The corner plot of the thermal profile and cloud parameters for this target can be seen in Fig. \ref{fig:SDSS_corner_tp}.

\subsubsection{Cloud Properties of GJ 499 C}

\begin{figure}
     \centering
     \begin{subfigure}[b]{\columnwidth}
         \centering
         \includegraphics[width=\textwidth]{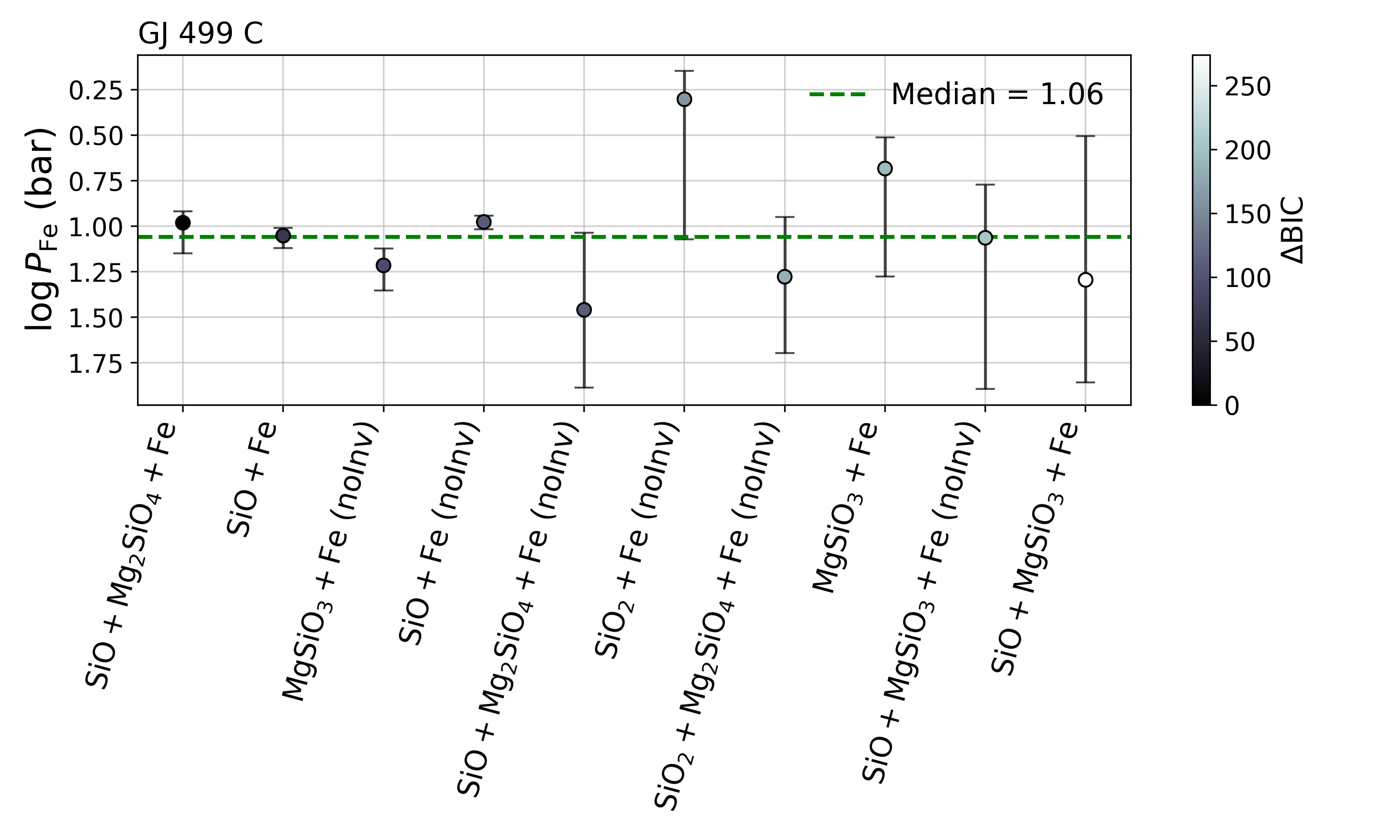}

     \end{subfigure}
     
     \begin{subfigure}[b]{\columnwidth}
         \centering
         \includegraphics[width=\textwidth]{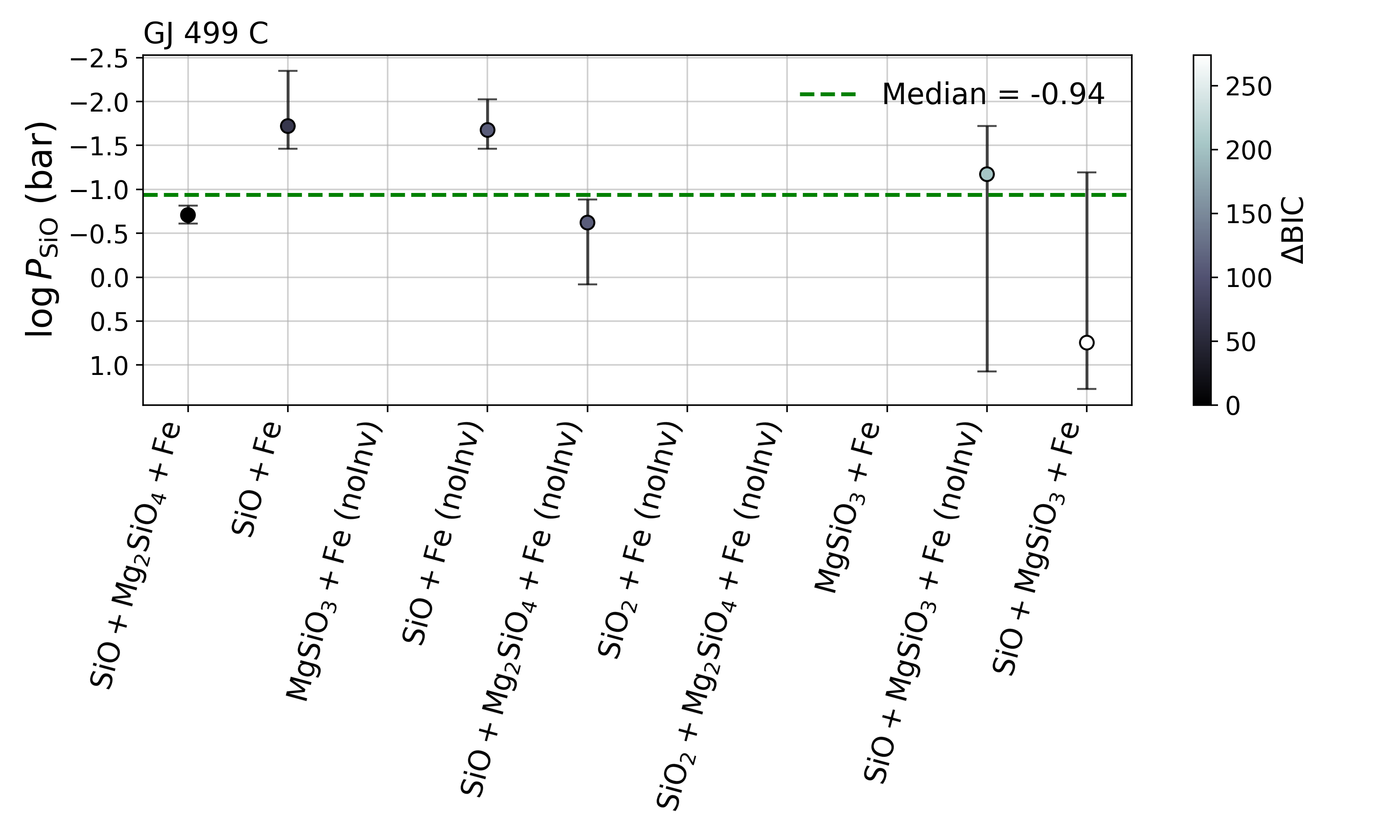}

     \end{subfigure}
     
    \begin{subfigure}[b]{\columnwidth}
         \centering
         \includegraphics[width=\textwidth]{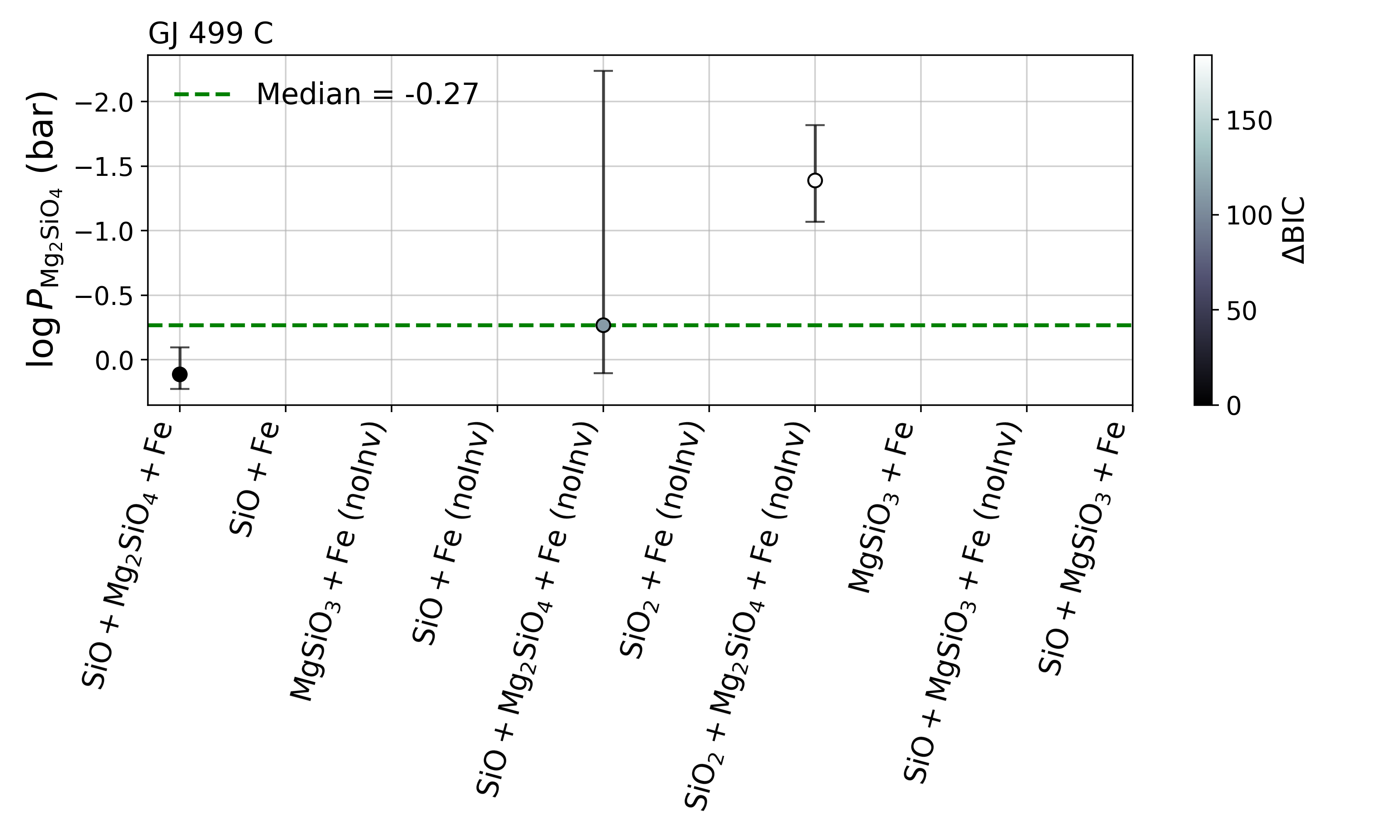}

     \end{subfigure}
     \caption{  The retrieved top pressure of the Fe deck (top figure), and the base pressure for SiO slab (middle figure), and  Mg$_2$SiO$_4$ slab (bottom figure) clouds of our top ten best-fitting models for GJ 499 C. Model results are colour coded based on their $\Delta$BIC values and are labeled on the x axis as mentioned in Table \ref{tab:summary}.}
     \label{fig:GJ_logp}
\end{figure}

\begin{figure}
     \centering
     \begin{subfigure}[b]{\columnwidth}
         \centering
         \includegraphics[width=\textwidth]{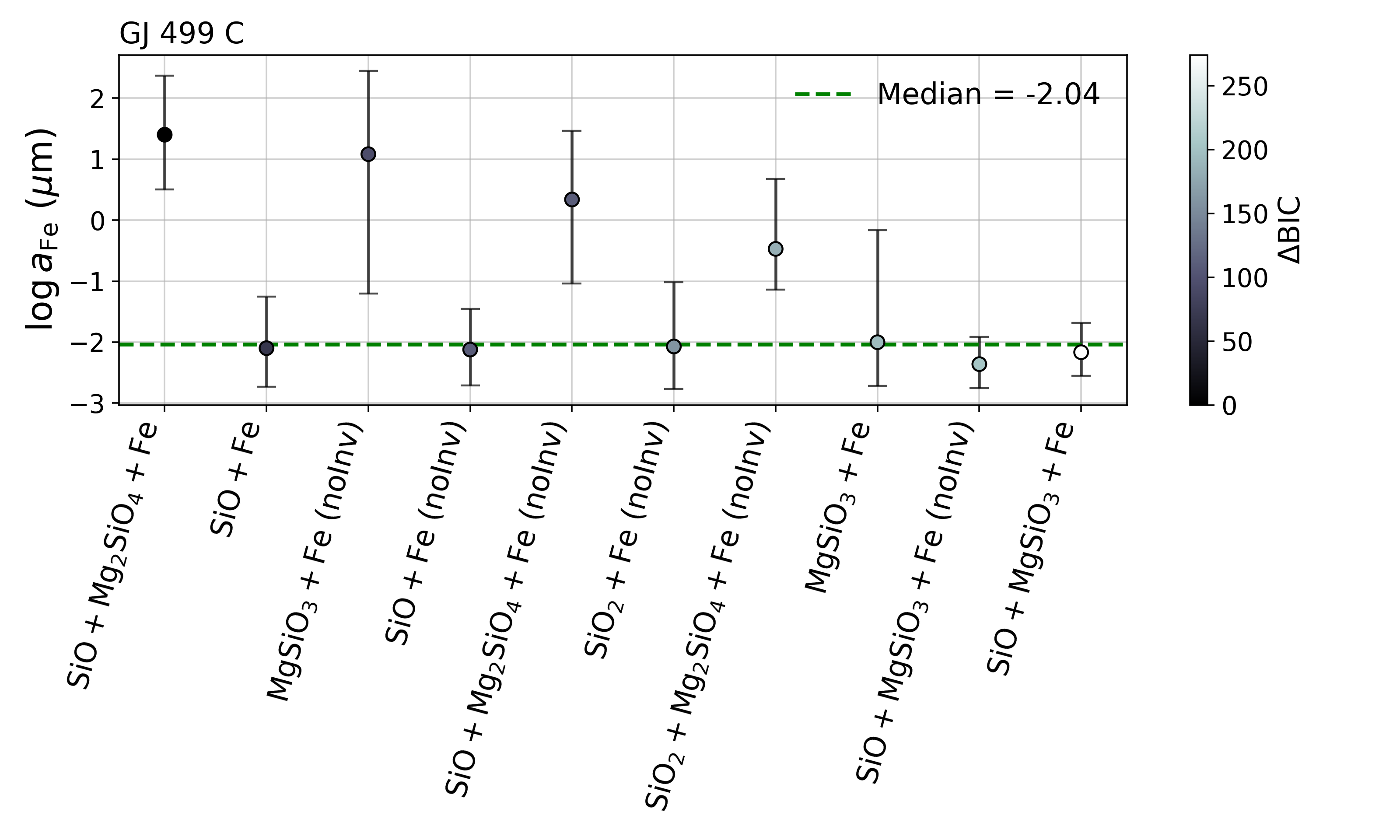}

     \end{subfigure}
     
     \begin{subfigure}[b]{\columnwidth}
         \centering
         \includegraphics[width=\textwidth]{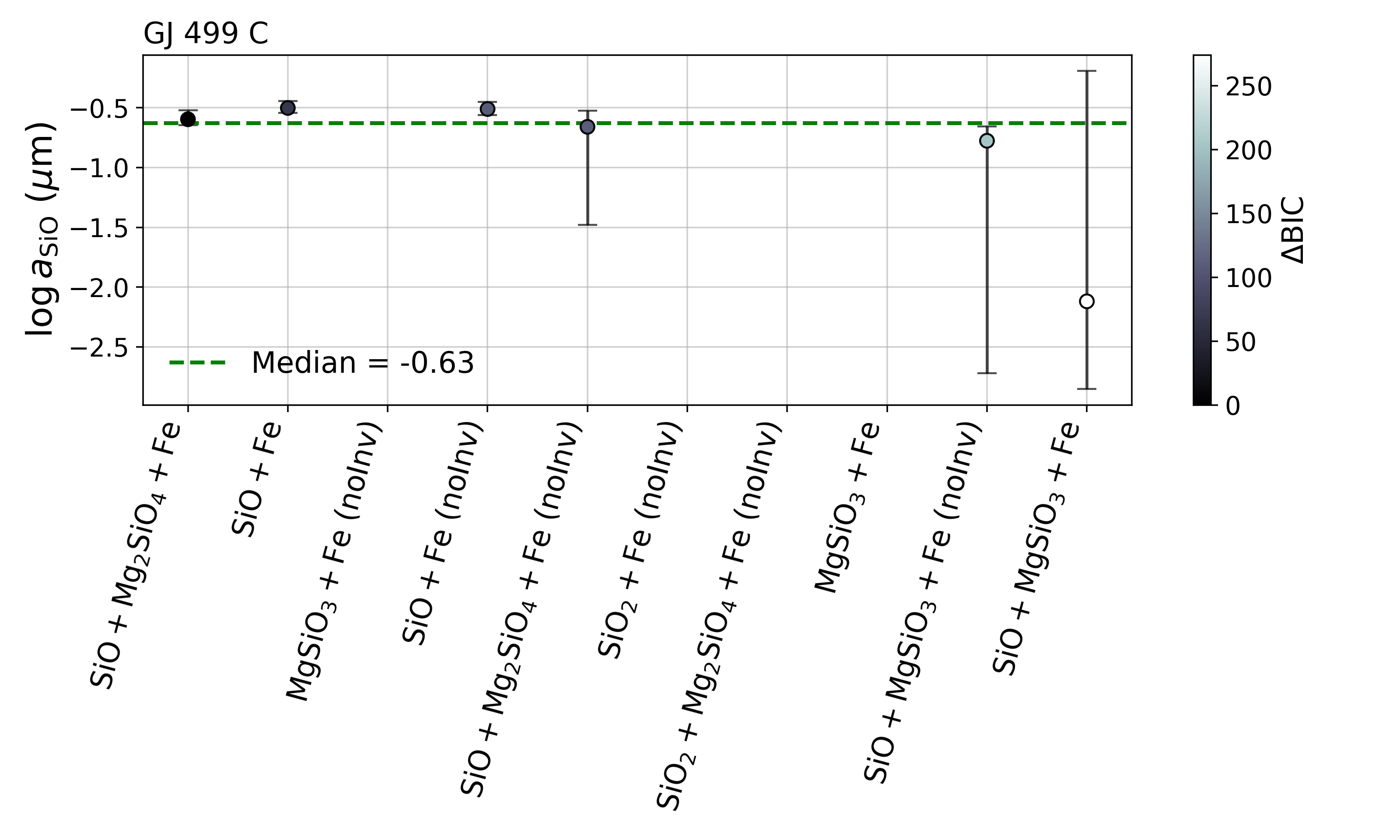}

     \end{subfigure}
     
    \begin{subfigure}[b]{\columnwidth}
         \centering
         \includegraphics[width=\textwidth]{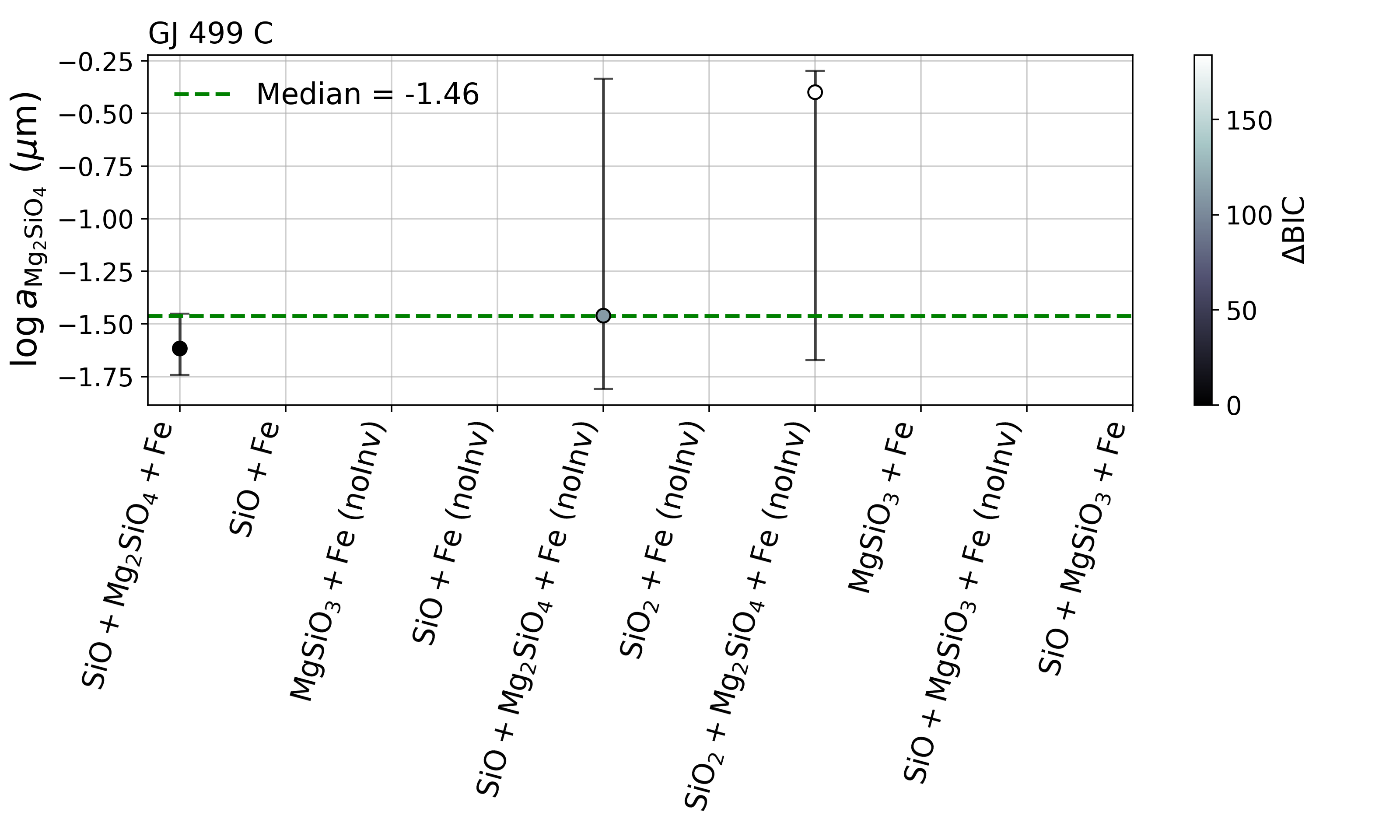}

     \end{subfigure}
     \caption{  The retrieved Fe deck (top figure), SiO slab (middle figure) and Mg$_2$SiO$_4$ slab (bottom figure) Hansen $a$ parameter of our top ten best-fitting models for GJ 499 C. Model results are colour coded based on their $\Delta$BIC values and are labeled on the x axis as mentioned in Table \ref{tab:summary}.}
     \label{fig:GJ_hansan_a}
\end{figure}

Within the atmosphere of GJ 499 C, the winning model's  median base pressure for SiO and Mg$_{2}$SiO$_{4}$ is log$P_{\mathrm{SiO}}$=-0.71
and log$P_{\mathrm{Mg_2SiO_4}}$=0.11 bar, respectively. Both clouds lie within the approximate location of the photosphere, well below the inversion. In the case of the winning model, the cloud locations are well constrained (see middle and bottom panel of Fig. \ref{fig:GJ_logp}), while other models struggle to recover these values with much larger uncertainties. The retrieved median Hansen $a$ particle effective radii for SiO is $\log a_\mathrm{SiO}$ -0.59 $\mu$m, which is well constrained and shares a similar value with the best four models (middle panel of Fig. \ref{fig:GJ_hansan_a}). The retrieved median Hansen $a$ parameter for Mg$_{2}$SiO$_{4}$ is $\log a_\mathrm{Mg_2SiO_4}$ = -1.62 $\mu$m. Only two other models have forsterite cloud with considerably different Hansen $a$ parameter and much larger uncertainties (bottom figure of Fig. \ref{fig:GJ_hansan_a}), however, they ranked poorly based on their BIC value.

For the top of the iron deck, where the cloud becomes optically thick ($\tau$=1.0), most models agreed on the retrieved value. The winning model has a median of log$P_\mathrm{Fe}$ of 0.98 bar. The average value of this retrieved parameter among the models is 1.03 bar, which is shown in the top panel of Fig. \ref{fig:GJ_logp}. The retrieved median particle effective radii of the iron deck is $\log a_\mathrm{Fe}$ = 1.40. The top panel of Fig. \ref{fig:GJ_hansan_a} shows that this retrieved parameter varies among the top ten models with the winning model retrieving the largest radii. The corner plot of the retrieved cloud and thermal profile parameters is presented in Fig. \ref{fig:GJ_corner_tp}.

\subsubsection{Contribution Functions}
\begin{figure*}
     \centering
     \begin{subfigure}[b]{0.48\textwidth}
         \centering
         \includegraphics[width=\textwidth]{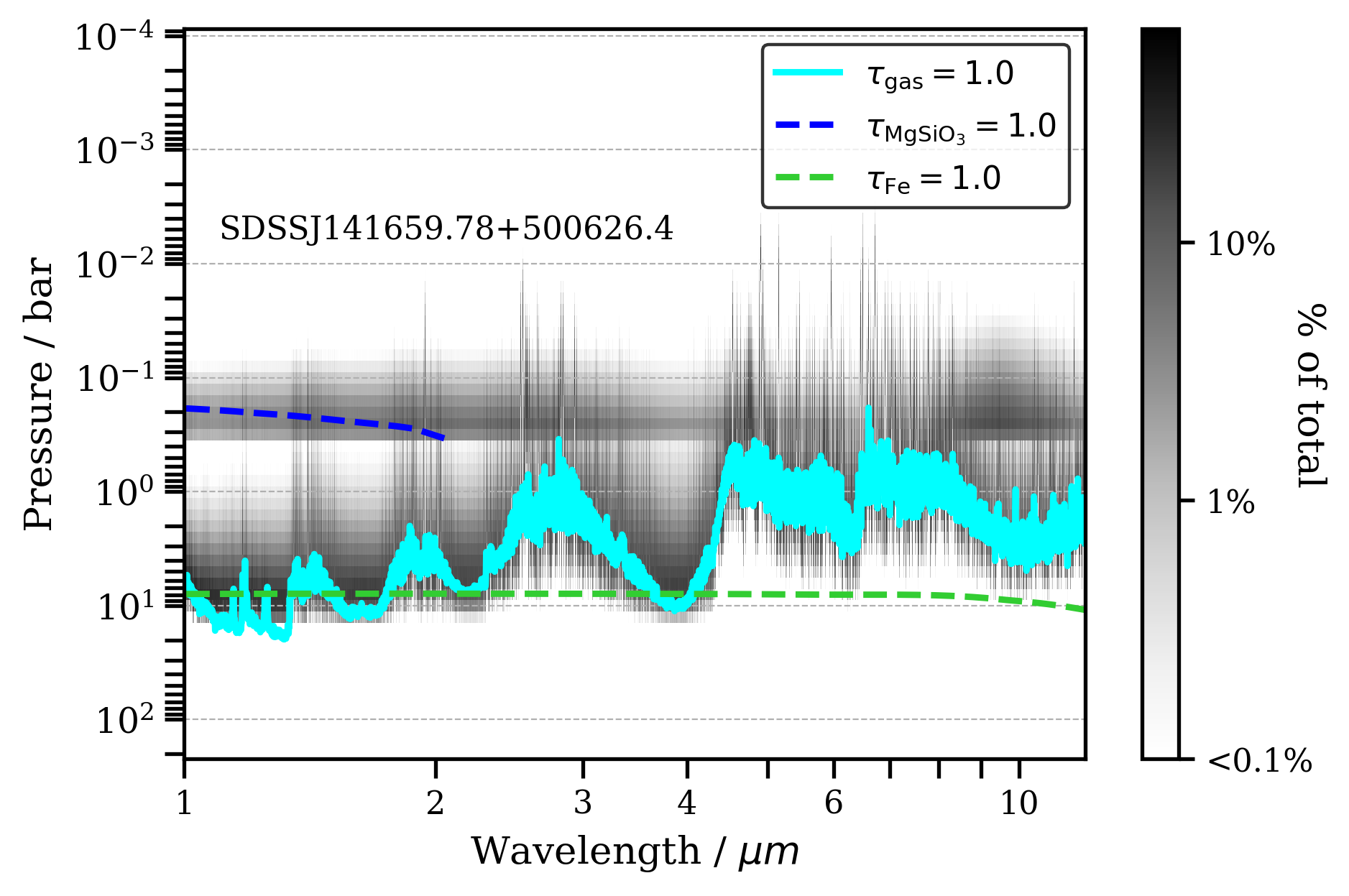}

     \end{subfigure}
     \hfill
     \begin{subfigure}[b]{0.48\textwidth}
         \centering
         \includegraphics[width=\textwidth]{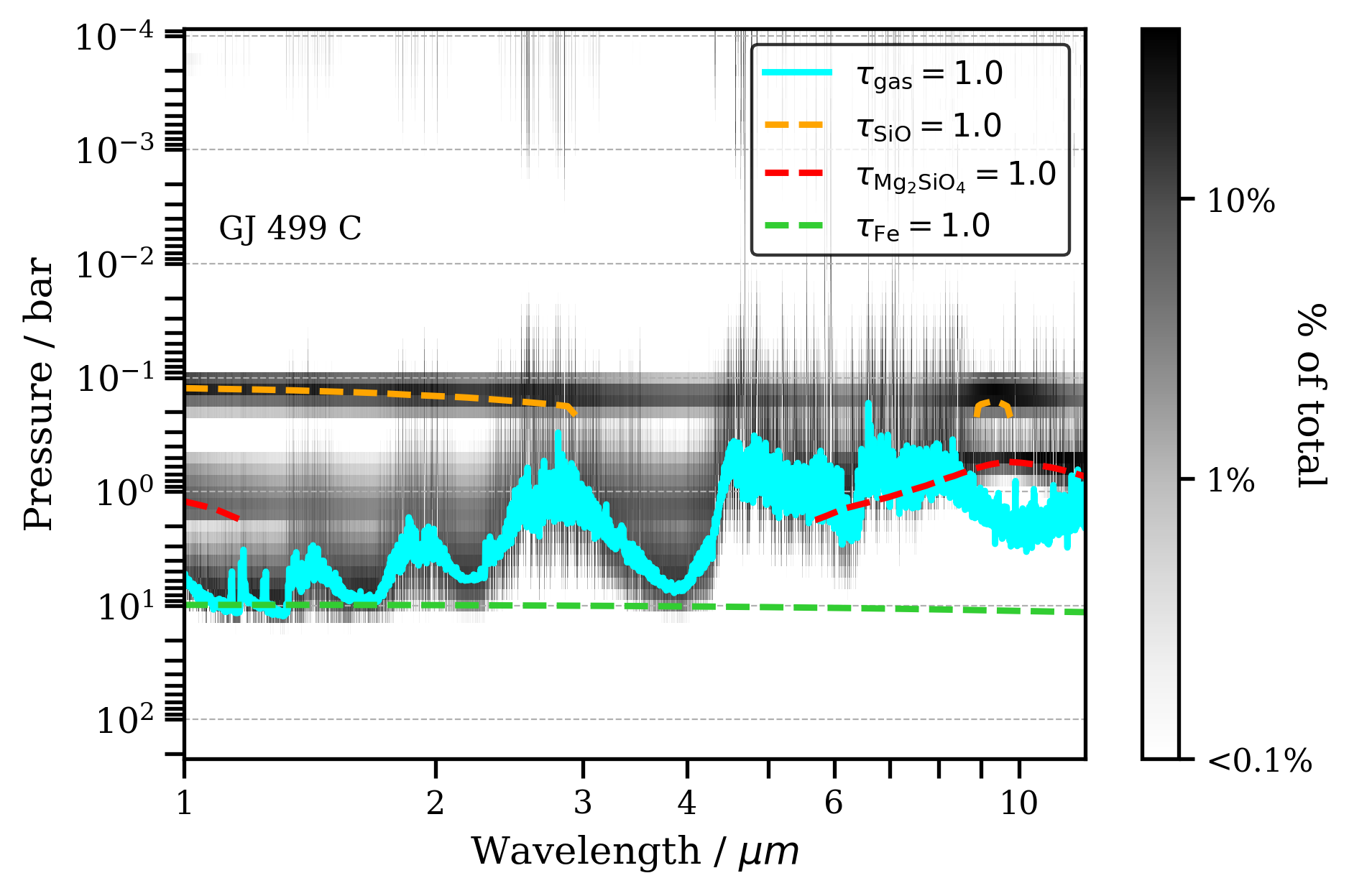}
     \end{subfigure}
     \caption{  Contribution functions of the winning retrieval model for SDSS1416 (left side) and GJ 499 C (right side). The pressure levels where the gas optical depth reaches $\tau_{\rm gas}=1$ are indicated in cyan. Dashed lines represent the $\tau$ = 1 surfaces of the individual cloud species as labelled. }
     \label{fig:contribution}
\end{figure*}

Fig. \ref{fig:contribution} shows the contribution functions of both targets based on the best fit models, which is defined as

\begin{equation}
C(\lambda, P)=\frac{B(\lambda, T(P)) \int_{P_1}^{P_2} d \tau}{\exp \int_0^{P_2} d \tau},
\end{equation}
where $B(\lambda, T(P))$ is the Planck function and $P_1$ and $P_2$ define the pressure boundaries of the layer. Darker black shading indicates a higher relative contribution to the emergent flux. It can be seen from the figures that the visible photosphere is roughly between $\log P_{\mathrm{photosphere}}$ $\sim$ -1.5 and 1.0 bar. The cyan line on the figure indicates where the gas optical depth is $\tau$ = 1.0. The blue, orange, red, and green dashed lines are the $\tau$ = 1.0 surfaces of MgSiO$_3$, SiO, Mg$_2$SiO$_4$, and Fe clouds, respectively.

In the case of SDSS1416, the emergent flux is coming from the MgSiO$_{3}$ cloud at $\sim$ 0.2 bar between 1.0 and 2.0 $\mathrm{\mu}$m. In the deeper levels at $\sim$ 8 bar the iron deck becomes an important contributor between $\sim$ 1.0 $-$ 2.2 $\mathrm{\mu}$m and  $\sim$ 3.5 $-$ 4.1 $\mathrm{\mu}$m. Over the rest of the spectral range, the iron cloud opacity is beneath the gas opacity, in agreement with previous studies of directly imaged exoplanets and L dwarfs \citep{2020A&A...640A.131M,2021MNRAS.506.1944B}.

The photosphere of GJ 499 C is similarly set by the  gas opacity across most wavelengths below which $\tau_{\mathrm{Fe}}$ lies at $\sim$ 10 bar, contributing minimally to the observed spectra below 2.2 $\mathrm{\mu}$m. The contribution function of GJ 499 C shows that the silicate clouds do contribute at shorter wavelengths (< 3.0 $\mathrm{\mu}$m) as well as around the silicate region at $\sim10$ $\mathrm{\mu}$m. While SiO is more important at shorter wavelengths  below 3.0 $\mathrm{\mu}$m, Mg$_{2}$SiO$_{4}$ contributes the most within the silicate region ($\sim$ 8.7 $\mathrm{\mu}$m $-$ 10.0 $\mathrm{\mu}$m). This phenomena has been seen in \citet{2021MNRAS.506.1944B}, where the shorter wavelengths of the L4.5 dwarf 2M2224-0158 was dominated by SiO$_{2}$ meanwhile MgSiO$_{3}$ played a more important role within the silicate feature. 

\subsection{Atmospheric Composition}
\label{sec:gasfrac}

Table \ref{tab:retrieved_abundances} shows the median and the maximum likelihood of the retrieved gas fractions and the parameters related to non-uniform abundances. Additionally, the corner plot of the relevant parameters for both targets are presented in Fig. \ref{fig:SDSS_corner_abundances} and \ref{fig:GJ_corner_abundances}. We discuss the chemical make-up properties of each target separately in the next subsections.

\begin{table*}
\caption{  Median values with 1-$\sigma$ confidence intervals for retrieved gas abundances and the non-uniform abundance parameters ($\log P_{\mathrm{ref}}$ and $\alpha$).}
\renewcommand{\arraystretch}{1.35}
\begin{tabular}{llc}
\hline
Target &            &            \\
\hline
SDSSJ141659.78+500626.4 &$\log f_{\mathrm{gas}}$  & Median  \\
\hline
&H$_2$O & $-3.50^{+0.02}_{-0.03}$  \\
&CO & $-3.08^{+0.03}_{-0.03}$  \\
&CO$_2$ & $-6.97^{+0.11}_{-0.12}$  \\
&CH$_4$ & $-5.53^{+0.05}_{-0.05}$  \\
&SiO & $-4.51^{+0.14}_{-0.14}$  \\
&TiO & $-7.58^{+0.19}_{-0.28}$  \\
&VO & $-8.73^{+0.10}_{-0.10}$  \\
&CrH & $-6.77^{+0.61}_{-0.55}$  \\
&FeH & $-6.12^{+1.47}_{-1.07}$  \\
&$\mathrm{K}+\mathrm{Na}$ & $-5.85^{+0.14}_{-0.19}$  \\
\hline
SDSSJ141659.78+500626.4  &Non-uniform abundance parameter& Median  \\
\hline

&$\log P_{\mathrm{ref,CrH}}$ & $1.03^{+0.38}_{-0.20}$  \\
&$\alpha_{\mathrm{CrH}}$ & $0.46^{+0.20}_{-0.13}$  \\
&$\log P_{\mathrm{ref,FeH}}$ & $1.44^{+0.57}_{-0.44}$ \\
&$\alpha_{\mathrm{FeH}}$ & $0.37^{+0.17}_{-0.07}$  \\

\hline
GJ 499 C &$\log f_{\mathrm{gas}}$ & Median  \\
\hline
&H$_2$O & $-3.58^{+0.03}_{-0.04}$  \\
&CO & $-3.16^{+0.05}_{-0.06}$  \\
&CO$_2$ & $-9.03^{+1.78}_{-2.12}$  \\
&CH$_4$ & $-5.73^{+0.10}_{-0.10}$  \\
&SiO & $-4.33^{+0.13}_{-0.13}$  \\
&TiO & $-9.76^{+1.10}_{-1.24}$  \\
&VO & $-10.51^{+0.86}_{-0.96}$  \\
&CrH & $-6.47^{+0.41}_{-1.24}$  \\
&FeH & $-5.37^{+0.90}_{-0.67}$ \\
&$\mathrm{K}+\mathrm{Na}$ & $-6.00^{+0.15}_{-0.23}$  \\
\hline
GJ 499 C &Non-uniform abundance parameter& Median  \\
\hline
&$\log P_{\mathrm{ref,CrH}}$ & $0.97^{+0.16}_{-0.11}$  \\
&$\alpha_{\mathrm{CrH}}$ & $0.28^{+0.24}_{-0.01}$  \\
&$\log P_{\mathrm{ref,FeH}}$ & $1.46^{+0.34}_{-0.32}$  \\
&$\alpha_{\mathrm{FeH}}$ & $0.32^{+0.14}_{-0.04}$  \\ \hline
\end{tabular}
\label{tab:retrieved_abundances}
\end{table*}

\subsubsection{Retrieved gas fractions of SDSS1416}
 
\begin{figure*}
\centering
    \begin{subfigure}{0.45\linewidth}
        \includegraphics[width=\linewidth]{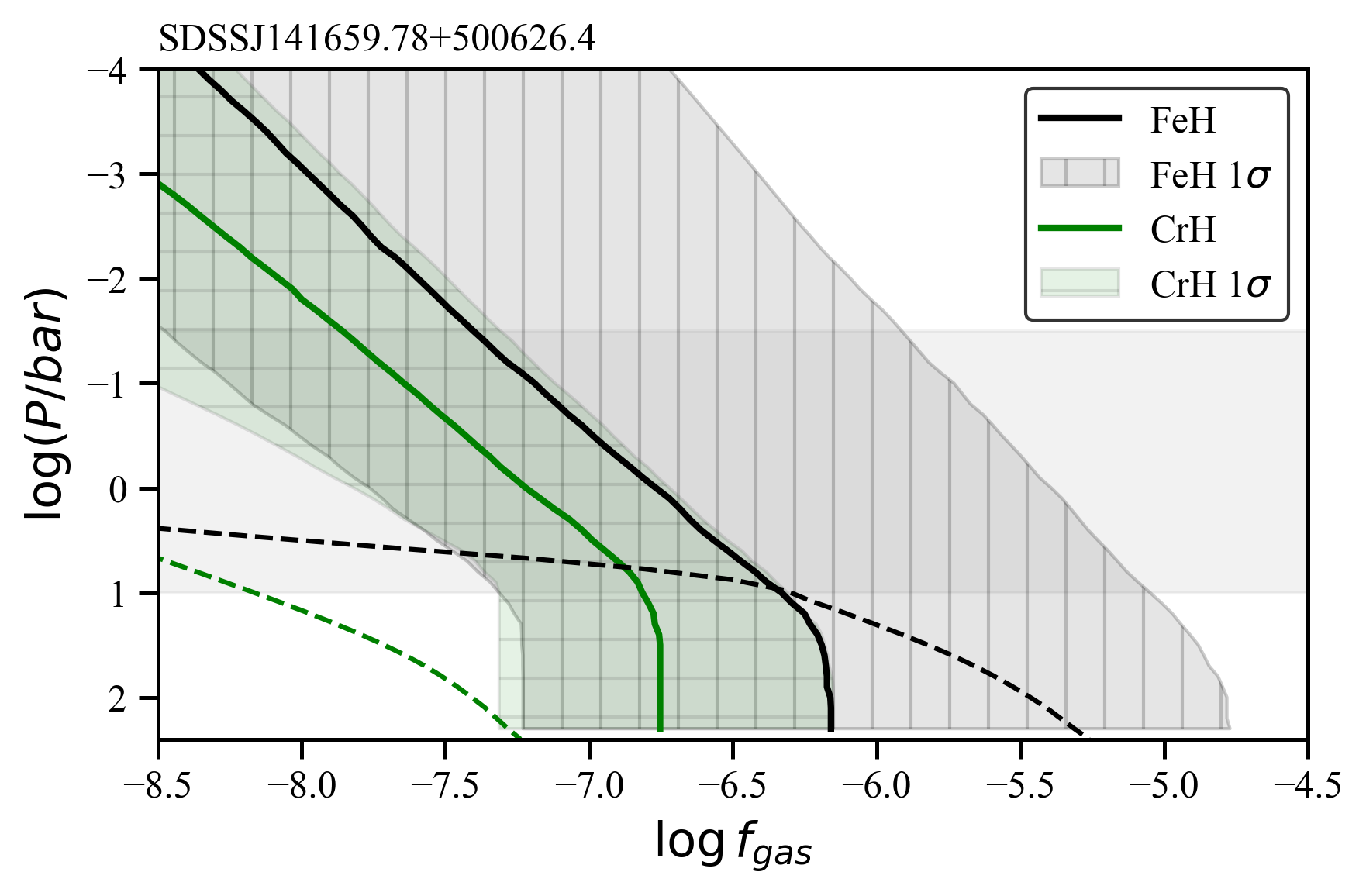}
    \end{subfigure}
\hfil
    \begin{subfigure}{0.45\linewidth}
        \includegraphics[width=\linewidth]{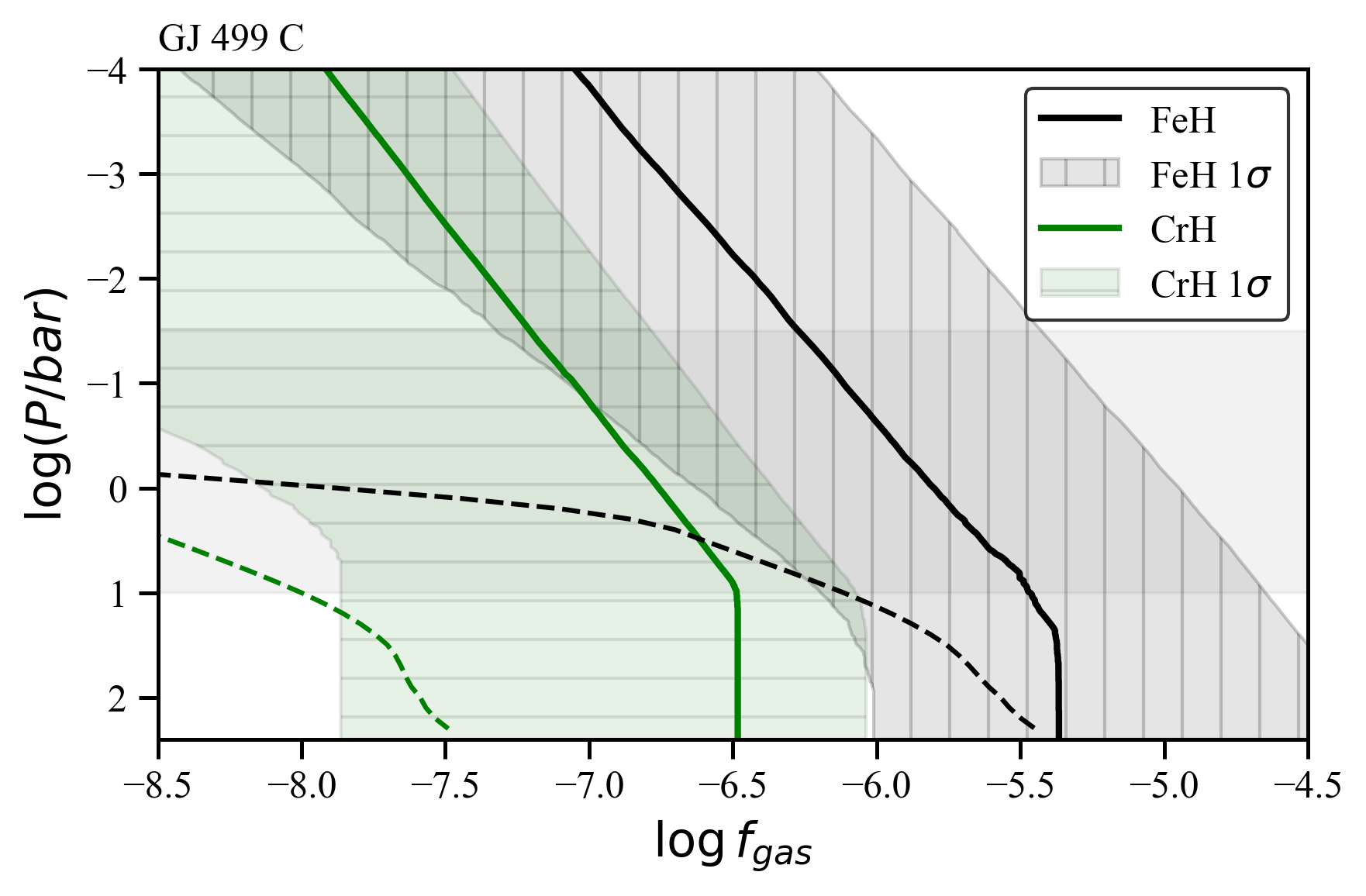}
    \end{subfigure}
    \begin{subfigure}{0.45\linewidth}
        \includegraphics[width=\linewidth]{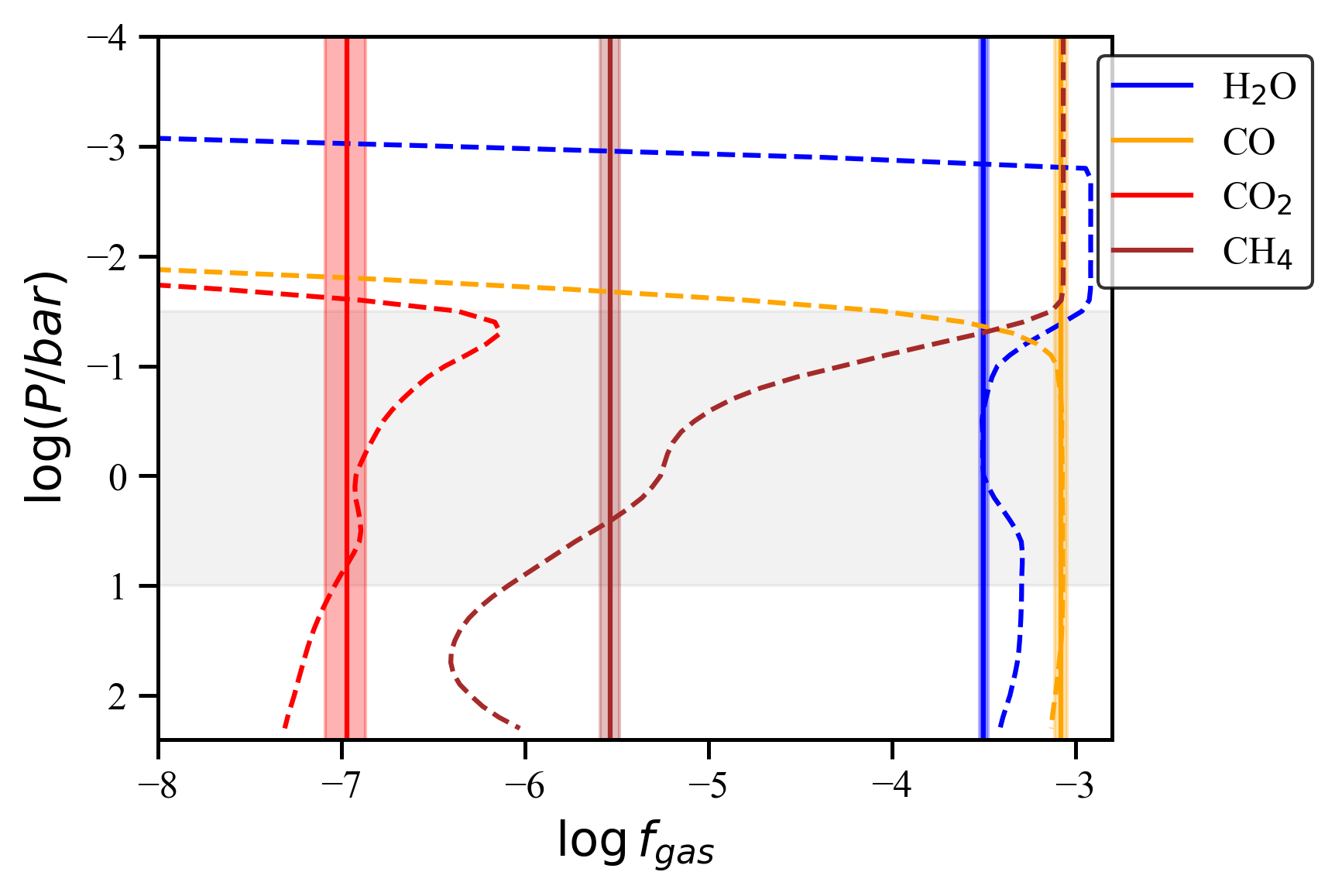}
    \end{subfigure}
\hfil
    \begin{subfigure}{0.45\linewidth}
        \includegraphics[width=\linewidth]{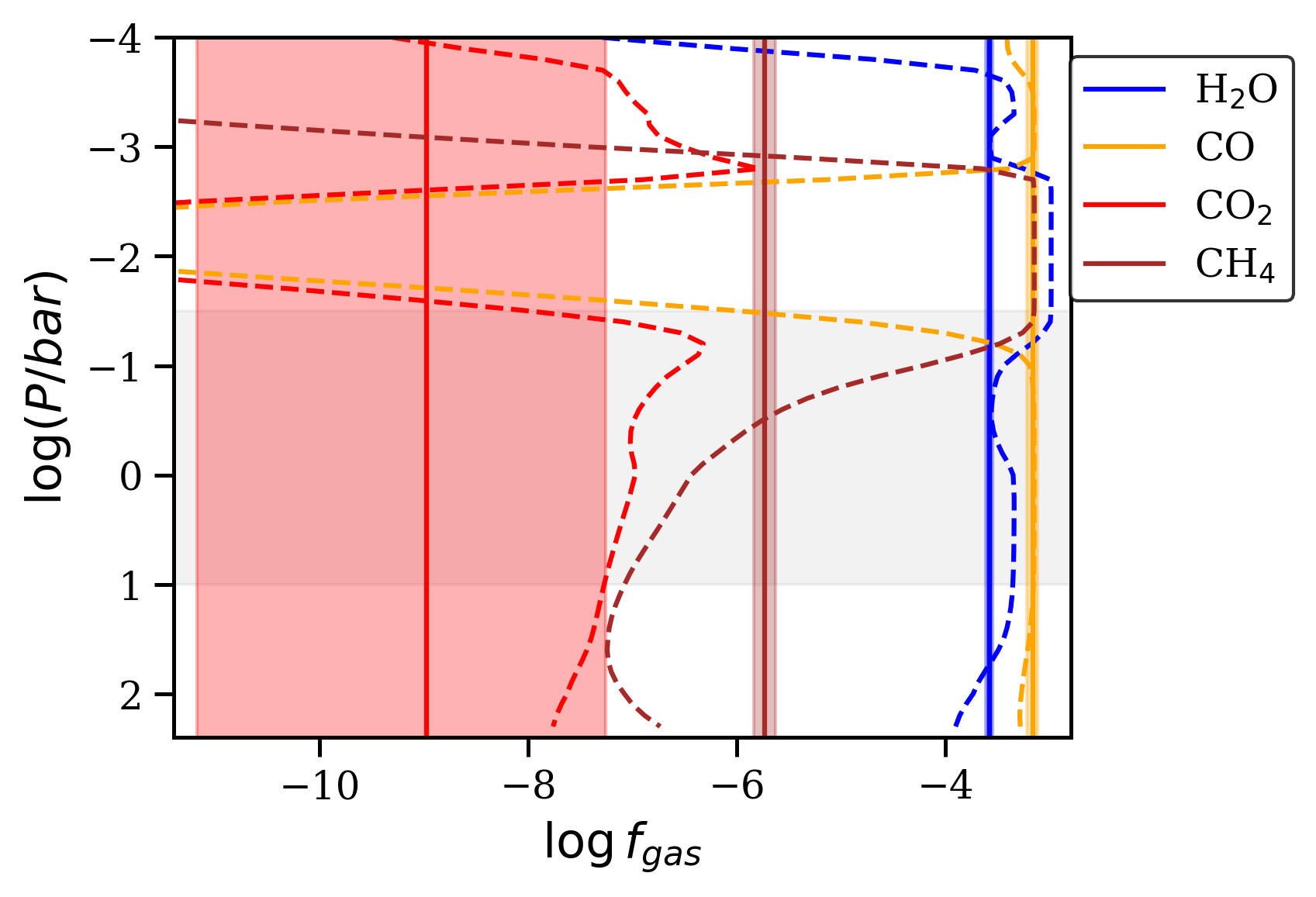}
    \end{subfigure}

    \begin{subfigure}{0.45\linewidth}
        \includegraphics[width=\linewidth]{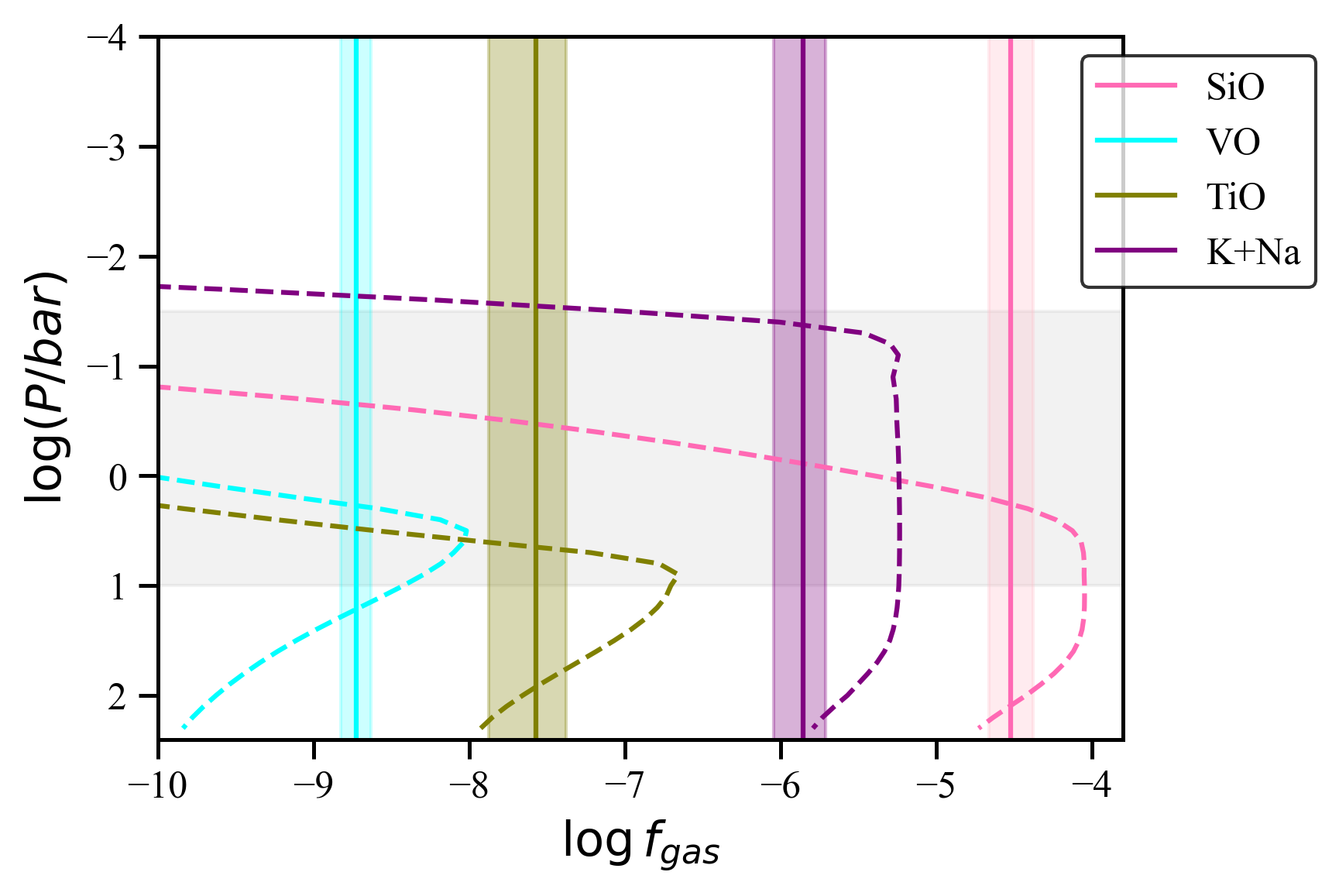}
    \end{subfigure}
\hfil
    \begin{subfigure}{0.45\linewidth}
        \includegraphics[width=\linewidth]{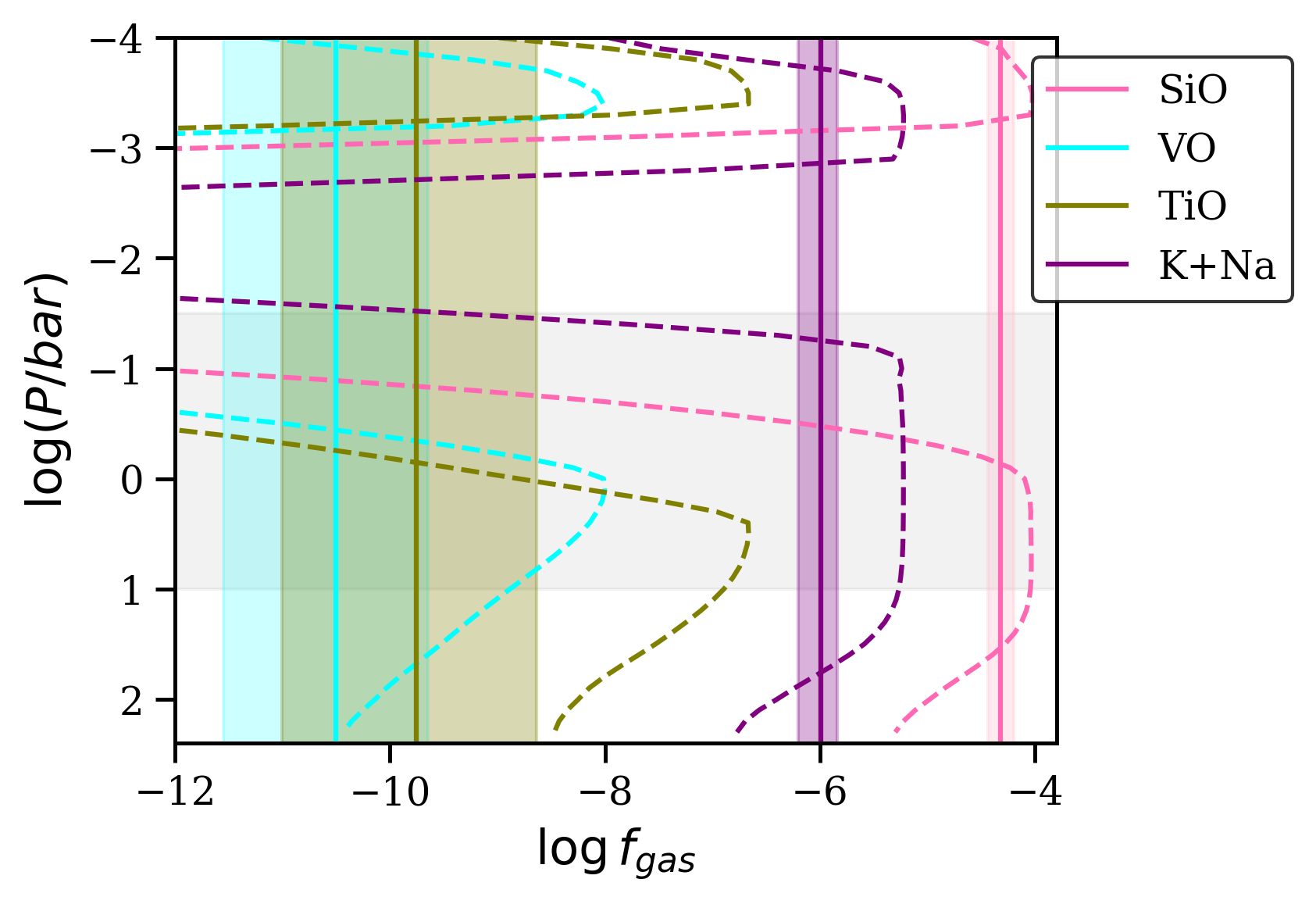}
    \end{subfigure}
\caption{  The logarithmic of the retrieved gas fractions as labeled versus the logarithmic of pressure. The shaded regions indicate the 1 $\sigma$ confidence interval. Dashed lines are equilibrium predictions based on our derived [M/H] and C/O values (see text for details). The figures on the left are the results of SDSS1416 whereas the figures on the right are the retrieved gas fractions of GJ 499 C.}
    \label{fig:gas_frac}
    \end{figure*}

The left side of Fig. \ref{fig:gas_frac} shows the winning model's median abundances of SDSS1416 along with thermochemical equilibrium predictions computed using the derived C/O ratio and metallicity [M/H] from the grid models \citep{2006ApJ...648.1181V, 2010Icar..209..602V,2012ApJ...757....5V, 2021ApJ...920...85M}. The gas fraction of FeH and CrH and their $P_{\mathrm{ref}}$ and $\alpha$ are not well constrained, which may explain the larger residuals observed in the J and H bands of the modelled spectra. The abundance of CrH is overestimated everywhere within the atmosphere by $\sim$ 1.0 dex, whereas the retrieved FeH has a $\sim$ 1.0 dex lower abundance below the photosphere. However, at $\sim$ 10 bar, where the photosphere starts, the modelled FeH and the predicted FeH are consistent. At shallower pressure levels, the modelled FeH is increasingly overestimated compared to the predicted FeH. Additionally, the rainout of FeH occurs  at the bottom of the photosphere, which is similar to our retrieved $P_{\mathrm{ref, FeH}}$.  Despite this partial agreement, the simple non-uniform abundance parameterisation adopted here — defined by a single reference pressure for rainout and its gradient — fails to reproduce the thermochemical predictions in both amplitude and shape. This suggests that a more physically motivated treatment of non-uniform gas abundances will be required to consistently match the predicted behaviour, e.g. adding a second reference pressure with its own gradient or using an exponential profile.

The other gases $-$ which are assumed to be uniform with altitude $-$ are well constrained. Our vertically constant CO mixing ratio matches the predicted CO mixing ratio in most part of the photosphere. Furthermore, in the middle of the photosphere between $\sim$ 0.1 and $\sim$ 1 bar, our retrieved H$_{2}$O  matches the predicted H$_2$O. In the other pressure levels, the predicted H$_2$O is slightly larger, by 0.1-0.5 dex. These trends with H$_2$O and CO were also visible in \citet{2021MNRAS.506.1944B}, who studied the atmosphere of a different L-dwarf. However, our retrieved thermal profile shows a cold upper atmosphere with temperatures lower than 500 K, which explains why the predicted H$_2$O and CO mixing ratio drops extremely, while the predicted CH$_4$ overtakes as the main carbon-bearing species. Similarly to our results, previous studies on L dwarfs also struggled to match the retrieved CH$_4$ and CO$_2$ mixing ratios with the predictions. While these previous findings indicated that the retrieved CO$_{2}$ is overestimated compared to the predictions (e.g., \citeauthor{2021MNRAS.506.1944B} \citeyear{2021MNRAS.506.1944B}, \citeauthor{2024ApJ...972..172P} \citeyear{2024ApJ...972..172P}), our results show that it is underestimated in the photosphere. Between $\sim$1 $-$ 10 bars our model somewhat match the CO$_2$ prediction. On the other hand, the modelled CH$_4$ only matches the prediction at one pressure level, where they cross at $\sim$ 4 bar.

As for SiO, VO, TiO and the combined K+Na fractions, the predictions indicate that they decrease significantly within the photosphere. For each species, within the photosphere, our modelled mixing ratios are lower than the prediction by $\sim$ 0.5 dex (in the case of TiO by $\sim$ 1.0 dex) below their rainout pressure levels.

\subsubsection{Retrieved gas fractions of GJ 499 C}

The right side of Fig. \ref{fig:gas_frac} represent the target's retrieved gas fractions with equilibrium predictions. The modelled FeH and CrH have a very similar issue, indicating that the existence of a temperature inversion does not change their result. The predicted CrH abundance is lower by $\sim$ 1.0 dex, whereas the retrieved FeH is somewhat similar to the prediction below the photosphere. Above that pressure level, our FeH is considerably higher than the prediction and their mismatch is increasing with shallower pressure levels.

Furthermore, we are unable to constrain CO$_{2}$, VO, and TiO, but receive very small uncertainties on CO, H$_{2}$O, SiO, and the combined Na+K fractions. Our model with the predicted CO shows almost perfect match, except where the inversion occurs, at $\sim$ 0.01 bar. We slightly underestimate the mixing ratio of H$_{2}$O by 0.1$-$0.5 dex, similarly to SDSS1416. Within the photosphere, we slightly underestimate SiO and K+Na by approximately 0.1 and 0.5 dex, respectively, except below $\sim$1 bar and $\sim$0.03 bar where the predictions show an extreme drop. Due to the inversion, their predicted mixing ratios increase in the top of the atmosphere, crossing our retrieved abundances at $\sim$ 10$^{-3}$ bar.

\section{Discussion}
\label{sec:disc}

\subsection{The non-uniform gas abundance parameterisation}
Our non-uniform gas abundance parameterisation consists of two extra parameters: a reference pressure and its gradient. This parameterisation is suitable for if the deep atmosphere is hot enough to maintain equilibrium chemistry, while the shallower pressure levels experience quenching due to vertical mixing. 

As it is shown in Fig. \ref{fig:gas_frac}, our model fails to constrain the parameters of FeH and CrH, which leads to a poorer fit in the H and J band in both targets' spectra. This suggests that a more physically motivated treatment of non-uniform gas abundances will be required to consistently match the predicted behaviour, e.g. adding a second reference pressure with its own gradient or using an exponential profile. 

Another possibility is that the vertical behaviour of these species cannot be described well by simple depletion profiles. If there are too few condensation seeds, a gas can remain in the atmosphere even when it should normally condense, which has been shown for water \citet{2024ApJ...974..190M}.

\subsection{C/O ratio and metallicity}

\begin{figure*}
     \centering
     \begin{subfigure}[b]{0.9\textwidth}
         \centering
         \includegraphics[width=\textwidth]{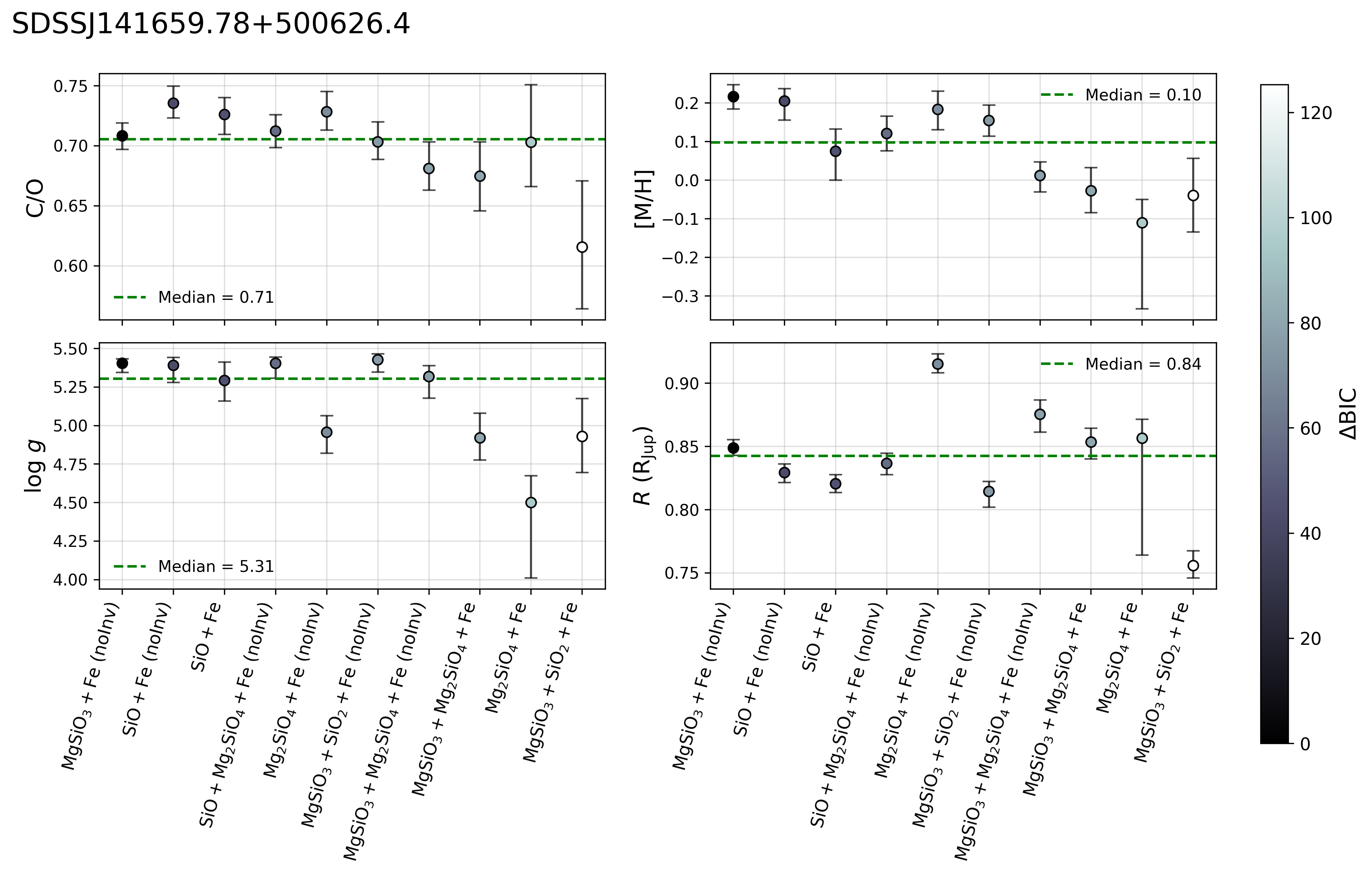}

     \end{subfigure}
     \caption{  Top row: inferred C/O ratio and [M/H] of the top ten best fitting models for SDSS1416. Bottom row: retrieved log $g$ and inferred radius of the same models. Red dotted line shows indicates the value for the winning model while green dashed line shows the mean value among the models.}
     \label{fig:SDSS_CO_MH_logg_R}
\end{figure*}

Using the formulas mentioned in Sec. \ref{sec:gasop}, we find that SDSS1416 has a super-solar C/O = 0.71$^{+0.01}_{-0.01}$. Top left panel of Fig. \ref{fig:SDSS_CO_MH_logg_R} shows that the C/O ratios inferred from the other models are consistent with, and closely clustered around, this value.
Additionally, we infer [M/H] = 0.22$^{+0.03}_{-0.03}$. The best fitting models lead to similar values regardless of cloud combination (top right panel of Fig. \ref{fig:SDSS_CO_MH_logg_R}), although this value is higher than that of the primary star ([M/H] = -0.084$\pm$0.10). From our retrieval we also obtain [C/H] = 0.31$\pm$0.03 and [O/H] = 0.19$\pm$0.03 for SDSS1416, while the host star has  [C/H] = 0.00$\pm$0.18 and [O/H] = 0.03$\pm$0.23 based on the findings by \citet{2026arXiv260709851P}. Although our estimated values for the companion are higher, the measurements remain consistent within the relatively large uncertainties of the stellar abundances. We also calculate [Fe/H] using the retrieved gas fraction of FeH, which yields [Fe/H] = -1.80$^{+1.47}_{-1.07}$. Since our FeH is not well constrained, the uncertainties on this value is large, and even the upper bound is smaller than the primary's [Fe/H], which is -0.08$\pm$0.04.

\begin{figure*}
     \centering
     \begin{subfigure}[b]{0.9\textwidth}
         \centering
         \includegraphics[width=\textwidth]{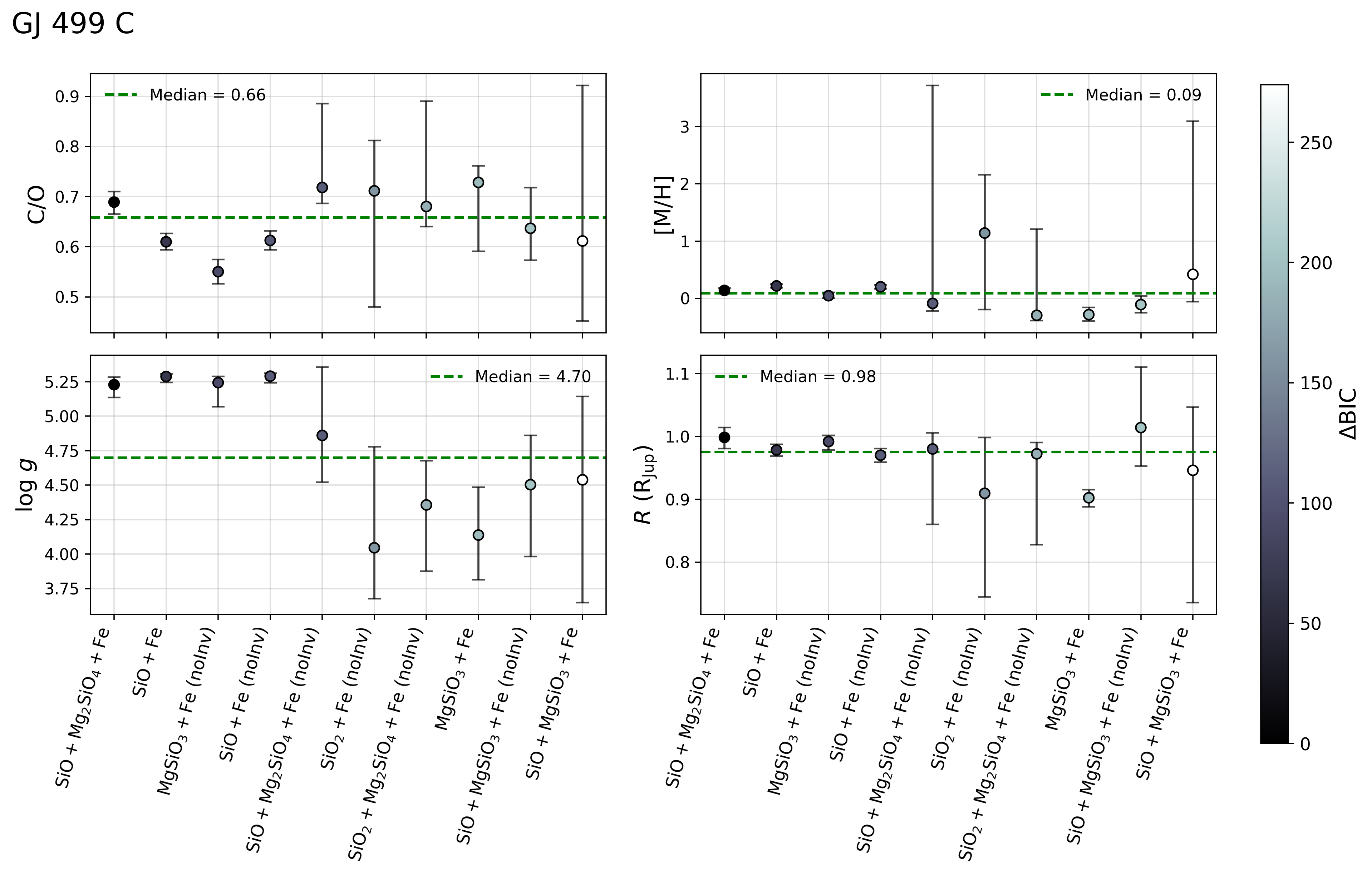}

     \end{subfigure}
     \caption{  Same as Fig. \ref{fig:SDSS_CO_MH_logg_R}, but for GJ 499 C.}
     \label{fig:GJ_CO_MH_logg_R}
\end{figure*}

Since the most important carbon-, and oxygen-bearing species are well constrained in GJ 499 C, we are able to calculate the carbon-to-oxygen ratio. We infer a slightly super-solar C/O ratio of 0.69$^{+0.02}_{-0.02}$ and a metallicity of [M/H] = 0.13$^{+0.04}_{-0.06}$ for this target. While our other retrieval models agree on the value of [M/H], there is slight disagreement in the inferred C/O ratio, ranging from 0.55 to 0.73 (Fig. \ref{fig:GJ_CO_MH_logg_R} top row). Furthermore, we estimate [C/H] = 0.22$^{+0.05}_{-0.06}$, [O/H] = 0.12$^{+0.04}_{-0.05}$, and [Fe/H] = -1.05$^{+0.90}_{-0.68}$. The latter's 1-$\sigma$ uncertainty overlaps with the error of the primary's [Fe/H] (Table \ref{tab:targets}).

Overall, both targets exhibit super-solar C/O ratios. In the case of GJ 499 C, it can be explained by gas-phase oxygen depletion due to condensation and rainout of silicate species that lie within the photosphere, whereby oxygen is sequestered into condensates and removed from the observable atmosphere \citep{1999ApJ...512..843B}.

The estimated C/O ratio of HD 125141 is 0.56$\pm$0.16  and only the upper bound of this range overlaps with the C/O ratio inferred for SDSS1416. However, \citet{2026arXiv260709851P} found an expected oxygen sequestration fraction of 20.5\% for the primary star following the theoretical framework presented by \citet{2024ApJ...963...67C}. If we assume SDSS1416 shares the same bulk C/O ratio as its primary (i.e. 0.56$\pm$0.16) and account for this oxygen sequestration, then the expected atmospheric C/O ratio should be 0.70, in excellent agreement with our estimated value of $0.71^{+0.01}_{-0.01}$. This suggests that the elevated atmospheric C/O ratio of SDSS1416 can be explained by oxygen rainout rather than a different bulk composition. However, comparing the retrieved MgSiO$_{3}$ cloud mass with the available oxygen budget shows that the MgSiO$_{3}$ slab accounts for only $\sim$1\% of the oxygen available within its pressure range. This suggests that the observed cloud may represent only part of the total oxygen sink, with additional condensate mass potentially residing lower in the atmosphere or our silicate cloud's vertical extent is larger than our retrieved value. It could also suggest that we are missing a  cloud below the enstatite cloud. Furthermore, our estimated 
C/O ratio might be incorrect due to our opacities being sampled at only $R$ = 10 000.

Similarly to \citet{2021MNRAS.506.1944B}, the C/O ratio of SDSS1416 does not depend on the model significantly, while the inferred C/O ratio of GJ 499 C exhibits a stronger model dependence. As can be seen in Fig. \ref{fig:contribution}, in the case of SDSS1416, the majority of the cloud opacity in our retrieval model occurs at pressures deeper than those probed by the dominant gas opacity across most of the spectral range considered. However, in the case of GJ 499 C, the same conclusion cannot be drawn, as the silicate clouds contribute at shallower pressure levels than the gas opacity in a significant portion of the spectra.

\subsection{Constraints on the age}
\label{sec:age}

By estimating the radius and mass of our targets, we can infer their ages using self-consistent evolutionary models. One of our goals is to assess whether these inferred ages are consistent with the independently determined ages of the primary stars. In the following subsection, we present these estimates for each BD and discuss the resulting age constraints. We note, however, that the upper mass prior is set to $80,M_{\mathrm{Jup}}$, as their luminosities are consistent with both BD and low-mass star classifications, as mentioned in Sec \ref{sec:priors}. 

\subsubsection{Age estimation of SDSS1416}
\label{sec:age_SDSS1416}

Using the known distance and the retrieved $R^{2}/D^{2}$ scale parameter, we infer a radius of $R$ = 0.85$^{+0.01}_{-0.01}$ $R_{\mathrm{Jup}}$. Then, using the newly obtained radius and the retrieved log $g$ parameter, we calculate the mass to be $M$ = 73.7$^{+4.9}_{-9.2}$ $M_{\mathrm{Jup}}$. We use these values to estimate the age of our target in Fig. \ref{fig:r_evo}. Here, we plot several Sonora Diamondback evolutionary models \citep{2024ApJ...975...59M}, showing how the radii change over time depending on various masses which are between our estimated mass range. In the case of SDSS1416, we underestimate the radius compared to the prediction of 73.3 $M_{\mathrm{Jup}}$, which is closest to the estimated median of our mass (73.7 $M_{\mathrm{Jup}}$). This phenomenon has been seen many times in L-dwarf retrieval studies regardless of observed instrument (e.g.; \citeauthor{2013ApJ...767...77S} \citeyear{2013ApJ...767...77S}, \citeauthor{2020ApJ...905...46G} \citeyear{2020ApJ...905...46G}, \citeauthor{2021MNRAS.506.1944B} \citeyear{2021MNRAS.506.1944B}, \citeauthor{2024ApJ...972..172P}, \citeyear{2024ApJ...972..172P}), and has been an issue for T- and Y-dwarfs, too \citep{2020ApJ...890..174K, 2019ApJ...877...24Z}, although these cases are not ubiquitous, indicating that the problem is not universal beyond the L-dwarf regime. Most of our best fitting retrievals on SDSS1416 predicted similarly low radius as shown on the bottom right panel of Fig. \ref{fig:SDSS_CO_MH_logg_R}.

As pointed out by \citet{2020ApJ...890..174K}, the "small radius problem" can be explained by the heterogeneous atmosphere, in which dark and weakly emitting regions, potentially associated with additional nearly gray cloud opacity at low pressure levels, reduce the inferred emitting area under homogeneous retrieval assumptions, indicating that it may be linked to limitations in the current cloud parameterisations. This may also be the reason why this issue is not common among T- and Y-dwarfs.

On the other hand, observations from the Transiting Exoplanet Survey Satellite (TESS) survey has revealed several transiting 40 $-$ 70 $M_{\mathrm{Jup}}$ and 3 $-$ 6 Gyr BDs with smaller radii than predicted for 10 Gyr objects \citep{2020AJ....160...53C, 2021AJ....161...97C}, indicating that higher mass (> 40 $M_{\mathrm{Jup}}$) L dwarfs reach their asymptotic minimum radius $\sim$ 4 - 7 Gyr earlier than predicted by evolutionary models. However, our estimated radius for SDSS1416 is significantly smaller than the asymptotic minimum radius of a BD with a similar mass. Other studies, however, showed that BD radii are larger what evolutionary models predict at 10 Gyr (e.g., see Fig. 4. of \citeauthor{2023MNRAS.519.5177C} \citeyear{2023MNRAS.519.5177C}  or Fig. 9. of \citeauthor{2024MNRAS.533.2823H} \citeyear{2024MNRAS.533.2823H}).

The problem could also arise from the high surface gravity which would imply a higher mass. High surface gravity could be plausible for  old BDs within our estimated mass \citep[e.g., see Fig. 14 of ][]{2024ApJ...975...59M}. Additionally, our models preferred higher log $g$ values with a mean above 5.15 (see bottom left of Fig. \ref{fig:SDSS_CO_MH_logg_R}).

Nevertheless, our estimated radius matches the predicted models of BDs within our mass range between the age of 4.3 and 12.9 Gyr. This is broadly consistent with the age of the primary star derived by \citet{2011A&A...530A.138C} using Bayesian isochrone fitting. They report a median age of 4.48 Gyr with a 1$\sigma$ interval of 1.77--7.28 Gyr. The lower bound of our inferred age range therefore falls within the 1$\sigma$ literature interval, while the upper bound extends beyond it. This level of agreement suggests that our retrieved radius and mass provide a physically consistent age estimate. 

The effective temperature ($T_{\mathrm{eff}}$) is derived from this flux using the retrieved radius via the Stefan–Boltzmann relation. We infer a $T_{\mathrm{eff}}$ of $\sim$ 1643 K and $\sim$ 1611 K for SDSS1416 and GJ 499 C, respectively. Fig. \ref{fig:teff_l} shows the effective temperature as a function of bolometric luminosity shown as $\log(L/L_{\odot})$. Note, that our original wavelength range only covers 0.55 -- 14.02 $\mu$m, therefore we might be missing some of the flux. To assess the impact of this limited coverage, we  calculate the effective temperatures from  Sonora Diamondback model spectra corresponding to the ones plotted on the thermal profile (Fig. \ref{fig:retrieved_TP}) within our wavelength range and within 0.5-20.0 $\mu$m, assuming $f_{\mathrm{sed}} =2$, solar metallicity and C/O ratio, and a log $g$ of $\sim$ 5.5. In all cases they only differed by $\sim 0.1 \%$, therefore we conclude that the missing flux outside our wavelength range has a negligible impact on our results, and we do not extrapolate the spectra to longer wavelengths. Our calculated $T_{\mathrm{eff}}$ for SDSS1416 is being overestimated compared to the evolutionary models which could be due to the lower-than-expected estimated radius. This could indicate that our retrieved mass is underestimated. The posterior distribution of the mass peaks near the upper prior limit (see Fig. \ref{fig:SDSS_corner_abundances}). However, the derived luminosity using the observed spectra indicates that SDSS1416 is very likely a brown dwarf as stellar evolutionary models show that stars do not cool to such low temperatures as our target \citep{1997ApJ...491..856B}, therefore, we believe that its mass is lower than 80.0 $M{_{\mathrm{Jup}}}$. A low mass star (M = 83.8 $M_{\mathrm{Jup}}$) should have a minimum luminosity of  $\log(L/L_{\odot})$ $\sim$ -3.65 based on the Sonora Diamondback models, which is much higher than our inferred luminosity for SDSS1416 (see Table \ref{tab:targets}).

As an independent consistency check of our derived fundamental parameters, we followed a standard approach that combines atmospheric and evolutionary models. We compared the observed spectral energy distribution (SED) with the cloudy Sonora Diamondback atmospheric models using the SEDA (SED analyser) package \citep{2021ApJ...920...99S,2025AAS...24531901S}. The best-fitting model spectrum was used to extend the observed SED to wavelengths not covered by the JWST spectra. Integrating the resulting hybrid SED, we found that the JWST observations capture approximately 93\% of the total bolometric luminosity of SDSS1416. The derived $\log(L/L_{\odot})$ and the radius inferred from the scaling factor of the best-fitting model spectrum are consistent within the $1\sigma$ region of the values obtained from our retrieval analysis. Interpolating this luminosity and radius onto the Sonora Diamondback evolutionary models assuming solar metallicity yields a mass of $73\pm3~M_{\mathrm{Jup}}$ and an age of $4.3\pm3.6$~Gyr, in agreement with the value derived from our retrieval analysis.

\subsubsection{Age estimation of GJ 499 C}

\begin{figure}
\centering

\begin{subfigure}{\columnwidth}
  \centering
  \includegraphics[width=\columnwidth]{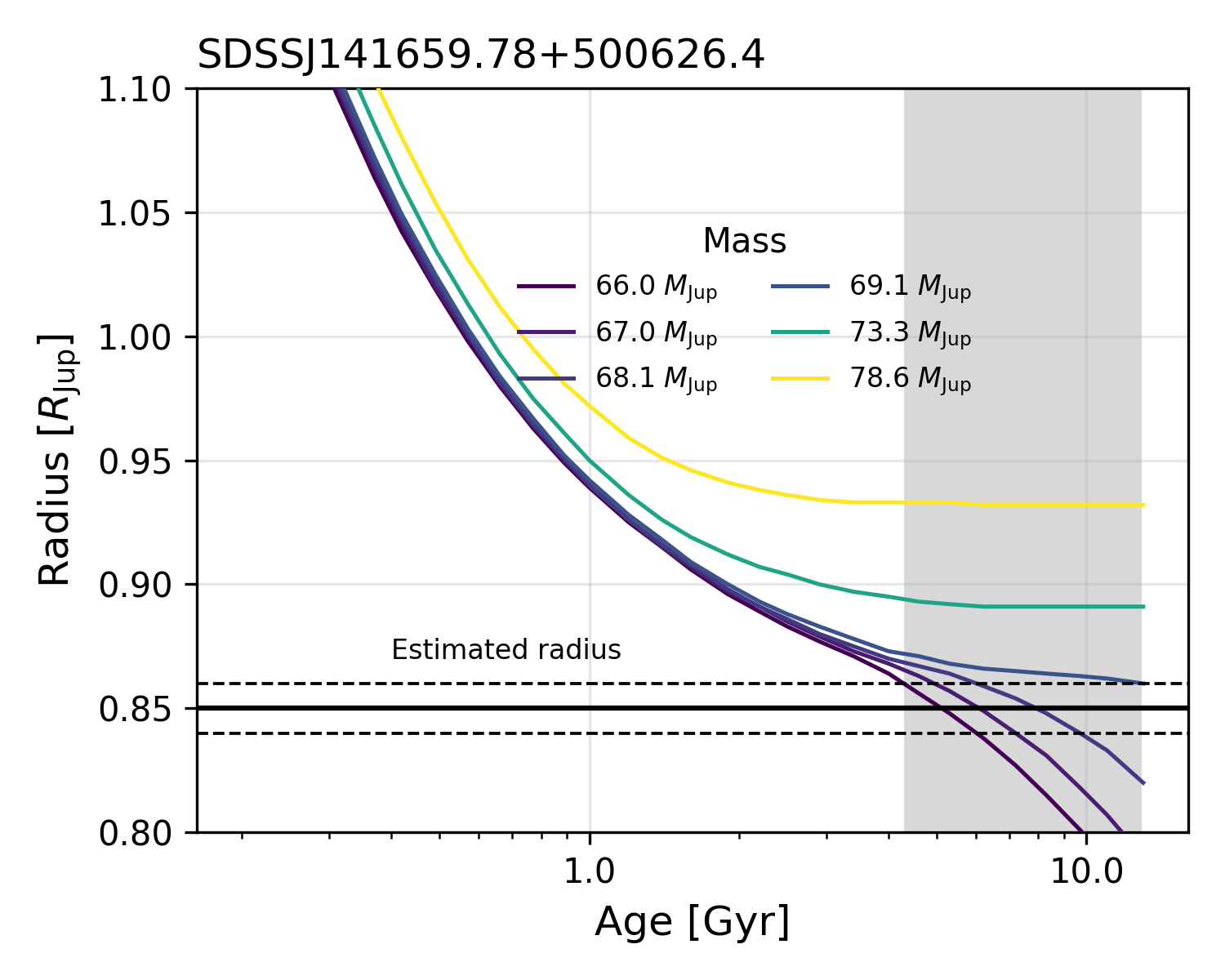}
\end{subfigure}


\begin{subfigure}{\columnwidth}
  \centering
  \includegraphics[width=\columnwidth]{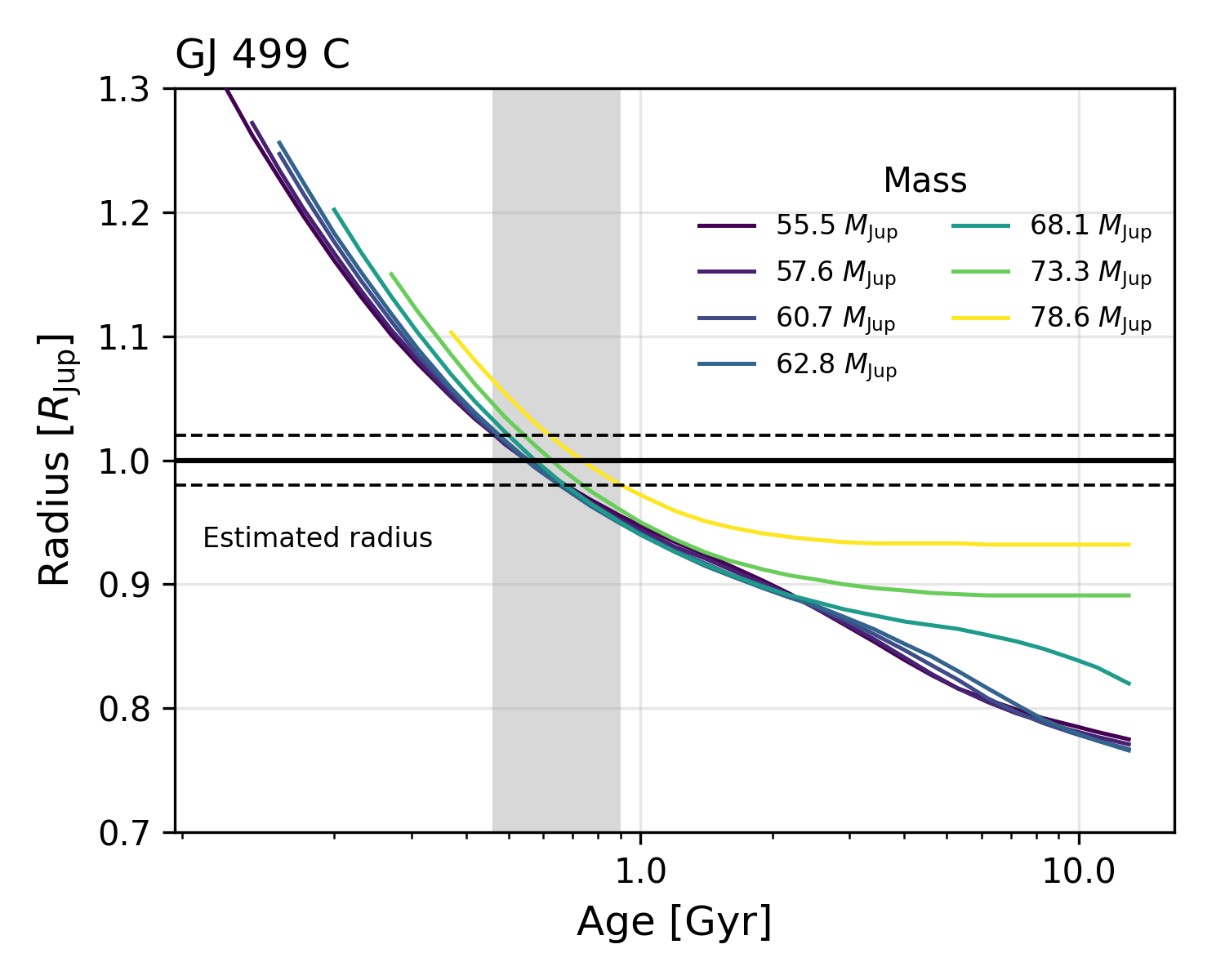}
\end{subfigure}
\caption{  The evolution of radius of brown dwarfs with masses $-$ that lie within our estimated mass range $-$ from the Sonora Diamondback cloudy evolutionary models of \citet{2024ApJ...975...59M}, assuming $f_{\mathrm{sed}}$ = 2 and solar metallicity and C/O ratio. The black horizontal lines represent our estimated radii along with the 1-$\sigma$ uncertainties of SDSS1416 (top) and GJ 499 C (bottom). The region where our inferred radius with uncertainties intersects the evolutionary models corresponding to the lowest and/or highest possible mass is highlighted in gray. This represents our estimated age of the target.}
\label{fig:r_evo}

\end{figure}

\begin{figure}
\centering

\begin{subfigure}{\columnwidth}
  \centering
  \includegraphics[width=\columnwidth]{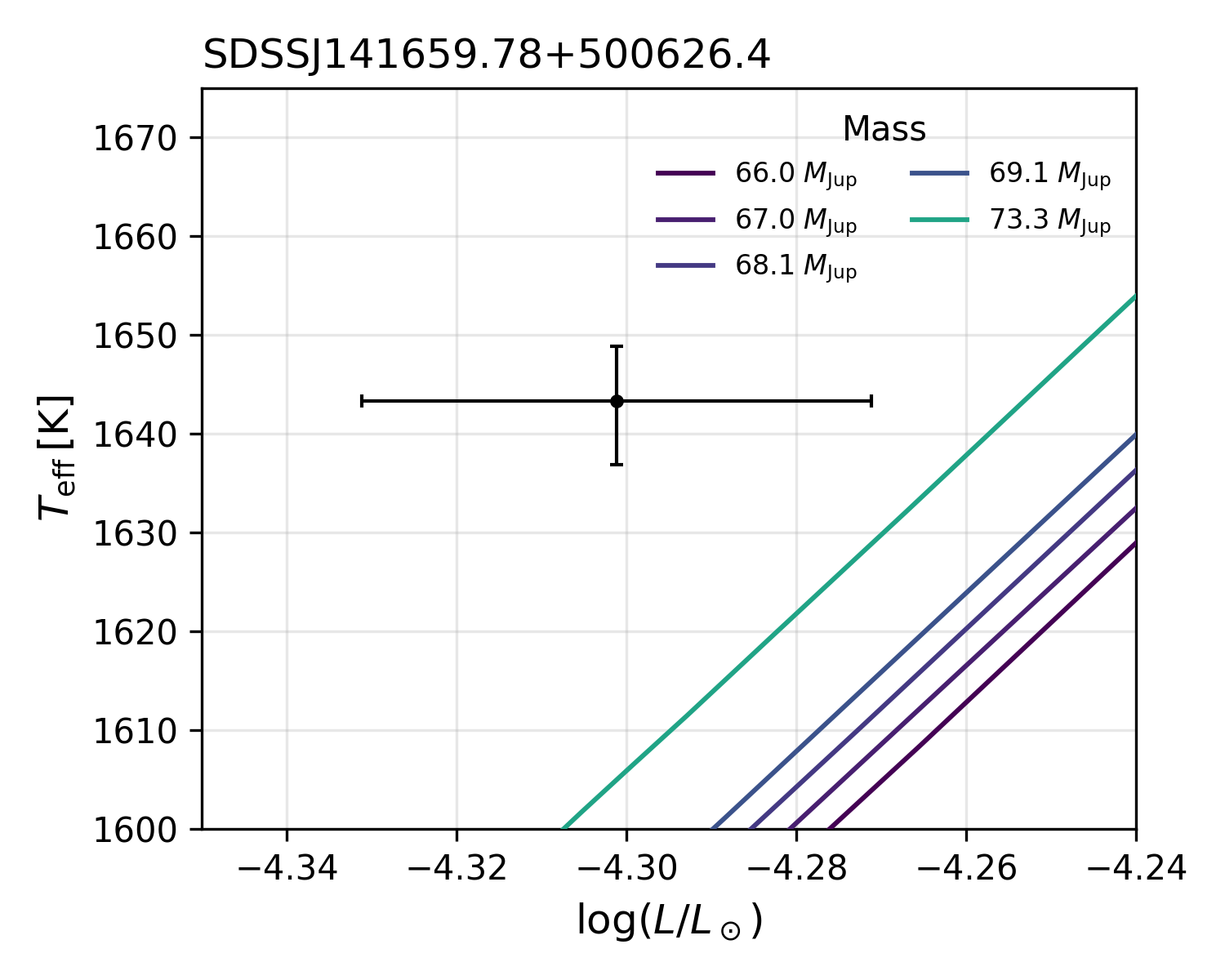}
\end{subfigure}


\begin{subfigure}{\columnwidth}
  \centering
  \includegraphics[width=\columnwidth]{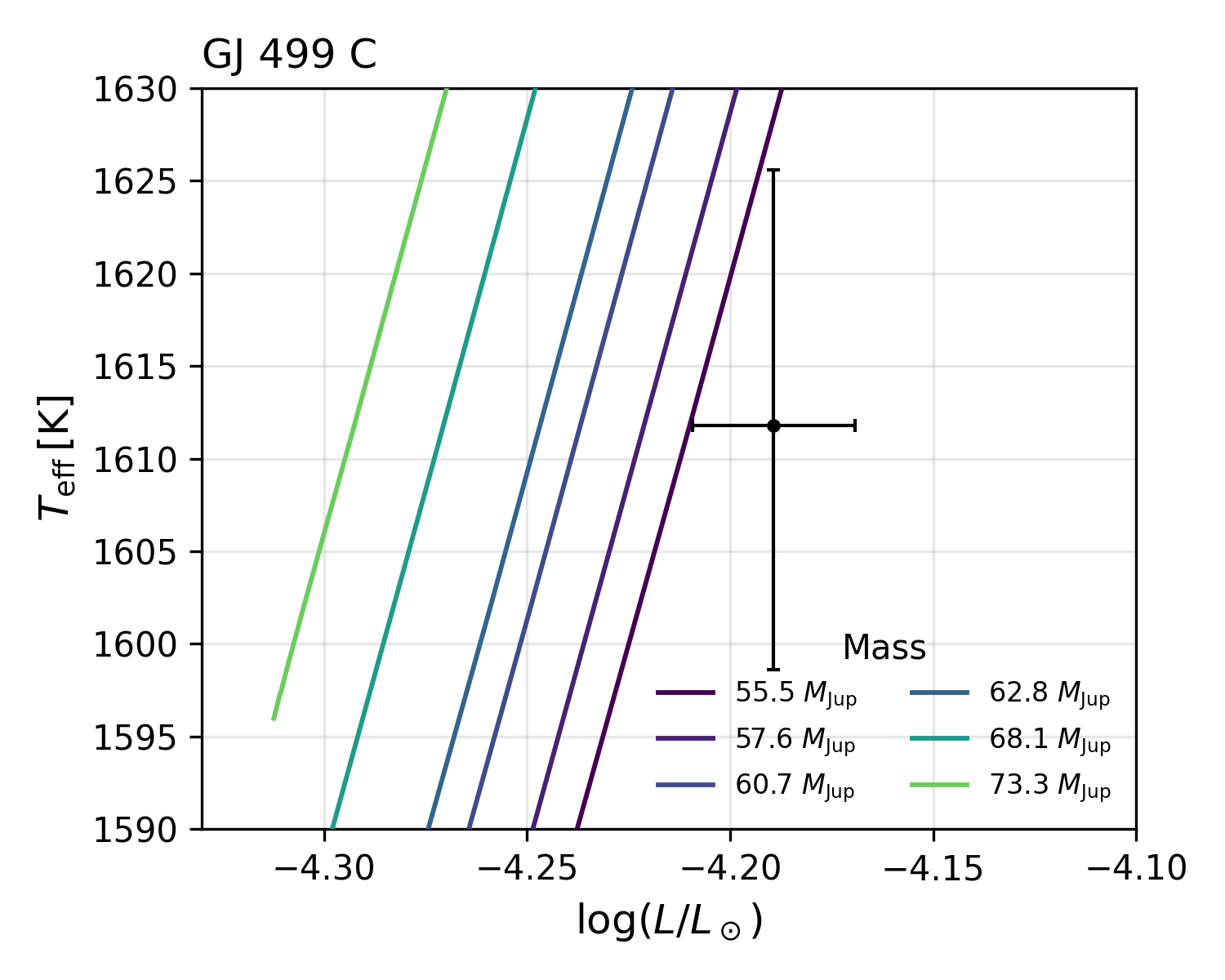}
\end{subfigure}
\caption{  The effective temperature as a function of the bolometric luminosity for SDSS1416 (top) and GJ 499 C (bottom). Black dot with error bars is the derived values from our observational data. Lines are the same Sonora Diamondback cloudy evolutionary models as in Fig. \ref{fig:r_evo}.}

\label{fig:teff_l}

\end{figure}

By using the retrieved $R^{2}/D^{2}$ scale parameter and log $g$, we calculate a radius and a mass of $R = 1.00^{+0.02}_{-0.02} R_{\mathrm{Jup}}$ and $M = 68.0^{+8.5}_{-12.7} M_{\mathrm{Jup}}$, which are consistent among the best-fitting retrieval models (see bottom row of Fig. \ref{fig:GJ_CO_MH_logg_R}). Although, the models that fit less well yields to a much lower log $g$ value. In the lower panel of Fig. \ref{fig:r_evo}, we estimate the age of the object using the Sonora Diamondback models. Unlike SDSS1416, our inferred radius and mass match the predicted radius of the median of our mass posterior. Propagating the 1$\sigma$ uncertainties on both the radius and mass yields to an age of 0.5$-$0.9 Gyr. We expect a young age due to the possible debris disk around the primary found by  \citet{2006ApJ...644L.125S} and considering the possibly lower-than-expected radius of SDSS1416, the age of GJ 499 C could be even younger. Even though we have limited prior knowledge about the GJ 499 system, the inferred mass and radius of GJ 499 C lead to physically plausible result with a young age of 0.5 -- 0.9 Gyr, supporting the reliability of our retrieval. Additionally, GJ 499 C is brighter than expected (see Fig. \ref{fig:teff_l}) indicating a relatively young system. 
Combining atmospheric models and evolutionary models, as described in Section~\ref{sec:age_SDSS1416}, we obtained that our observed SED for GJ 499 C is also $\approx93\%$ complete and derived fundamental parameters that are consistent with those derived from our retrieval analysis. The mass inferred from this approach is slightly lower ($56\pm3~M_{\mathrm{Jup}}$) but remains consistent within our retrieved $1\sigma$ uncertainties. The inferred age ($0.62\pm0.08$~Gyr) also supports the relatively young age of GJ 499 C discussed above. In addition, based on the Montreal Open Clusters and Associations (MOCA) database \citep{2018ApJ...856...23G, 2024PASP..136f3001G,2026arXiv260215695G}, the K primary has an age estimate of 0.414 $\pm$ 0.023 Gyr. Together with the known debris disk around the system \citep{2006ApJ...644L.125S} we conclude that our estimated young age is physically plausible. However, we note that our retrieved $\log g$ is higher than expected for BDs of such young ages \citep{2024ApJ...975...59M}.

\subsection{The Mg/Si ratio}

Compositional benchmark BDs provide a key test of phase equilibrium cloud formation assumptions, as they are expected to share bulk composition with their primary stars. This can be tested by for example measuring the ratio of rock-forming elements, such as magnesium (Mg) and silicon (Si). Recently, \citet{2025arXiv251210904S} found that the Mg/Si ratio of hot gaseous exoplanets are consistent with those of their primaries, and a similar consistency is expected for BDs.

The Mg/Si ratio of the host star is an important indicator of the main cloud species in the atmosphere of their BD companion. Based on the findings of \citet{2026arXiv260709851P}, the primary of SDSS1416 has a Mg/Si of 0.95 $\pm$ 0.07, hence we expect the dominant cloud species to be MgSiO$_{3}$ and possibly Mg$_{2}$SiO$_{4}$ \citep{2024ApJ...963...67C}. This is consistent with our findings as our winning model consists of an MgSiO$_{3}$ slab and an Fe deck clouds.

\citet{2021MNRAS.506.1944B} used the same atmospheric retrieval framework, \textit{Brewster}, to study the atmosphere of an L dwarf. Their winning model consisted of three clouds: MgSiO$_{3}$ slab, SiO$_{2}$ slab, and an Fe deck. They did not measure the Mg/Si ratio directly, instead they inferred a naive Mg/Si ratio from the relative amounts of their two silicate condensates as:

\begin{equation}
    \mathrm{Mg/Si} \approx \frac{N (\mathrm{MgSiO_{3}})}{N(\mathrm{MgSiO_{3}}) + N(\mathrm{SiO_{2}})}
\end{equation}

However, for SDSS1416 we only retrieved one silicate cloud, MgSiO$_{3}$. Using the same methodology yields to Mg/Si = 1.0. This is extremely close to the primary's Mg/Si ratio, which is $\sim$0.95, suggesting similar chemical composition. Additionally, we infer a column density of 2.03 $\times$ 10$^{-4}$ g cm$^{-2}$ for MgSiO$_{3}$.

As for GJ 499 C, we do not know this ratio of its primary. For this target our winning model has two silicate clouds: SiO and Mg$_{2}$SiO$_{4}$. Using the same methodology, we obtain the Mg/Si following:

\begin{equation}
    \mathrm{Mg/Si} = \frac{2 N(\mathrm{Mg_2SiO_{4}})}{N(\mathrm{SiO}) + N(\mathrm{Mg_2SiO_{4}})},
\end{equation}
where  $N$ denotes the total number of condensate molecules, obtained by vertically integrating the retrieved condensate number density profiles from the atmospheric retrieval. This yields to a very high Mg/Si ratio of $\sim$1.9. We find that around 95\% of the silicon bearing condensate moles are in Mg$_{2}$SiO$_{4}$ and the rest is in SiO. Therefore, this result is dominated by Mg$_{2}$SiO$_{4}$. Since we estimated a large ratio with a forsterite dominance, we expect that the primary has Mg/Si $\gtrsim$ 1.4 \citep{2024ApJ...963...67C}. The inferred column density for SiO is 1.7$\times$10$^{-4}$ g cm$^{2}$, whereas for Mg$_{2}$SiO$_{4}$ it is 1.1$\times$10$^{-2}$ g cm$^{2}$, $\sim$ 60 times higher.

While the absence of the primary's Mg/Si ratio in the case of GJ 499 C prevents us to do a direct comparison, for SDSS1416, the Mg/Si ratio derived from the retrieved cloud composition closely matches that of the primary star. This supports phase equilibrium cloud formation assumptions and also indicates shared bulk composition, as expected for compositional benchmark systems. On the other hand, the cloud combination found in the atmosphere of GJ 499 C is an argument against phase equilibrium (Sec. \ref{sec:sio_q}).

\subsection{Is there an SiO cloud in the atmosphere of GJ 499 C?}
\label{sec:sio_q}
An uncertainty in our study is the presence of the SiO cloud within the atmosphere of GJ 499 C. Based on phase equilibrium assumptions, BDs with our estimated Mg/Si ratio should only contain Mg$_{2}$SiO$_{4}$ clouds within their atmosphere. On the other hand, having a magnesium-to-silicon ratio of  $\sim$ 1.9 is surprisingly high and unexpected for L dwarfs \citep{2024ApJ...963...67C}. Therefore, our results might be missing silicons. 

\begin{figure*}
     \centering
     \begin{subfigure}[b]{0.48\textwidth}
         \centering
         \includegraphics[width=\textwidth]{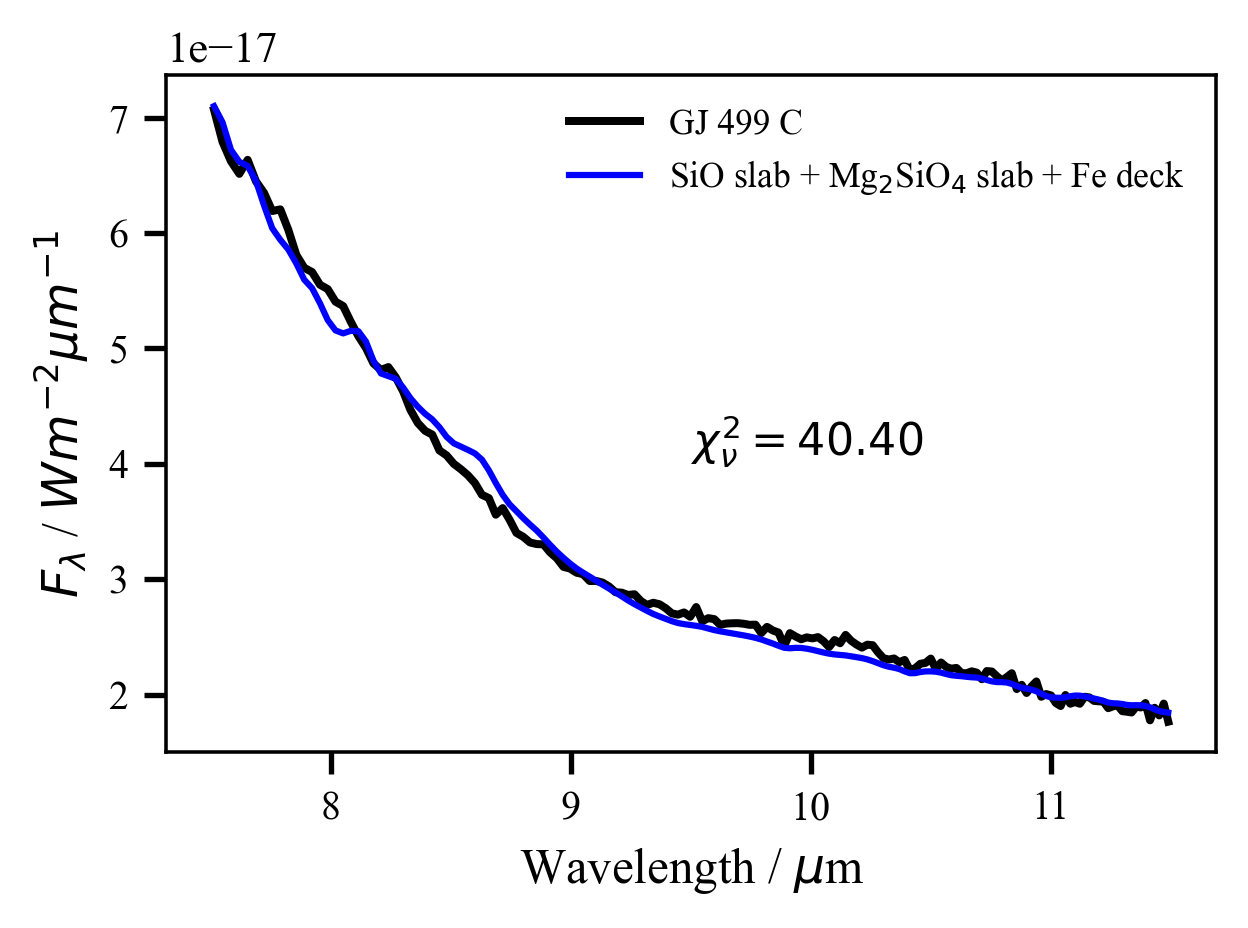}

     \end{subfigure}
     \hfill
     \begin{subfigure}[b]{0.48\textwidth}
         \centering
         \includegraphics[width=\textwidth]{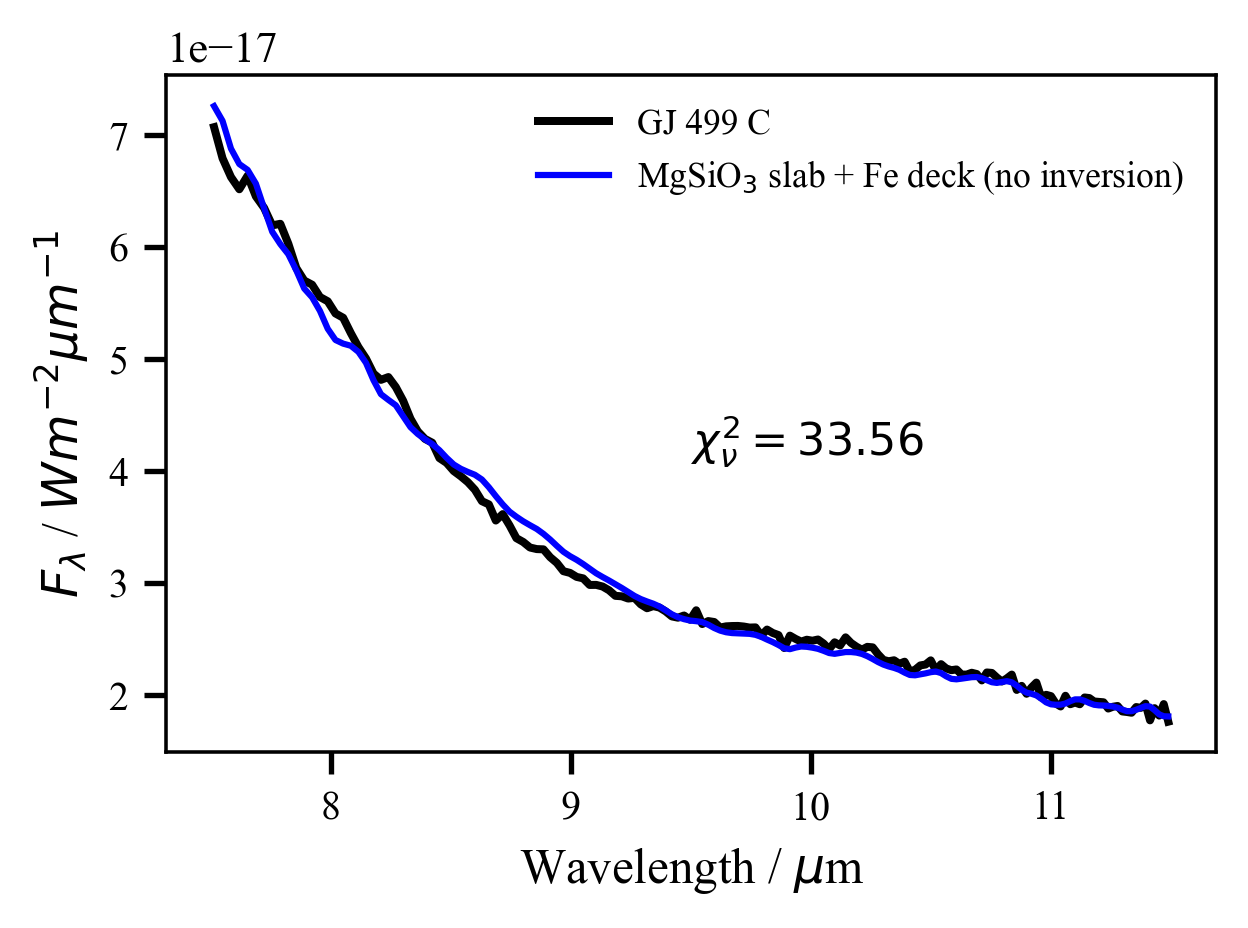}
     \end{subfigure}
     \caption{  The BIC-preferred model (left side) and the model that best fits the silicate region based on the reduced $\chi^{2}$ (right side) of GJ 499 C.}
     \label{fig:silicate_fit}
\end{figure*}

To further investigate the best cloud combinations for our targets, we analyse how well our model fits the data within the silicate index region (7.5 -- 11 $\mu$m) by calculating the reduced $\chi^{2}$, which we summarise in Table \ref{tab:summary}. While in SDSS1416, there is no significant difference between the best BIC-based model and the one that fitted the silicate region the best, GJ 499 C shows a more prominent difference (see Fig. \ref{fig:silicate_fit}). The model with SiO and Mg$_{2}$SiO$_{4}$ slabs and an iron deck has a reduced $\chi^{2}$ of 40.40 within 7.5$-$11.5 $\mu$m. The model that has MgSiO$_{3}$ slab and Fe deck clouds provides a reduced $\chi^{2}$ of 33.56 within the same silicate feature region.  While the latter reproduces the silicate feature more closely, it is strongly disfavoured by model comparison, with a $\Delta$BIC of 94.02 (see Table \ref{tab:summary}). It is worth noting that in both cases, there is a slight wavelength offset between the model and the data within the silicate feature regardless of cloud composition, even though it is expected that the forsterite's absorption efficiency occurs at slightly longer wavelength than the enstatite's absorption efficiency \citep{2009ApJ...701..571R, 2010ARA&A..48...21H, 2021ApJ...920..146L}. 
In the case of the winning model, it could be due to the large Hansen $a$ particle size of the SiO. Interestingly, \citet{2023MNRAS.523.4739S} found that high-gravity mid-L dwarfs, such as GJ 499 C, are better reproduced by small ($\lesssim$0.1 $\mu$m) amorphous MgSiO$_{3}$ or SiO grains, which could also indicate that smaller particles are preferred in our case.

In a recent study by \citet{2025A&A...703A..79M}, the authors retrieved SiO cloud for a late-L dwarf, called PSO J318.5338-22.8603, which is estimated to be 12$^{+8}_{-4}$ Myr old. Their approach involved first assessing the goodness of fit of the silicate feature and then adopting the best-fitting cloud model for the retrieval analysis of the full spectrum. However, just by analysing the silicate feature only, we cannot confidently predict the best cloud combinations. As we show in Fig. \ref{fig:contribution}, in both targets the silicate clouds contribute to the shorter wavelengths as well. This further supports our choice of choosing the BIC-preferred model. To investigate whether smaller particle radii are preferred, we rerun our winning model on GJ 499 C by initialising the Hansan $a$ particle radius at log $a$ = -1.5 $\mu$m with a tight Gaussian of 0.2 centered around it. The model with much smaller particle radii yields a worse overall fit with a $\Delta$BIC of 38.

The study by \citet{2025A&A...703A..79M} concluded that the SiO cloud in their target provides evidence for high-altitude SiO nucleation, with small condensation nuclei (< 0.1 $\mu$m) forming in the upper atmosphere and acting as cloud-seeding particles. However, in our case this scenario is unlikely, as our retrievals favour somewhat larger sub-micron SiO particles ($a \approx 0.25\mu$m) located much deeper in the atmosphere, near the photosphere ($\log P \approx -0.71$ bar). Fig. \ref{fig:GJ_logp} shows that in all of our tested models that included SiO prefer to locate the SiO cloud in or below the photosphere with large particle radii. To place these results in the context of physically motivated cloud models, we compared them with predictions using Virga cloud models \citep{2026AJ....171...98B}, assuming $T_{\mathrm{eff}}$ = 1600 K, log $g$ = 5.5, and [M/H] = 0.0. The predictions yield a mean particle radii of $\sim0.5 \mu$m even for low sedimentation efficiency of $f_{\mathrm{sed}}$ = 0.5. While larger than the very small ($<0.1 \mu$m) high-altitude SiO grain inferred by \citet{2025A&A...703A..79M}, the particles retrieved here remain smaller than those predicted by equilibrium cloud models at similar pressure levels.

While Mg$_{2}$SiO$_{4}$ and SiO would not necessarily coexist assuming phase-equilibrium chemistry \citep{2024ApJ...963...67C}, it can still be a physically plausible result. Different condensates may dominate at different pressure levels. First, Mg$_{2}$SiO$_{4}$ condenses in the deeper hotter layers, whereas higher in the atmosphere where the temperature is lower, Mg becomes depleted. Hence, these colder pressure levels might be Mg-poor but Si-bearing, leading to SiO clouds. Additionally, it has been shown that the viewing geometry correlates with the strength of the silicate absorption feature and that equatorial latitudes are more cloudy than the polar latitudes \citep{2017ApJ...842...78V, 2023ApJ...954L...6S}. The two silicate clouds, therefore, could be located at different latitudes. These explanations, however, require future modelling for validation.

\subsection{The thermal inversion in the atmosphere of GJ 499 C}
\label{sec:inv}

\begin{figure}
     \centering
     \begin{subfigure}[b]{0.4\textwidth}
         \centering
         \includegraphics[width=\textwidth]{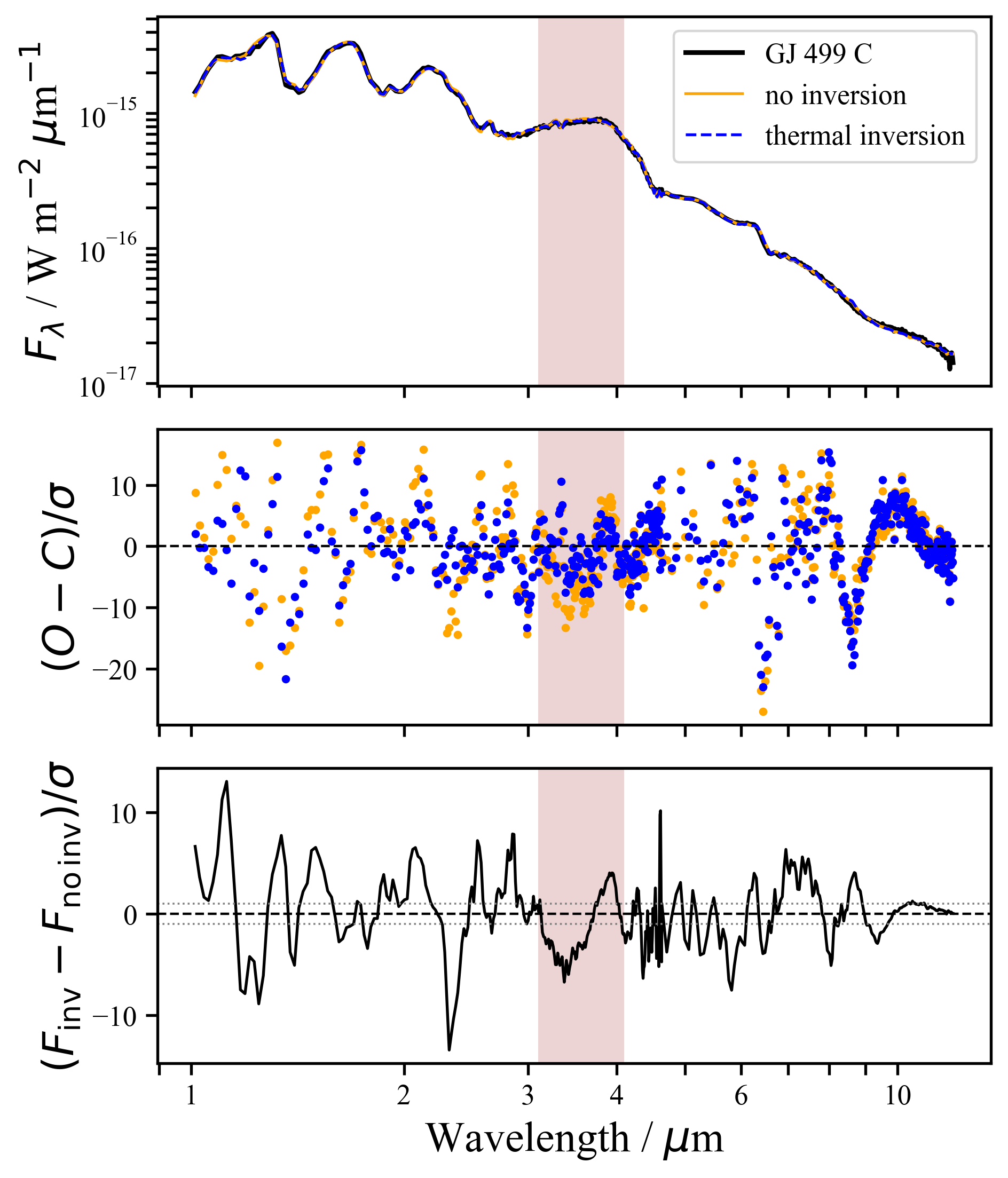}

     \end{subfigure}
     \caption{Comparison of the "winning" model of GJ 499 C in blue and the same model without thermal inversion in orange. The brown shaded area represents the wavelength region of the CH$_{4}$ opacity. Top panel: the emission spectra. Middle panel: residuals divided by the observational errors. Bottom panel: the difference between the winning thermal-inversion model and the no-inversion model, measured in units of the observational error. Dotted lines indicate wavelengths where the two models differ by more than 1$\sigma$.}
     \label{fig:inv_comp}
\end{figure}

Thermal inversion has been found in only a few brown dwarfs so far. \citet{2024Natur.628..511F} showed that the CH$_{4}$ emission feature on the spectrum of CWISEP J193518.59-154620.3 can only be fitted with a model that includes thermal inversion. In Fig. \ref{fig:inv_comp} we show the spectra comparison of our 'winning' model for GJ 499 C that include thermal inversion with one that has the same parameters in which thermal inversion is disallowed. The model without inversion provided a significantly worse fit with a $\Delta$BIC value of 110.54. However, this figure indicates that the two models have more mismatched data points within the wavelength region that is dominated by the CH$_{4}$ opacity, which is shown in light brown and is between 3.1 $-$ 4.1 $\mu$m, than elsewhere in the spectrum, suggesting that the upper atmosphere temperature structure might influence the methane-sensitive part of the spectrum.

We then repeated the inversion retrieval using a vertically variable CH$_{4}$ abundance profile. However, in this non-uniform parameterisation, we assume a constant methane abundance above a retrieved reference pressure, while below that pressure the abundance varies with depth according to a retrieved gradient. This model also retrieved thermal inversion confidently, with $\Delta$BIC = 28. This indicates that the thermal inversion is not an artefact due to our original vertically-constant CH$_{4}$ abundance.

\citet{2025A&A...702A...1N} also retrieved thermal inversion above 10 mbar with non-uniform CH$_{4}$ abundance in a T2.5 brown dwarf, called SIMP-0136. They found a correlation between the observed thermal inversion and the measured CH$_{4}$ abundance, suggesting that the methane-sensitive spectral regions probe the upper-atmosphere thermal structure.

Motivated by \citet{2025A&A...702A...1N}  we perform a statistical test using the parameters relevant to our thermal profile and CH$_{4}$ parameterisations.  We compute the Pearson correlations between our CH$_{4}$ parameters and the upper-atmosphere temperature parameters T$_1$, T$_2$ and T$_{3}$ as well as the derived temperature contrasts T$_1$-T$_2$, T$_2$-T$_3$, and T$_1$-T$_3$. This can be read in Table \ref{tab:ch4_inversion_correlations}. We find negative weak to moderate negative correlation between the reference pressure and the temperatures. The largest correlations are found for $P{\rm ref, CH_4}$, with $r \sim -0.44$ for $T_1$ and $T_1-T_2$. This means that the deeper the reference pressure is, the colder $T_1$ and $T_1-T_2$ are. However, this pressure point also shows a weak positive correlation with $r \sim 0.32$ and $r \sim 0.34$ for $T_2$ and $T_2-T_3$, respectively. This indicates that the inversion is is moderately coupled to the pressure at which the CH$_4$ abundance profile changes, although the correlations are not strong enough to suggest that the inversion is primarily driven by the CH$_4$ parameterisation. However, the lack of a stong spectral feature associated with the inversion -- as in CWISEP J193518.59-154620.3 -- suggests deeper investigation is required to firmly establish its presence and ascertain its origin. Retrievals across a larger sample may reveal if some as yet unidentified bias in the retrieval parameterisation is responsible. However, the presence of an inversion in only one of the two targets investigated here does not make such a conclusion obvious.

\begin{table}
\centering
\caption{Pearson correlation coefficients between the non-uniform CH$_4$ abundance parameters and the higher altitude temperature parameters, including the derived temperature contrasts used as proxies for the inversion strength.}
\label{tab:ch4_inversion_correlations}
\begin{tabular}{lrrrrrr}
\hline
Parameter & T$_1$ & T$_2$ & T$_3$ & T$_1$--T$_2$ & T$_2$--T$_3$ & T$_1$--T$_3$ \\
\hline
CH$_4$ 
&  0.13 & -0.03 &  0.13 &  0.10 & -0.08 &  0.08 \\

$\log P_{\rm ref, CH_4}$ 
& -0.44 &  0.32 & -0.12 & -0.44 &  0.34 & -0.37 \\

$\alpha_{\rm CH_4}$ 
& -0.24 &  0.12 & -0.03 & -0.22 &  0.12 & -0.21 \\
\hline
\end{tabular}
\end{table}

\section{Conclusions}
\label{sec:conc}
We presented the retrieval analysis of two compositional benchmark mid-L dwarfs $-$ SDSS1416 and GJ 499 C $-$ using the \textit{Brewster} retrieval framework. These objects allowed us to test the plausibility of our retrieval results and place robust constraints on their atmospheric structure, chemistry, and cloud properties using independent information from their primary stars. Both targets had large silicate indices (SDSS1416: 1.19$\pm$0.01 and GJ 499 C: 1.32$\pm$0.01), indicating the presence of silicate clouds in their photospheres.

Our findings on SDSS1416 based on its winning retrieval model are:

\begin{itemize}
    \item An MgSiO$_{3}$ slab cloud with an Fe deck cloud that match phase-equilibrium assumptions.
    \item A retrieved log $g$ of 5.40$^{+0.03}_{-0.06}$ with and estimated mass of 73.7$^{+4.9}_{-9.2}$ $M_{\mathrm{Jup}}$.
    \item An estimated radius of 0.85$^{+0.01}_{-0.01}$ $R_{\mathrm{Jup}}$. This value is lower than what the Sonora Diamondback evolutionary models predict for our median retrieved mass by $\sim$ 0.04 $R_{\mathrm{Jup}}$, however, it is still consistent with model predictions within the inferred mass range leading to an age of between $\sim$4.3$-$12.9 Gyr. This estimation roughly falls within the range of the findings of previous studies on the primary star.  
    \item A super-solar atmospheric C/O ratio of 0.71$^{+0.01}_{-0.01}$, suggesting oxygen sequestering, and a slightly metal rich [M/H] = 0.22$^{+0.03}_{-0.03}$ atmosphere. The enstatite slab cloud might not represent the total oxygen sink and there might be an additional cloud within the photosphere.
    \item A host-consistent Mg/Si ratio (BD Mg/Si$\sim$1.0 vs primary Mg/Si$\sim$0.95), consistent with the expected silicate condensation at this magnesium-to-silicon ratio.
\end{itemize}

Our results on GJ 499 C:

\begin{itemize}
    \item Thermal inversion in our retrieval model at $\sim \log P$ = - 2.2 bar.
    \item Cloud combination of an Mg$_{2}$SiO$_4$ and SiO slab with an Fe deck cloud, challenging phase-equilibrium assumptions. 
    \item A retrieved log $g$ of 5.23$^{+0.05}_{-0.09}$ with and estimated mass of 68.0$^{+8.5}_{-12.7}$ $M_{\mathrm{Jup}}$.
    \item An estimated radius of 1.00$^{+0.02}_{-0.02}$ $R_{\mathrm{Jup}}$. This value matches the predicted evolutionary models between 0.5 $-$ 0.9 Gyr, without indicating that our radius is significantly underestimated. The estimated age is consistent with the age of the primary star.
    \item A super-solar atmospheric C/O ratio of 0.69$^{+0.02}_{-0.02}$, consistent with oxygen sequestration into the retrieved photospheric silicate condensates, and a slightly metal rich [M/H] = 0.14$^{+0.04}_{-0.06}$ atmosphere.
    \item An Mg/Si ratio of 1.9.
\end{itemize}
Our results showed compositional consistency between the target SDSS1416 and its primary, considering the large uncertainties of the host star, lending confidence to the retrieval results. However, future observations and analyses on the host stars are needed to obtain their metallicity and carbon-to-oxygen ratio to confirm the shared composition and learn more about their formation. Additionally, knowing Mg/Si ratio of the primary of GJ 499 C, which we expect to be high (>1.4), will be essential for testing whether the inferred Mg-rich composition of the BD reflects its natal environment.

We also find that thermochemical equilibrium alone may not fully describe BD atmospheres. With an Mg/Si ratio of 1.9, one would expect Mg$_2$SiO$_4$ to be the dominant silicate condensate, however, our retrieval results yielded the combination of Mg$_{2}$SiO$_4$ and SiO silicate clouds, suggesting condensation behavior not fully described by simple equilibrium models.

\section*{Acknowledgements}
We thank the anonymous referee for their useful comments, which improved the quality of this paper. VK acknowledges support from a UK Science and Technology
Facilities Council studentship. This work has made use of the
University of Hertfordshire’s high-performance computing facility.
BB and FW acknowledge support from UK Research and Innovation Science and Technology Facilities Council [ST/X001091/1].
JMV acknowledges support from the European Union through the Exo-PEA ERC project (grant number 101164652). Views and opinions expressed are however those of the author(s) only and do not necessarily reflect those of the European Union or the European Research Council Executive Agency. Neither the European Union nor the granting authority can be held responsible for them. The authors declare that they have no conflicts of interest.

\section*{Data Availability}
The code and data products necessary to reproduce the results of this study are available on GitHub (https://github.com/fwang23nh/brewster\_v2) within the branch \textit{Sinking\_Silicates\_Retrieval\_Publication}. This includes the Brewster driver files and analysis notebooks used in this work.



\bibliographystyle{mnras}
\bibliography{example} 




\appendix

\section{The corner plots}

\begin{figure*}
     \centering
     \begin{subfigure}[b]{\textwidth}
         \centering
         \includegraphics[width=\textwidth]{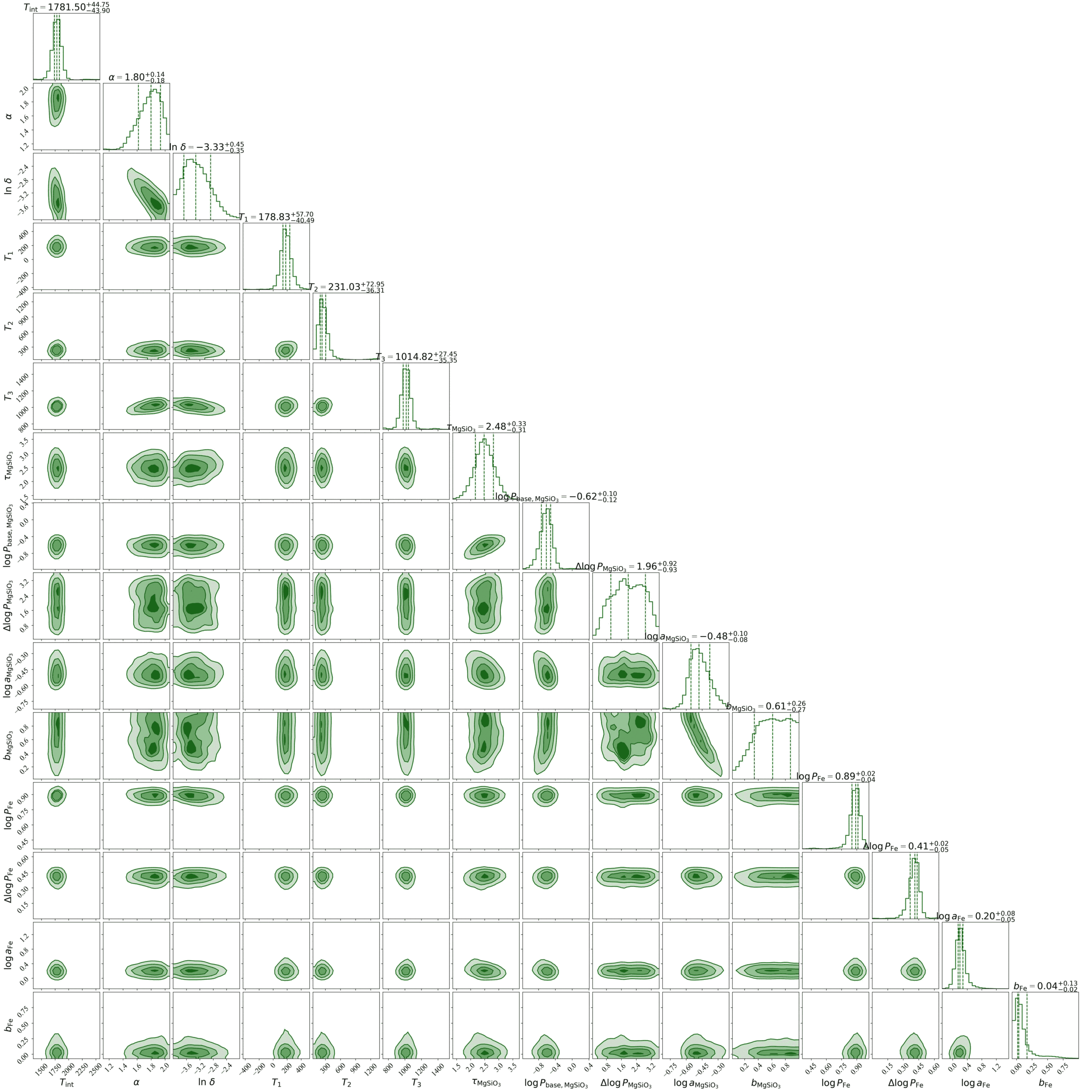}

     \end{subfigure}
     \caption{  Corner plot of the posterior distributions and parameter correlations for the thermal and cloud properties of the winning retrieval model for SDSS1416.}
     \label{fig:SDSS_corner_tp}
\end{figure*}

\begin{figure*}
     \centering
     \begin{subfigure}[b]{\textwidth}
         \centering
         \includegraphics[width=\textwidth]{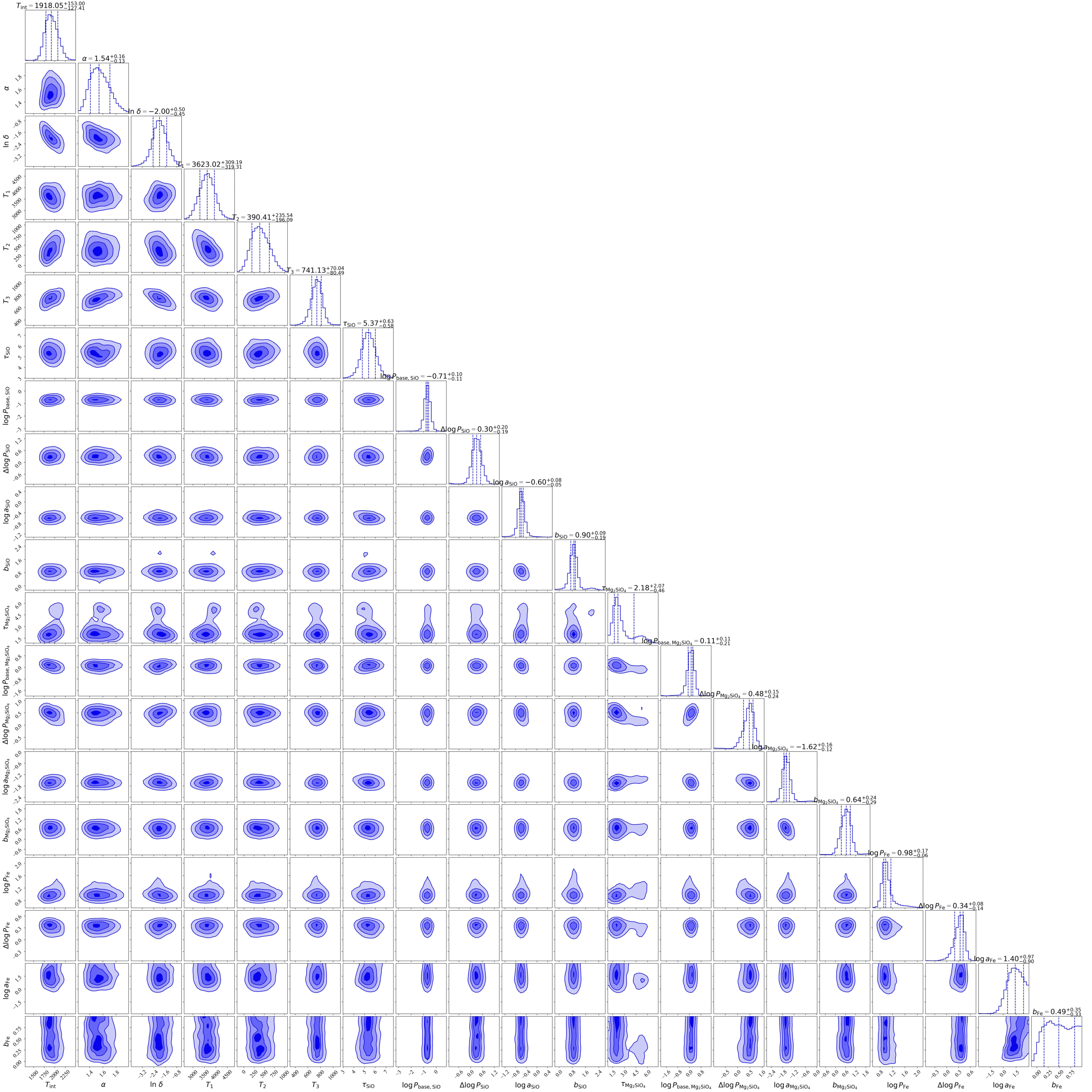}

     \end{subfigure}
     \caption{  Corner plot of the posterior distributions and parameter correlations for the thermal and cloud properties of the winning retrieval model for GJ 499 C. Dashed lines indicate the median and the 1$\sigma$ uncertainties.}
     \label{fig:GJ_corner_tp}
\end{figure*}

\begin{figure*}
     \centering
     \begin{subfigure}[b]{\textwidth}
         \centering
         \includegraphics[width=\textwidth]{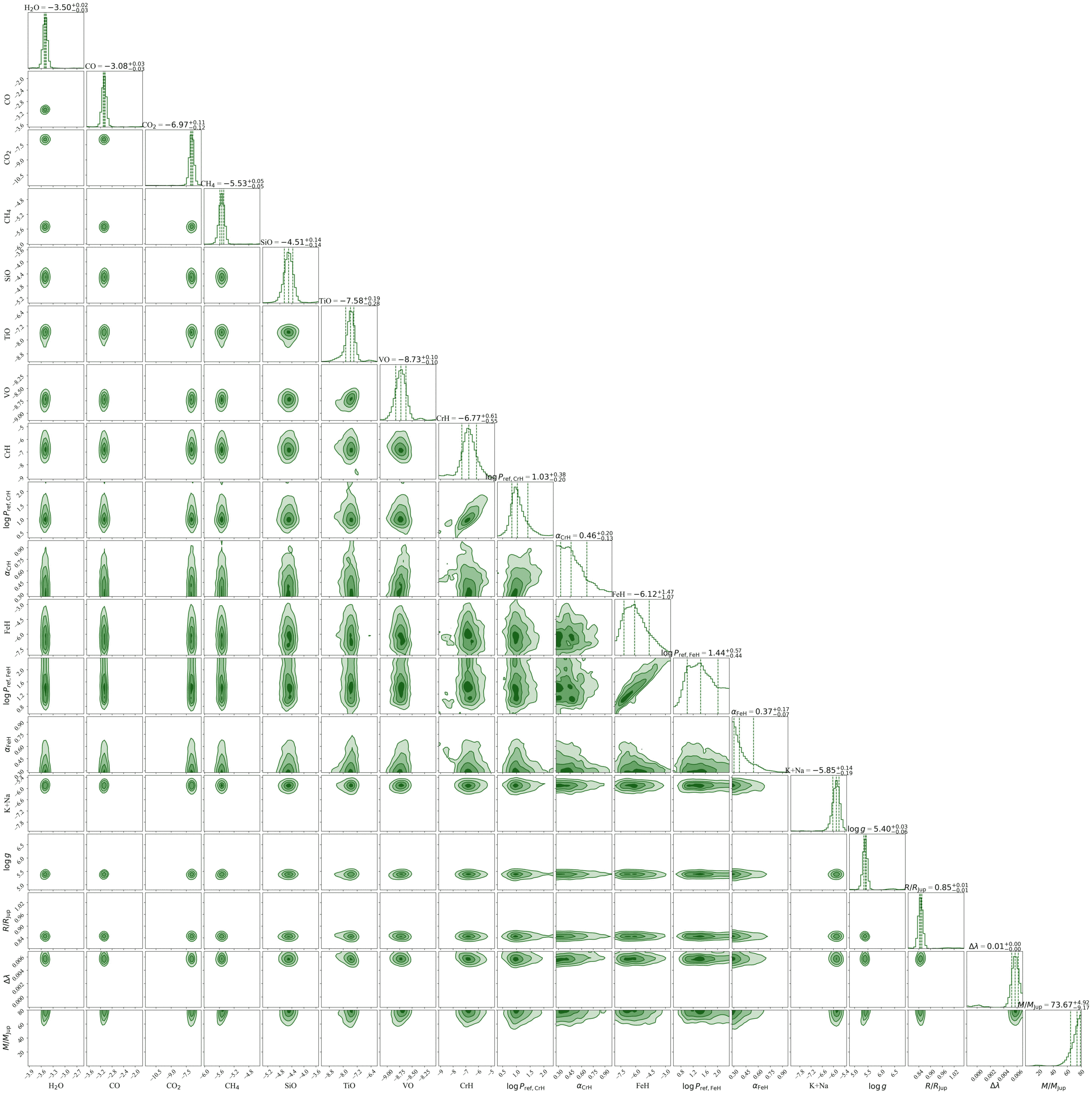}

     \end{subfigure}
     \caption{  Same as Fig. \ref{fig:SDSS_corner_tp} for SDSS1416 but for the abundances and other fundamental parameters such as log $g$, radius, and mass. }
     \label{fig:SDSS_corner_abundances}
\end{figure*}

\begin{figure*}
     \centering
     \begin{subfigure}[b]{\textwidth}
         \centering
         \includegraphics[width=\textwidth]{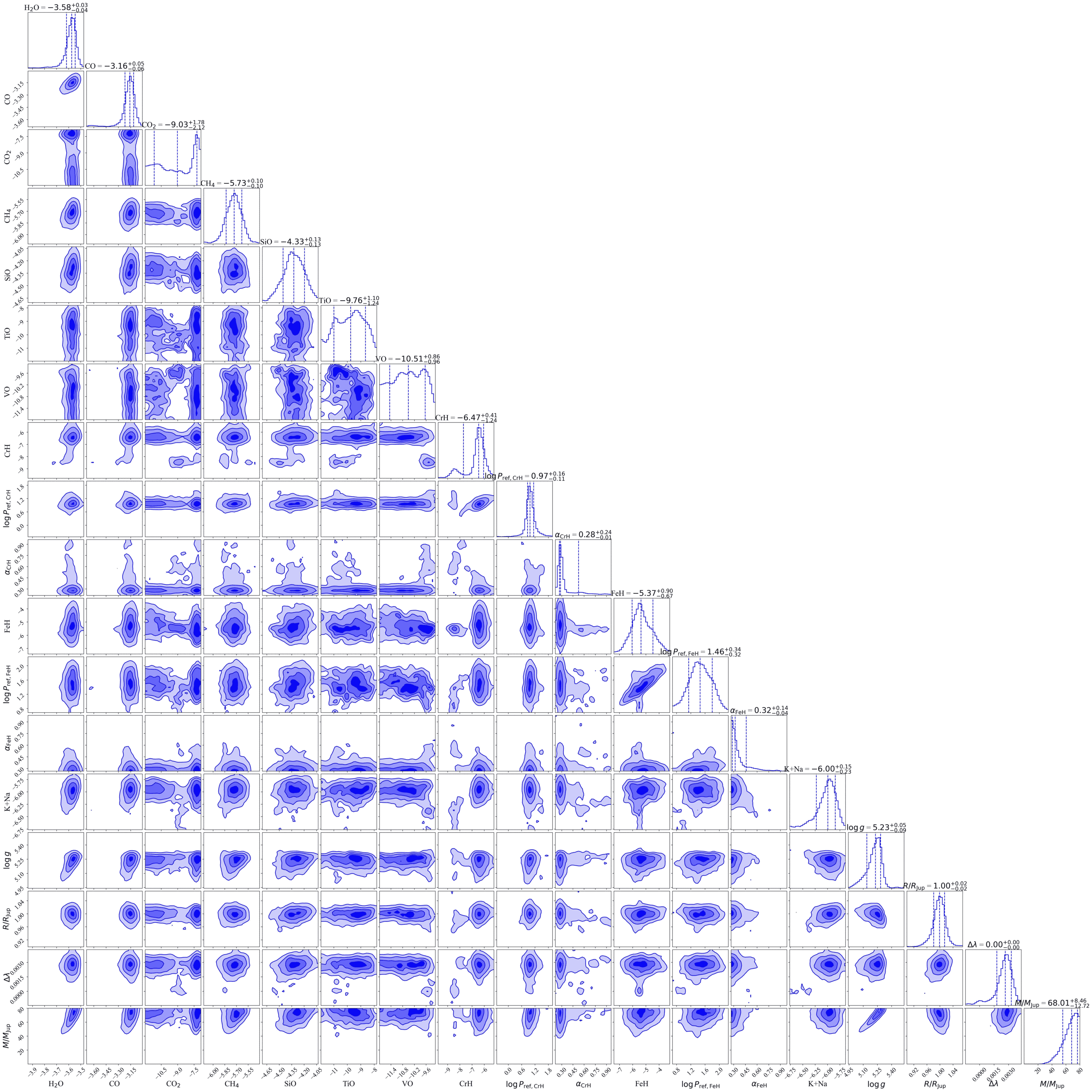}

     \end{subfigure}
     \caption{  Same as Fig. \ref{fig:GJ_corner_tp} for GJ499 C  but for the abundances and other fundamental parameters such as log $g$, radius, and mass. }
     \label{fig:GJ_corner_abundances}
\end{figure*}



\bsp	
\label{lastpage}
\end{document}